\documentclass[usenatbib,useAMS]{mn2e}
\usepackage{epsfig}
\begin{document}
\title[Testing MOG with rotation curve of galaxies]{The MOG weak field approximation and observational test of galaxy rotation curves}
\author[J. W. Moffat and S. Rahvar]
{J. W. Moffat$^{1,2}$\thanks{jmoffat@perimeterinstitute.ca}, S. Rahvar$^{1,3}$\thanks{rahvar@sharif.edu}  \\
$^1$ Perimeter Institute for Theoretical Physics, 31 Caroline St. N., Waterloo, Ontario, N2L 2Y5,Canada  \\
$^2$ Department of Physics and Astronomy, University of Waterloo,
Waterloo, Ontario N2L 3G1, Canada\\
$^3$ Department of Physics, Sharif University of Technology, PO
Box 11155-9161, Tehran, Iran }

\maketitle
\begin{abstract}
As an alternative to dark matter models, MOdified Gravity (MOG)
theory is a covariant modification of
Einstein gravity. The theory introduces two additional scalar fields
and one vector field. The aim is to explain the dynamics of
astronomical systems based only on their baryonic matter. The effect
of the vector field in the theory resembles a Lorentz force where
each particle has a charge proportional to its inertial mass. The weak 
field approximation of MOG is derived by perturbing
the metric and the fields around Minkowski space--time. We obtain an
effective gravitational potential which yields the Newtonian
attractive force plus a repulsive Yukawa force. This potential, in
addition to the Newtonian gravitational constant, $G_N$, has two
additional constant parameters $\alpha$ and $\mu$. We use the THe HI Nearby Galaxy
Survey catalogue of galaxies and fix the two parameters $\alpha$ and $\mu$ of
the theory to be $\alpha =8.89 \pm 0.34$ and $\mu =0.042 \pm
0.004~{\rm kpc}^{-1}$. We then apply the effective potential with
the fixed universal parameters to the Ursa-Major catalogue of galaxies
and obtain good fits to galaxy rotation curve data with an average
value of $\overline{\chi^2} = 1.07 $. In the fitting process, only
the stellar mass-to-light ratio $(M/L)$ of the galaxies is a free
parameter. As predictions of MOG, our derived $M/L$ is shown to be
correlated with the colour of galaxies. We also fit the Tully-Fisher
relation for galaxies. As an alternative to dark matter, introducing
an effective weak field potential for MOG opens a new window to the
astrophysical applications of the theory. \\

{ \bf keywords}: {gravitation, galaxies: Spiral, galaxies: Kinematics and Dynamics, cosmology: dark matter, cosmology: theory}\\

\end{abstract}

\section{Introduction}
Observations of the dynamics of galaxies as well as the dynamics of the
whole universe reveal that a main part of the universe's mass must
be missing or, in modern terminology, this missing mass is made of
dark matter~\cite{bertone}. One of the important astronomical
systems which is the subject of dark matter studies is galactic
scale dynamics. The observations of galaxies reveal that there
is a discrepancy between the observed dynamics and the mass inferred
from luminous matter~\cite{rubin1,rubin2}.

An alternative approach to the problem of missing mass is to replace
dark matter by a modified gravity theory.  There are various approaches to modifying gravity, such as MOdified Newtonian Dynamics (MOND) or its relativistic version 
the so-called Tensor--Vector--Scalar (TeVeS) theory~\cite{milgrom,beken}. In some modified gravity theories, the dark energy responsible for the accelerated expansion of the universe is 
described by a generic function of the Einstein--Hilbert action, as in $f(R)$-gravity models~\cite{sobouti,saffari}. Asymptotically, safe quantum gravity can produce quantum 
corrections derived from a renormalization group calculation, which can generate galaxy rotation curves compatible with observation~\cite{RGGR}. There has also been an attempt 
to interpret missing mass by introducing non-local gravity~\cite{mashhoon}.

The generally covariant MOdified Gravity (MOG) theory is a Scalar Tensor Vector Gravity
(STVG)~\cite{moffat06}.  In the following, we will study its astrophysical applications and possible predictions~\cite{moffat06}. In this theory, the dynamics of a test particle 
are given by a modified equation of motion in which, in addition to the curvature of
space--time, a massive vector field couples to the charge of a fifth force
and produces a Lorentz-type force. Since the metric field is coupled
to scalar fields and a massive vector field, the solution of the
field equations for a point mass is different from the point mass
Schwarzschild solution of General Relativity~\cite{moffat09}. The predictions of MOG for the rotation curves of galaxies have been
compared to the data~\cite{brown2006,brown2009}, using a static spherically symmetric
point mass solution derived from the field equations. The same
approach has also been applied to the dynamics of globular
clusters~\cite{moffattoth}, clusters of galaxies~\cite{brownMoffat2006} and the Bullet Cluster~\cite{brownMoffat2007}.

We obtain the weak field approximation of the MOG field equations by
perturbing the fields around Minkowski space--time. The result
for the dynamics of a test particle is that the acceleration of a
particle is driven by an effective potential. This potential
contains Newtonian gravity with a larger gravitational constant
$G_{\infty}$ and a repulsive force with a length scale $\mu^{-1}$
associated with a massive vector field. For length scales shorter
than $\mu^{-1}$, when the Yukawa exponent is of the order of unity, the
repulsive force cancels the strong attractive force and we recover
Newtonian gravity. On the other hand, at larger length scales the
repulsive vector field force becomes weaker and we obtain a
Newtonian potential with a larger Newtonian constant $G_{\infty}$.
The advantage of the weak field approximation, in contrast to the
exact spherically symmetric point particle solution of the MOG field
equations, is that we can use it to describe extended objects such as
galaxies and clusters of galaxies. The agreement of general
relativity with Solar system data is retained by the modified
acceleration law for massive test particles, derived in the weak
field approximation from the MOG field equations.

In Section (\ref{mogfield}), we review the action and the field
equations of MOG. In Section (\ref{weakapprox}), we obtain the weak
field approximation of MOG and derive an effective potential for an
arbitrary distribution of matter. In Section (\ref{rotcurve}), we
apply the results of the weak field approximation to two classes of
galaxies. For the first class, we use The HI Nearby Galaxy Survey
(THINGS\footnote{http://www.mpia-hd.mpg.de/THINGS/Overview.html})
catalogue of nearby galaxies to fix the two free parameters $\alpha$
and $\mu$ of the theory. Then we apply the effective potential predictions
to fit the observed galaxy rotation curves in the Ursa--Major catalogue of galaxies with the only free parameter,
the stellar mass-to-light ratio $M/L$. The $M/L$ used to fit the
galaxy data are shown to be correlated with the coluors of galaxies.
Based on the dynamics of galaxies derived from MOG and the
luminosities of galaxies obtained from observation, we obtain results in good 
agreement with the Tully--Fisher relation. Section (\ref{conc}) contains a summary of our results.

\section{Field equations in MOG}
\label{mogfield} We use the metric signature convention
$(+,-,-,-)$. The generic form of the MOG action is given by \cite{moffat06}:
\begin{equation}
\label{action1}
S=S_G+S_\phi+S_S+S_M.
\end{equation}
It is composed of the Einstein gravity action:
\begin{equation}
S_G=-\frac{1}{16\pi}\int\frac{1}{G}\left({\it R}+2\Lambda\right)\sqrt{-g}~d^4x,
\end{equation}
the massive vector field $\phi_\mu$ action:
\begin{eqnarray}
S_\phi&=&-\frac{1}{4\pi}\int\omega\Big[\frac{1}{4}{\bf\it B^{\mu\nu}B_{\mu\nu}}-\frac{1}{2}\mu^2\phi_\mu\phi^\mu\nonumber\\
&+&V_\phi(\phi_\mu\phi^\mu)\Big]\sqrt{-g}~d^4x,
\end{eqnarray}
and the action for the scalar fields:
\begin{eqnarray}
S_S=-&\int\frac{1}{G}\Big[\frac{1}{2}g^{\alpha\beta}\biggl(\frac{\nabla_\alpha G\nabla_\beta G}{G^2}
+\frac{\nabla_\alpha\mu\nabla_\beta\mu}{\mu^2}\biggr)\nonumber\\
&+\frac{V_G(G)}{G^2}+\frac{V_\mu(\mu)}{\mu^2}\Big]\sqrt{-g}~d^4x.
\label{scalar}
\end{eqnarray}
Here, $\nabla_\nu$ is the covariant derivative with respect to
the metric $g_{\mu\nu}$, the Faraday tensor of the vector field is defined by $B_{\mu\nu}=\partial_\mu\phi_\nu-\partial_\nu\phi_\mu$, $\omega$ is
a dimensionless coupling constant, $G$ is a scalar field representing the gravitational coupling strength and $\mu$ is a scalar field corresponding to the mass of the vector field. Moreover, $V_\phi(\phi_\mu\phi^\mu)$, $V_G(G)$ and $V_\mu(\mu)$ are the self-interaction potentials associated with the
vector field and the scalar fields, respectively.  The action for pressureless dust can be written as
\begin{equation}
S_M = \int(- \rho \sqrt{u^\mu u_\mu} - \omega
Q_5u^\mu\phi_\mu)\sqrt{-g} dx^4.
\label{SM}
\end{equation}
Here, $\rho$ is the density of matter and $Q_5$ is the fifth force source charge, which is related to the mass density, $Q_5=\kappa\rho$, where $\kappa$ is a constant.

Varying the action with respect to the fields results in the MOG
field equations. We start by varying the matter action $S_M$ with
respect to the metric, which yields the energy-momentum tensor:
\begin{equation}
T_{\mu\nu} = -\frac{2}{\sqrt{-g}}\frac{\delta (S_M + S_\phi +
S_S)}{\delta g^{\mu\nu}}.
\end{equation}
The variation of $S_M$ with respect to the vector field $\phi_\mu$ results in the fifth force current:
\begin{equation}
J^\mu = -\frac{1}{\sqrt{-g}}\frac{\delta S_M}{\delta \phi_\mu}.
\end{equation}
For the dynamics of a test particle, we adopt the action of a point
particle~\cite{moffat06,moffat09} which can also be obtained by substituting $\rho(x) = m \delta^3(x) $ in equation (\ref{SM}):
\begin{equation}
S_{tp} = \int(-m - \omega q_5 \phi_\mu u^\mu)d\tau .
\end{equation}
Here $q_5$ is the fifth force charge of the test particle, which is related to the
inertial mass of the particle, $q_5 =\kappa m$. Varying this action
results in the equation of motion of a test particle with an extra force
on the right-hand side:
\begin{equation}
\frac{du^\mu}{d\tau} + \Gamma^\mu{}_{\alpha\beta}u^\alpha u^\beta =
\omega\kappa B^\mu{}_\alpha u^\alpha. \label{geo}
\end{equation}
As in general relativity, the equation of motion is independent of
the mass of the test particle, so that the particle motion satisfies the weak equivalence principle. A main difference between this
equation of motion and the standard geodesic equation in general relativity is
that the fifth force contributes an extra repulsive force which depends on the velocity of the particle.
We set for simplicity the potentials of the fields to zero i.e., $V_\phi(\phi_\mu\phi^\mu) = V(G) = V(\mu) = 0$. 

\section{Weak Field approximation in MOG}
\label{weakapprox} As we noted in the introduction, the exact static
spherically symmetric solution of the MOG field equations has been
obtained for a point-like mass \cite{moffat09}. For extended
physical systems, we must use numerical calculations to solve the nonlinear MOG
field equations.

A natural way to study the behaviour of MOG on
astrophysical scales is to derive a weak field approximation for the dynamics of the fields. Our aim is to obtain the field equations for such a weak field approximation by
perturbing the fields around Minkowski space--time for an arbitrary distribution of non-relativistic matter. 

We perform a perturbation of the metric around the Minkowski metric $\eta_{\mu\nu}$:
\begin{equation}
g_{\mu\nu} = \eta_{\mu\nu} + h_{\mu\nu}.
\end{equation}
For the vector field, we have
\begin{equation}
\phi_\mu = \phi_{\mu(0)} + \phi_{\mu (1) },
\end{equation}
where $\phi_{\mu(0)}$ is the zeroth order and $\phi_{\mu(1)}$ the first order perturbations.  For Minkowski space--time, we set $\phi_{\mu(0)}$ equal to zero, for in the absence of matter there is no gravity source for the vector field $\phi_\mu$. We write for convenience $\phi_{\mu(1)}=\phi_\mu$. For the scalar field $G$, we perturb it around the Minkowski metric background:
\begin{equation}
G = G_{(0)} + G_{(1)},
\end{equation}
where $G_{(0)}$ is a constant in Minkowski space.  We perturb the scalar field $\mu$ around the Minkowski space background:
\begin{equation}
\mu=\mu_{(0)}+\mu_{(1)},
\end{equation}
where $\mu_{(0)}$ is a constant which we for convenience will label as $\mu$. We assume that $\mu_{(1)}$ is negligibly small and fix the scalar field $\mu$ in equation (\ref{scalar}) to be the constant value $\mu$, representing the mass of the vector field.  Finally, we perturb the energy-momentum tensor about the Minkowski background:
\begin{equation}
T_{\mu\nu}=T_{\mu\nu(0)}+T_{\mu\nu(1)}.
\end{equation}

We vary the action with respect to the three fields $g_{\mu\nu}$, $G$ and $\phi_\mu$, taking into account the perturbations around
flat space. Varying the action with respect to the $G$ field gives
\begin{equation}
\sq G_{(1)} = -\frac{G_{(0)}}{16\pi} R_{(1)}, \label{g1}
\end{equation}
where $R_{(1)}$ is the first--order perturbation of the Ricci scalar. Again, for convenience, we replace the background value of $G_{(0)}$ by $G_0$.

Varying the action with respect to the metric and ignoring the higher orders of perturbation, we get
\begin{equation}
R_{\mu\nu(1)} - \frac{1}{2} R_{(1)} \eta_{\mu\nu} =  - 8\pi G_0 T_{\mu\nu(1)}^{(M)}
- 8\pi G_0 T_{\mu\nu(1)}^{(\phi)}, \label{field1}
\label{ein1}
\end{equation}
where the first term on the right-hand side of this equation
represents the energy-momentum tensor of matter, and the second term
corresponds to the energy--momentum tensor of the vector field given by
\begin{eqnarray}
T_{\mu\nu}^{(\phi)} &=& \frac{\omega}{4\pi}(B_\mu{}^{\alpha}
B_{\nu\alpha} - \frac{1}{4}
g_{\mu\nu}B^{\alpha\beta}B_{\alpha\beta})\nonumber \\
&-&\frac{\mu^2\omega}{4\pi}(\phi_\mu\phi_\nu -
\frac{1}{2}\phi_\alpha\phi^\alpha g_{\mu\nu}).
\label{tphi}
\end{eqnarray}
By taking the trace of equation (\ref{ein1}), we obtain on the left-hand side of the equation $-R_{(1)}$ and on the right-hand side the trace of the energy-momentum tensor of matter and the vector field. We will ignore the higher order perturbation of the vector field $\phi_\mu$, and ignore the density of the vector field compared to the density of matter (i.e., $ T_{\mu\nu(1)}^{(\phi)}\ll  T_{\mu\nu(1)}^{(M)}$). The trace of equation (\ref{ein1}) can be written as 
\begin{equation}
R_{(1)} =  8\pi G_0 T^\mu{}_{\mu(1)}^{(M)},
\label{perturbR}
\end{equation}
where on the right-hand side we have for pressureless matter $T^\mu{}_{\mu(1)}^{(M)} = \rho$.  Substituting equation (\ref{perturbR}) into equation (\ref{g1}), for the static solution, we get
\begin{equation}
\nabla^2\biggl(\frac{G_{(1)}}{G_0}\biggr) =\frac{1}{2}G_0\rho.
\end{equation}
From the solution of this equation, we know that
$G_{(1)}/G_0$ is of the order of the gravitational potential or
$(v/c)^2$, where $v$ is the internal velocity of the system. Hence, for systems such as galaxies or clusters of
galaxies, the deviation from the constant $G_0$ is of the order of
$G_1/G_0\simeq 10^{-7}-10^{-5}$. In what follows, we only keep the background value of $G_0$ in our equations.

In the weak field approximation, for the $(0,0)$ component we have
\begin{eqnarray}
R_{00(1)} =\frac{1}{2}\nabla^2 h_{00},
 \end{eqnarray}
and we obtain the field equation:
\begin{equation}
\frac{1}{2}\nabla^2( h_{00}) = -4\pi G_0 \rho.
\label{effpois}
\end{equation}

For the vector field, we obtain the field equation by varying the action with respect to $\phi^\mu$:
\begin{equation}
\nabla_\nu B^{\mu\nu} - \mu^2 \phi^\mu = -\frac{4\pi}{\omega} J^\mu.
\end{equation}
Let us assume that the vector matter current $J^\mu$ is conserved,
$\nabla_\mu J^\mu=0$, then we can impose in the weak field approximation the gauge condition,
$\phi^\mu{}_{,\mu} = 0$.  For the static case, we obtain 
\begin{equation}
\nabla^2\phi^0 -\mu^2\phi^0 = -\frac{4\pi}{\omega}J^0,
\end{equation}
which has the solution
\begin{equation}
\phi^0(x) = \frac{1}{\omega}\int\frac{e^{-\mu|{\bf x}-{\bf x'}|}}{|{\bf x}-{\bf x'}|}J^0({\bf x'})d^3{\bf x'}. \label{phi0}
\end{equation}

\begin{figure*}
\begin{center}
\begin{tabular}{ccc}
\includegraphics[width=60mm]{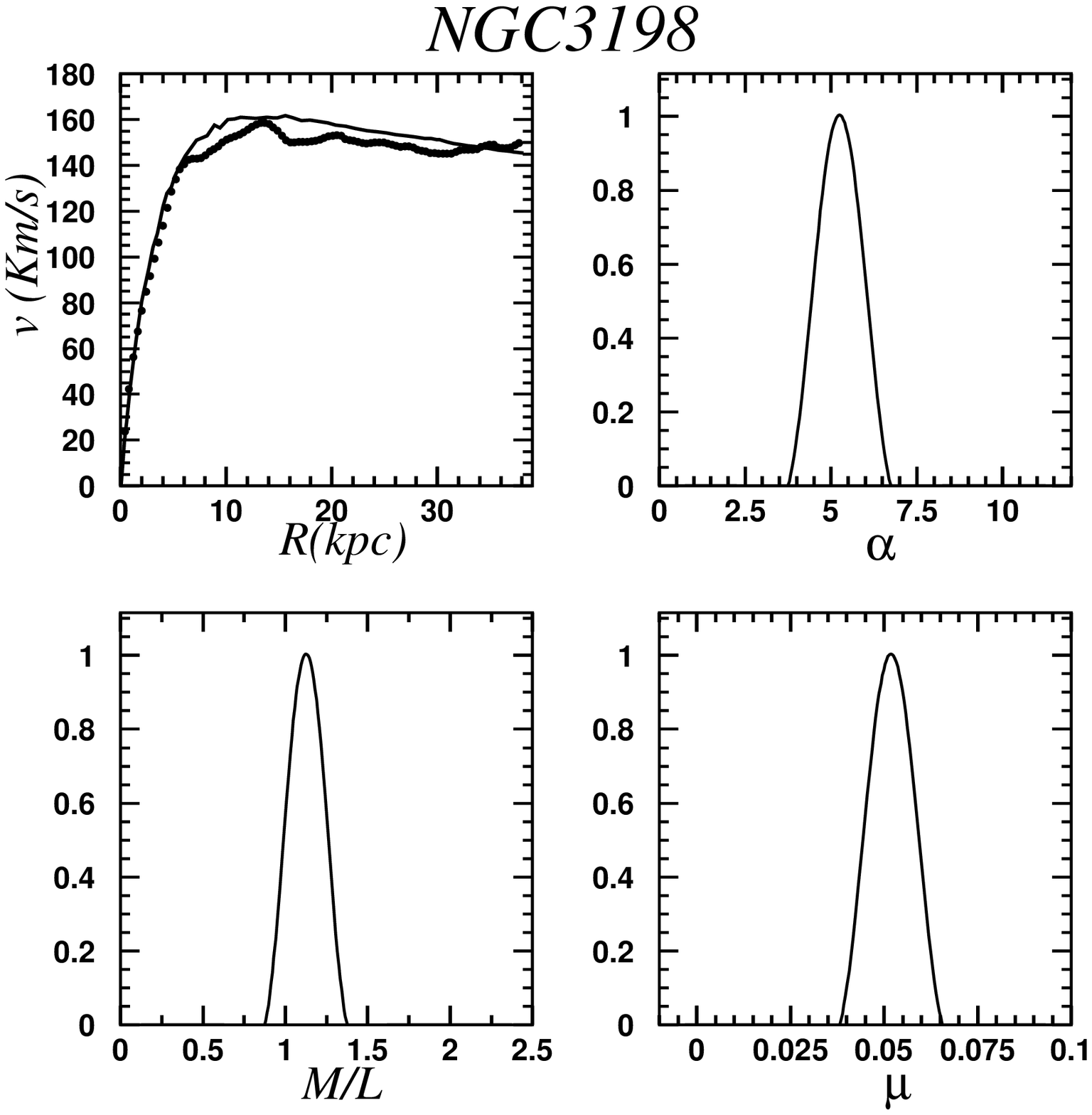} &
\includegraphics[width=60mm]{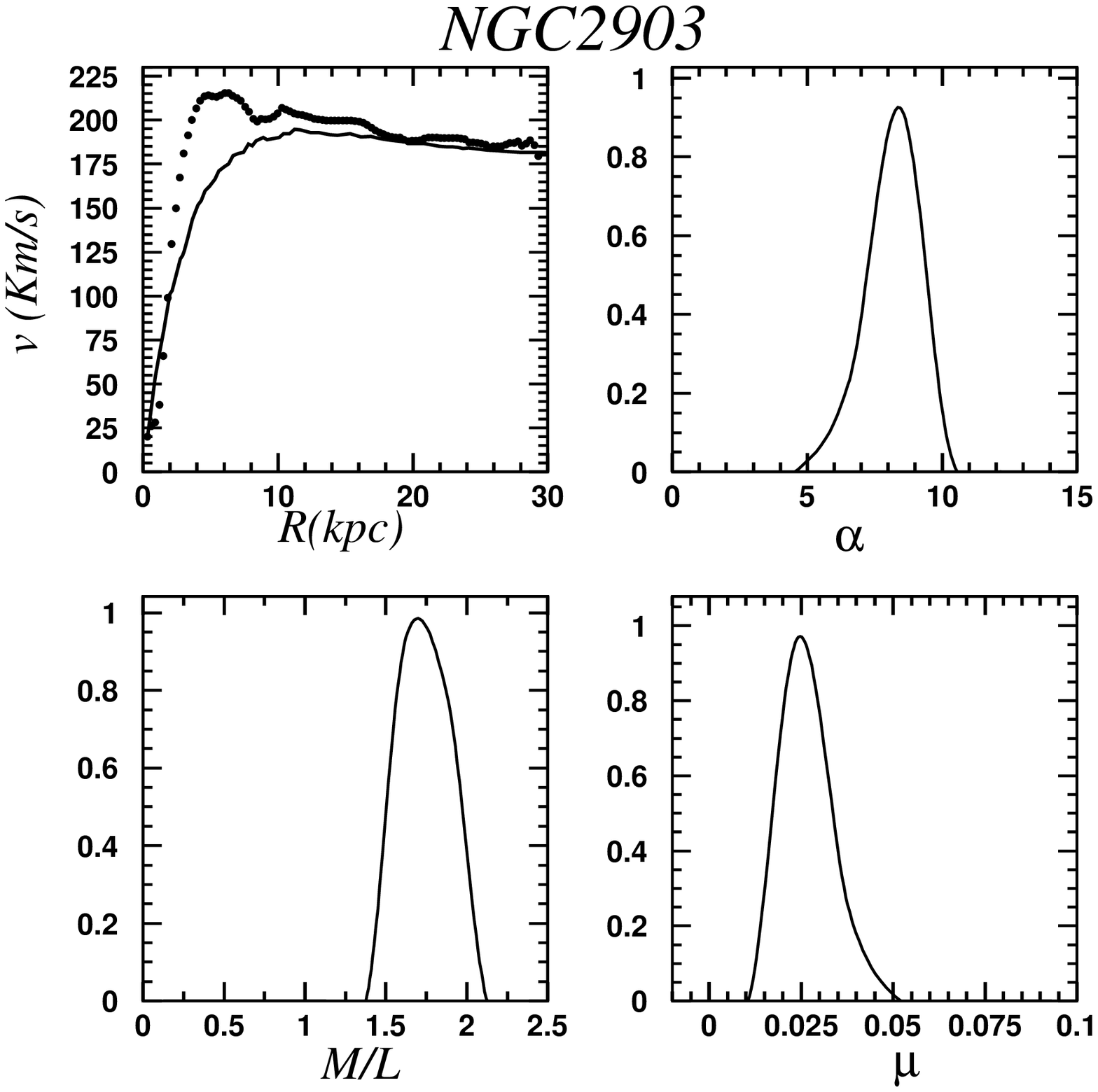}&
\includegraphics[width=60mm]{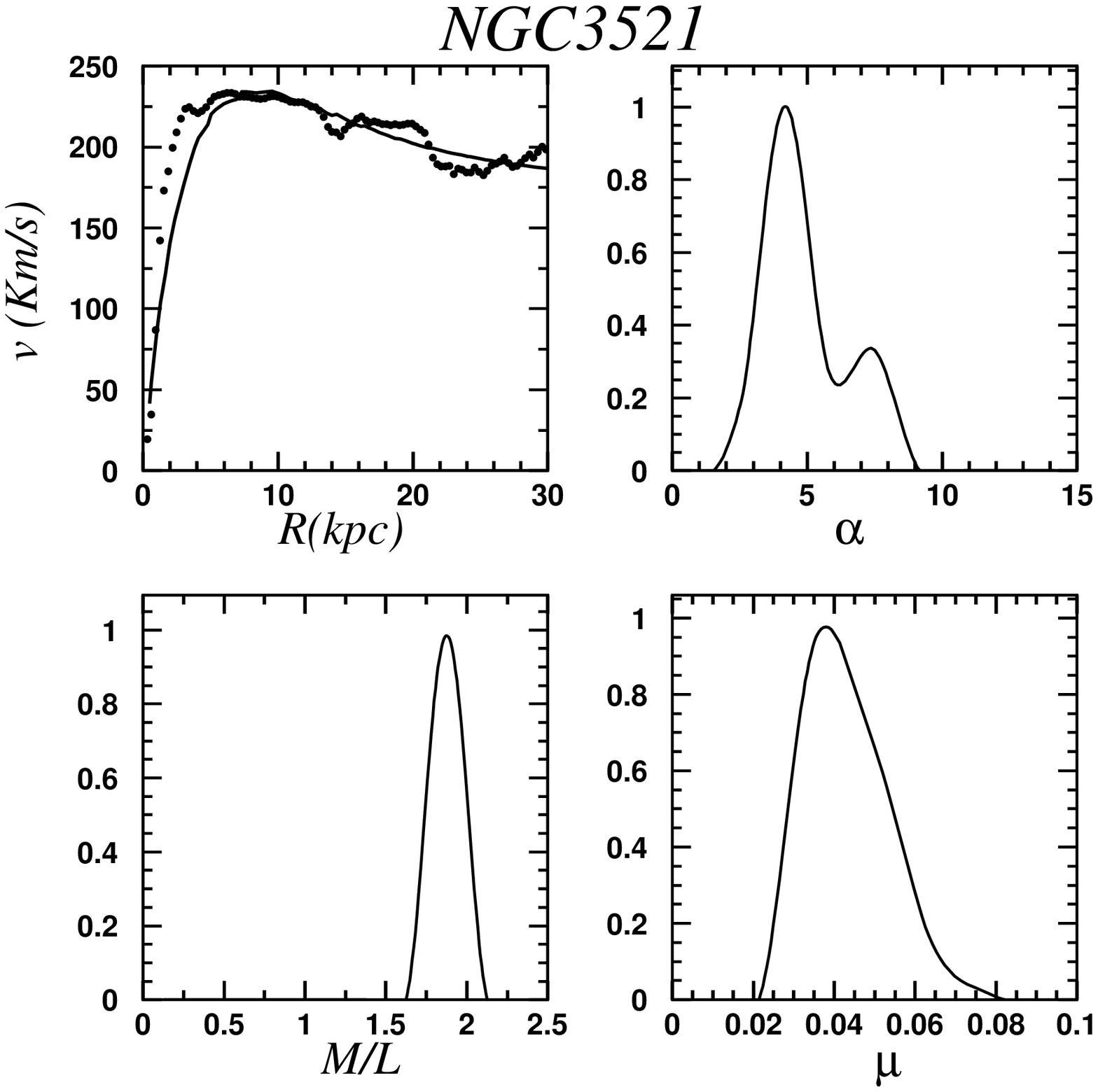}\\
\includegraphics[width=60mm]{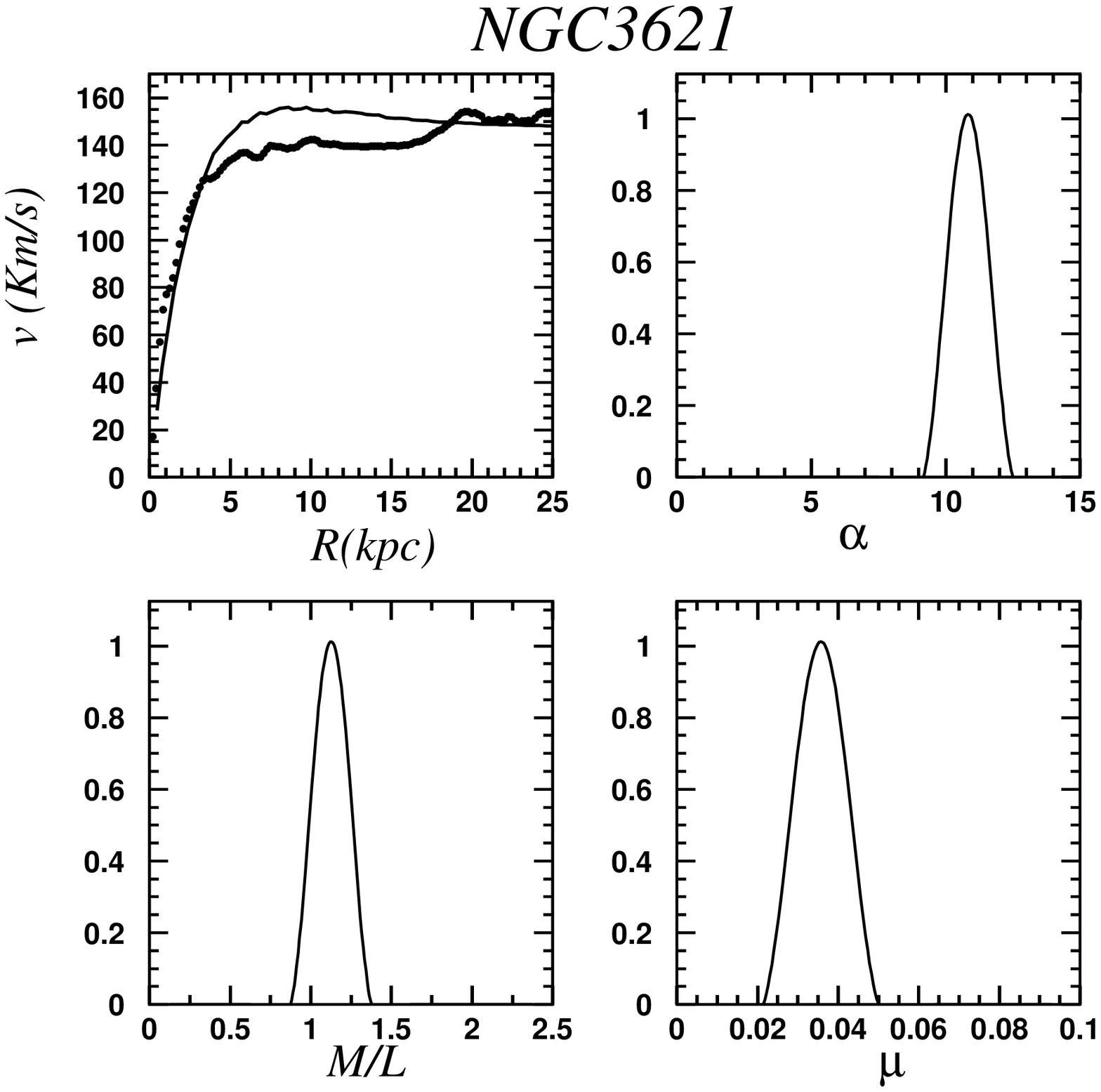} &
\includegraphics[width=60mm]{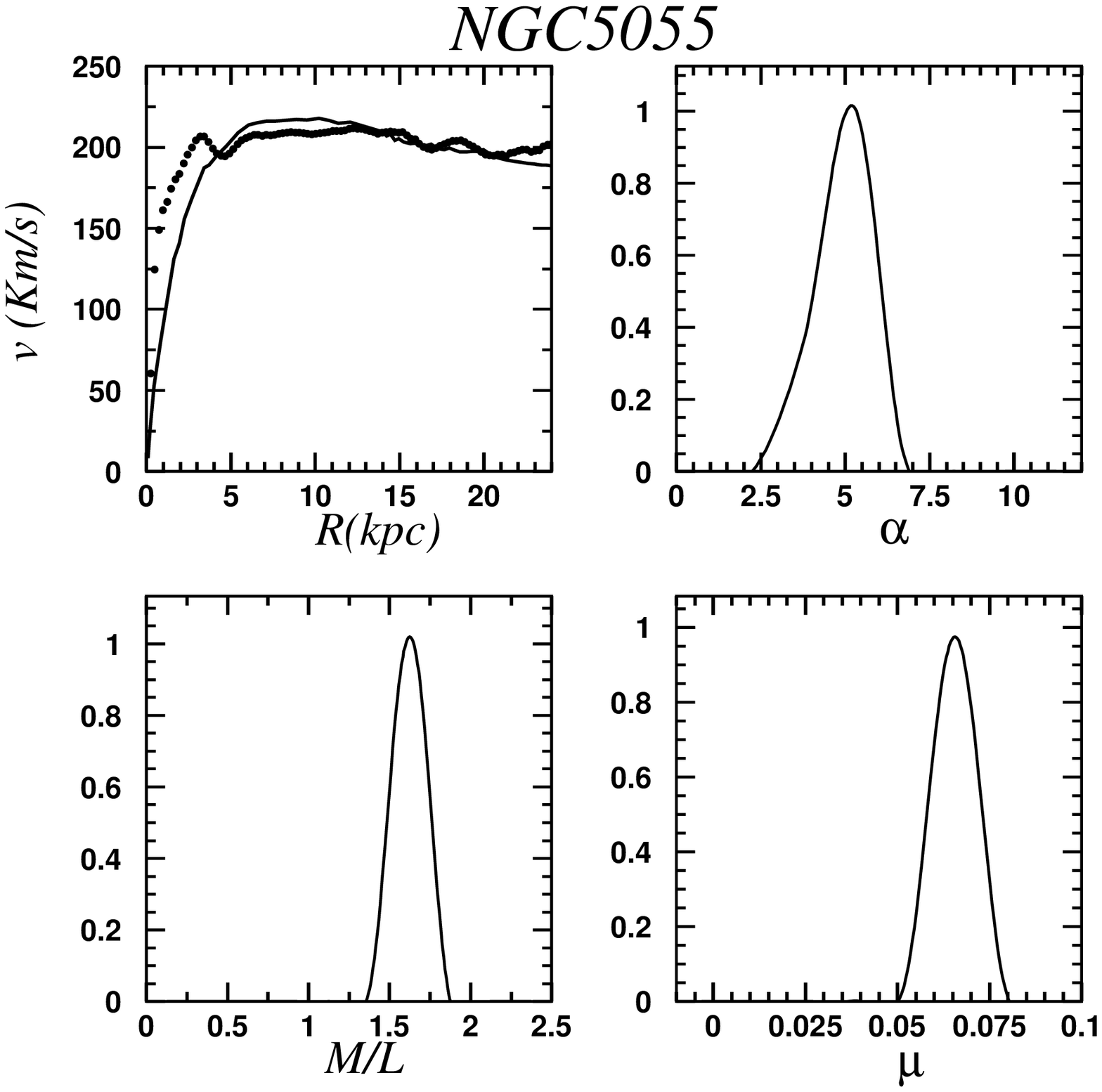}&
\includegraphics[width=60mm]{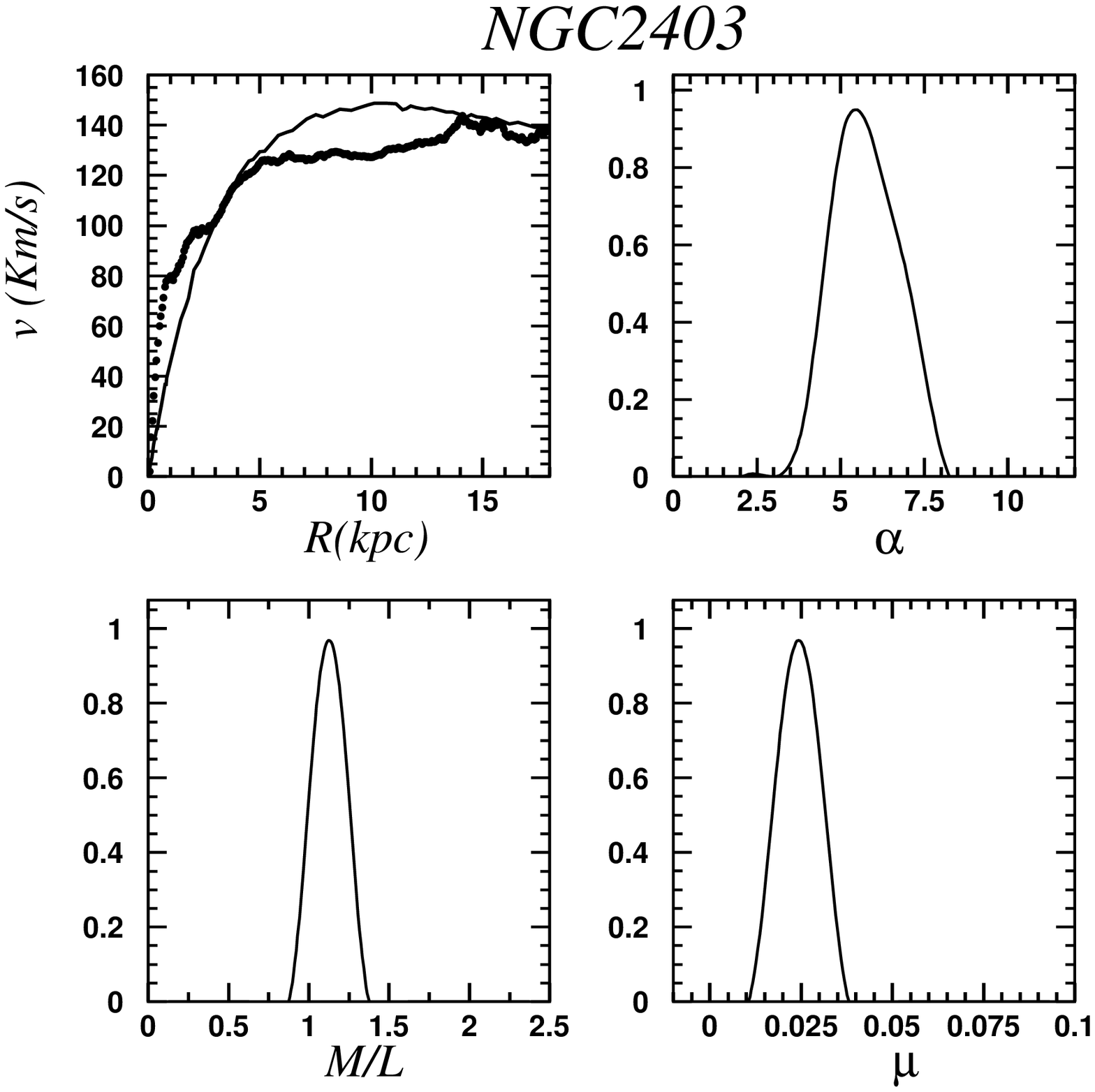}\\
\includegraphics[width=60mm]{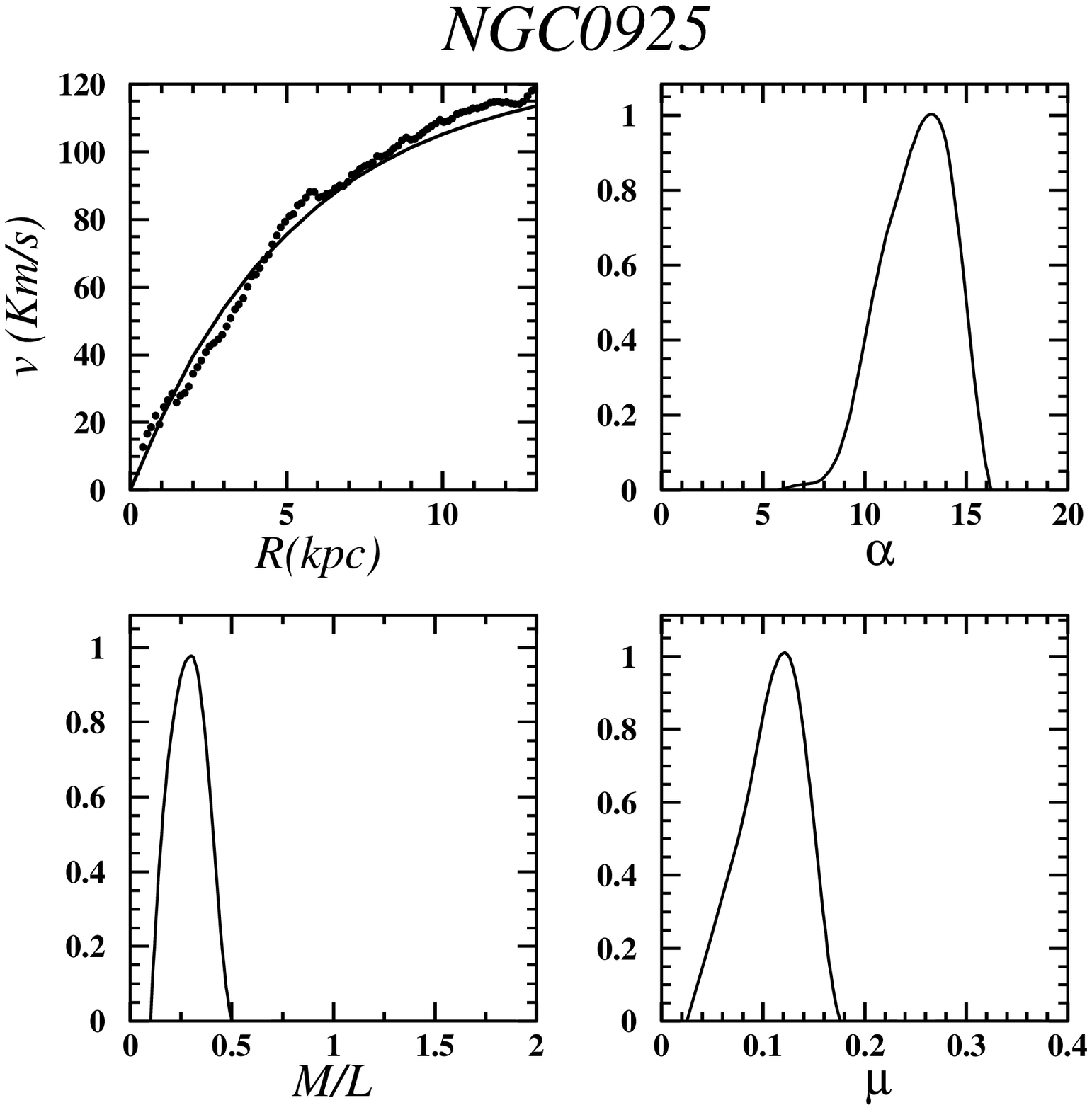} &
\includegraphics[width=60mm]{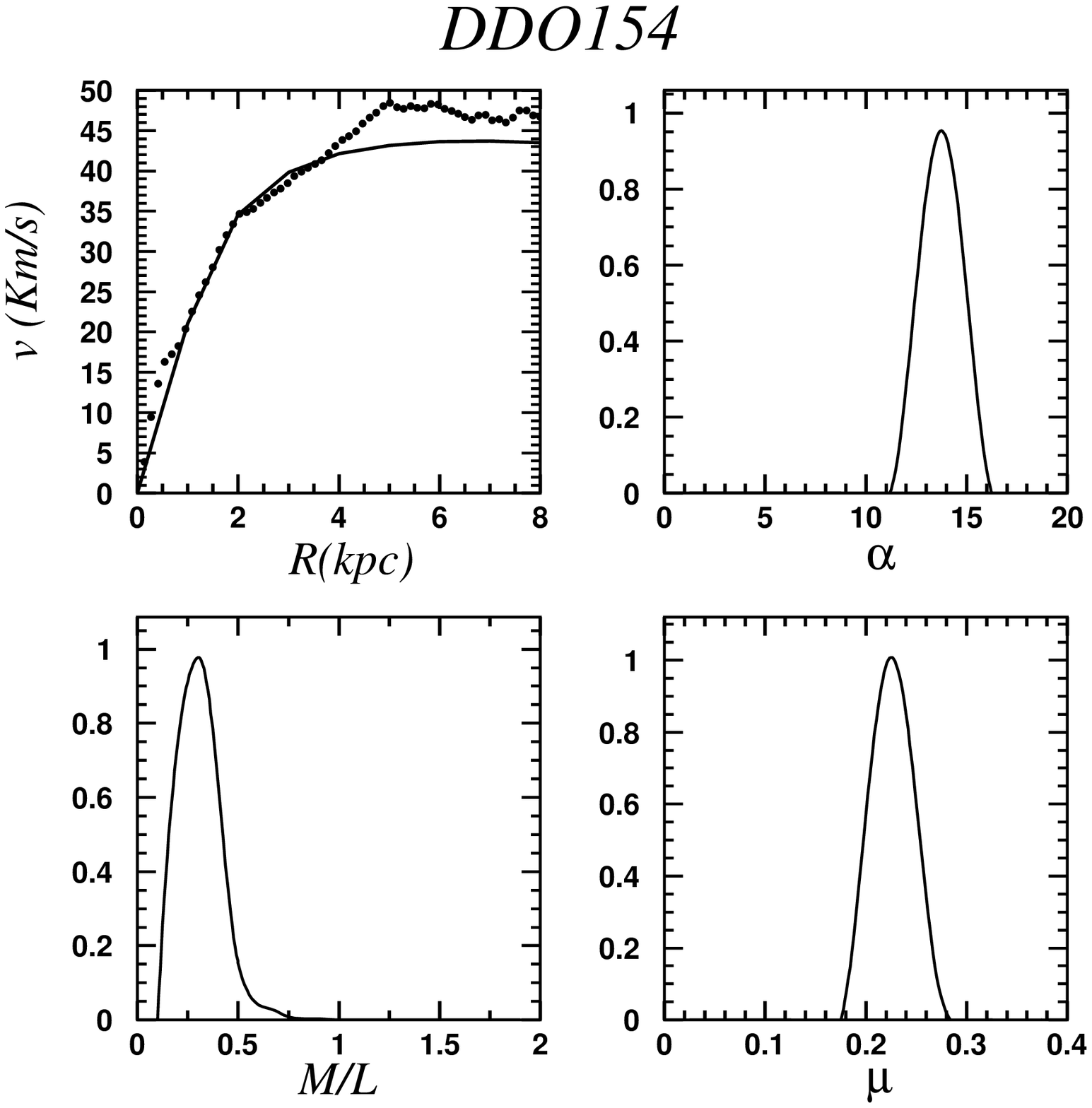} &
\includegraphics[width=60mm]{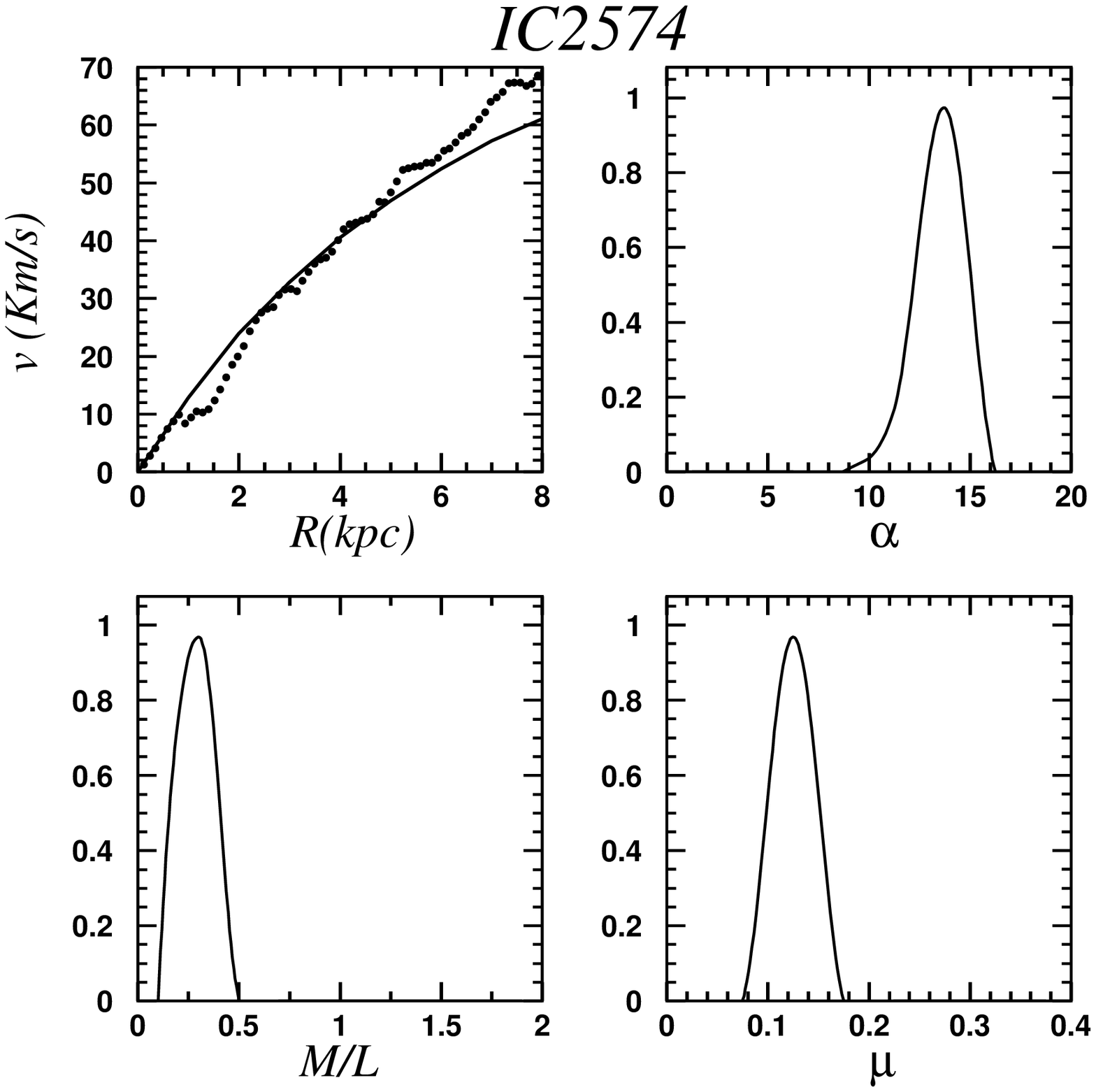}\\
\end{tabular}
\end{center}
\caption {The best fit for the sub-sample of THINGS galaxies with
the corresponding marginalized likelihood functions of $\alpha$, $\mu$ and $M/L$.
Here in this list we have both HBS and LSB galaxies. Table
(\ref{tabl}) provides the best values of the parameters with the
corresponding error bars. \label{fig1}}
\end{figure*}

In order to obtain the field equation for an effective potential in the weak field
approximation, we take the divergence of the spatial component of
the equation of motion (\ref{geo}):
\begin{equation}
\nabla\cdot {\bf a}  - \frac12\nabla^2 h_{00} = - \omega\kappa
\nabla^2\phi^0, \label{p2}
\end{equation}
where $\vec{a}$ represents the acceleration of the test particle. We
substitute $\nabla^2h_{00}$ from equation (\ref{effpois}) into equation (\ref{p2}) and
relate directly the acceleration of the test particle to the
distribution of matter. We define the effective potential for the
test particle by, ${\bf a} = -\nabla\Phi_{eff}$, and relate it to the distribution of matter:
\begin{equation}
\nabla\cdot(\nabla\Phi_{eff} - \kappa\omega\nabla\phi^0) = 4\pi G_0\rho. \label{po3}
\end{equation}

On the left-hand side of this equation, we define $\Phi_N$ as the solution to the Poisson equation:
\begin{equation}
\Phi_{N} = \Phi_{eff} - \kappa\omega\phi^{0}. \label{poteffect}
\end{equation}
Substituting the solution for $\phi^0$ from equation (\ref{phi0}) by
replacing $J^0$ with $\kappa \omega \rho$, and using the solution of
$\Phi_N$ from equation (\ref{poteffect}), the effective potential becomes
\begin{equation}
\Phi_{eff}({\bf x}) = - \int\frac{G_0 \rho({\bf x'})}{|{\bf x}-{\bf x'}|}d^3{\bf x'} +\kappa^2\int
\frac{{\rm e}^{-\mu|{\bf x}-{\bf x'}|}}{|{\bf x}-{\bf x'}|}\rho({\bf x'})d^3{\bf x'}. \label{potential}
\end{equation}
Here, the first term corresponds to the attractive gravitational
force, while the second term is the repulsive Yukawa force. For a
point mass particle, using the Dirac-delta function $\rho({\bf x'})= M
\delta^3({\bf x'})$, the effective potential reduces to
\begin{equation}
\Phi_{eff}(x) = - \frac{G_0 M}{x} + \kappa^2\frac{M {\rm e}^{-\mu x}}{x},
\label{phieffpointmass}
\end{equation}
where $x=\vert{\bf x}\vert$. For small distances compared to $\mu^{-1}$, we can expand the
exponential term yielding the effective potential:
\begin{equation}
\Phi_{eff}(x) = - \frac{(G_0 - \kappa^2)M}{x} - \mu\kappa^2 M.
\end{equation}
The constant second term on the right-hand side does not enter into
the dynamics of the test particle; the first term should be the
Newtonian gravitational contribution and we obtain $$ G_0 - \kappa^2 =
G_N.$$ On the other hand, at large distances (i.e. $\mu x\rightarrow
\infty$), we just have the first term of equation
(\ref{phieffpointmass}). Hereafter, we rename $G_0$ as $G_{\infty}$
which corresponds to the effective gravitational constant at
infinity. 

Substituting $G_0$ and $\kappa^2$ into equation (\ref{potential}),
the effective potential for an extended distribution of matter in
MOG in the weak field approximation is given by
\begin{equation}
\Phi_{eff}({\bf x}) = - G_\infty \left[\int\frac{\rho({\bf x'})}{|{\bf x}-{\bf x'}|}\biggl(1
-\frac{G_\infty - G_N}{G_\infty}{\rm e}^{-\mu|{\bf x}-{\bf x'}|}\biggr)d^3{\bf x'} \right].  
\label{potential2}
\end{equation}
Now we use the same notation as used by Moffat and Toth (2009),
defining $\alpha = (G_\infty - G_N)/G_N$. Then the effective
potential can be written as
\begin{equation}
\Phi_{eff}(\vec x) = - G_N \left[\int\frac{\rho(\vec x')}{|\vec x-\vec x'|}(1+\alpha
-\alpha e^{-\mu|\vec x-\vec x'|})d^3x' \right], \label{potential2}
\end{equation}
and the acceleration of the test particle can be obtained from the
gradient of the potential, $\vec{a} = -\vec{\nabla}\Phi_{eff}$, yielding the
result
\begin{eqnarray}
\vec{a}({\bf x}) &=& - G_{\rm N}
\int\frac{\rho({\bf x'})({\bf x}-{\bf x'})}{|{\bf x}-{\bf x'}|^3}[1+\alpha \nonumber
\\
&-&\alpha {\rm e}^{-\mu|{\bf x}-{\bf x'}|}(1+\mu|{\bf x}-{\bf x'}|) ]d^3{\bf x'}. \label{acceleration}
\end{eqnarray}
In the weak field approximation, we shall treat $\alpha$ and $\mu$ as constant parameters. However, in the exact static spherically symmetry solution, $\mu$ and $\alpha$ depend on the mass of the source~\cite{moffat09}. In the weak
approximation limit, any deviation from Newtonian gravity depends on
the size of the system. 

In the next section, by means of a numerical
calculation of the potential for spiral galaxies, we compare the
weak field approximation limit of MOG to observational data for the rotation curves of galaxies.

\section{Rotation curves of spiral galaxies}

\label{rotcurve} In this section, we investigate the rotation curves
of spiral galaxies determined by the distribution of baryonic matter,
which is made of stars and interstellar gas without exotic dark
matter. For a galaxy with cylindrical symmetry, the radial component
of acceleration can be calculated by discretizing space into small
elements and adding the acceleration of each element as follows
\begin{eqnarray}
&& a_r(r) = G_N\sum_{r'=~0}^\infty \sum_{\theta'=~
0}^{2\pi}\frac{\Sigma(r')}{|r-r'|^3}(-r +
r'cos\theta')(1+\alpha \nonumber\\
& &-\alpha e^{-\mu|r-r'|} -\mu\alpha|r-r'|{\rm e}^{-\mu|r-r'|} )r'\Delta
r'\Delta\theta',
\end{eqnarray}
where $\Sigma(r)$ represents the column density of a spiral galaxy.

From the observations, we have the column density of stars in a
given colour band as well as the column density of hydrogen. Fitting the
distribution of matter with an exponential function, we can identify
any spiral galaxy with the following parameters, (i) the total
luminosity of the galaxy in a given filter (here we will use data
from the B--band and the infrared band), (ii) the total hydrogen of
the galaxy from $21$cm observations and, (iii) the characteristic
length scale of the galaxy $R$, which is the length scale occurring
in the exponential law for the column density of a galaxy~
\cite{fathi10}:
\begin{equation}
\Sigma(r) = \Sigma_0 \exp(-r/R). \label{sigma}
\end{equation}
Integrating over the surface to infinity, the total mass of the disc
is related to $R$ and the central column density, $M_{disk} =
2\pi\Sigma_0 R^2$. Thus, knowing $M_{disk}$ and $R$, we can
calculate $\Sigma_0$. For the gaseous component of the galaxy,
we obtain the mass of gas from hydrogen and helium abundance from big
bang nucleosynthesis, $M_{\rm gas} = (4/3) M_{H}$. In the calculation of
the mass of the stars, we assume a stellar mass-to-light ratio,
$\Upsilon_\star$, and the total mass of the stars can be obtained
from $M_{\rm stars} = \Upsilon_\star \times L$, where $L$ is the overall
luminosity of the galaxy in a given filter.

We choose a subsample of nearby galaxies, from the THINGS
catalogue with high resolution measurements of velocity and density of
hydrogen profile~\cite{Blok2008}. For this set of galaxies, we adopt
in the weak field approximation $\mu$ and $\alpha$ as well as the stellar mass-to-light ratio
$M/L$ when fitting the rotation curves of galaxies to the data. We then find the best values of  $\alpha$ and $\mu$ and fix these two parameters. Then we fit the
observed rotation curves of the larger Ursa-Major sample of galaxies,
letting the stellar mass-to-light ratio $M/L$ be the only free parameter.

\subsection{THINGS catalogue}
\label{things}
\begin{table*}
\begin{center}
\caption[]{\label{tabl} The sub-sample of galaxies from the THINGS
catalogue with the best fit parameters obtained from fitting the
observed rotation curves to the MOG theoretical rotational curves.
The description of the columns is given by, (1) the name of the
galaxy, (2) type of galaxy, (3) distances of the galaxies, (4) the
overall luminosity of the galaxies in the B-band, (5) the
characteristic size of the galaxy in equation (\ref{sigma}), (6) the
overall hydrogen mass of the galaxy, (7) the overall mass of the
galaxy calculated by $M_{disk} = \frac43 M_{HI} +
L_B\times\Upsilon_{\star}$, (8) the best fitting value of $\alpha$, (9)
the best fitting value of $\mu$ and, (10) the stellar mass-to-light
ratio for each galaxy. The error bars are obtained from the
likelihood functions given in Figure (1). The observational data are
taken from the THINGS publications
in~\cite{Blok2008,walter2008,leroy2008}. }
\begin{tabular}{|c|c|c|c|c|c|c|c|c|c|}
\hline\hline
   & & Distance & $L_B$ & $R_0$ & $M_{HI}$ & $M_{disk}$ &$\alpha$& $\mu$ & $\Upsilon_\star$  \\
Galaxy & Type & (Mpc) & ($10^{10}L_B$) &(kpc)& $(10^{10}M_\odot)$& $(10^{10}M_\odot)$ &  & $(kpc^{-1})$ & $(M_\odot/L_\odot)$  \\
 (1)   & (2)  & (3)   & (4)            & (5) &(6)                &(7)                  &(8)          & (9)                 & (10) \\
\hline
NGC 3198 & HSB & $13.8$ & $3.241$ & $4.0$  &  $1.06$ & 4.72 & $5.94 \pm 1.01$ & $0.051\pm 0.012$ & $1.02\pm 0.13$  \\
NGC 2903 & HSB & $8.9$ & $4.088$  & $3.0$ &   $0.49$  & 7.35 &$8.02\pm 1.78$ &  $0.032\pm0.007$& $1.64\pm0.14$ \\
NGC 3521 & HSB & $10.7$ & $4.769$ & $3.3$ & $1.03$   & 6.23 &$4.31\pm1.03$   & $0.037\pm0.013$ & $1.02\pm 0.06$    \\
NGC 3621 & HSB & $6.6$ & $2.048$ & $2.9$  &  $0.89$ & 3.48 &$9.82\pm 0.31$   & $0.027\pm 0.011$ & $1.12\pm0.08$  \\
NGC 5055 & HSB & $10.1$ & $3.622$ & $2.9$ & $0.76$   & 6.59 &$5.40\pm0.060$ & $0.057\pm0.006$ & $1.54\pm0.16$\\
NGC 2403 & LSB & $3.2$ & $1.647$ & $2.7$ & $0.46$ &  2.45&$5.60\pm0.61$ & $ 0.018\pm0.007$& $1.12\pm 0.13$ \\
DDO 0154 &LSB  &4.3    & 0.007   &0.8    & 0.03   & 0.04 &$13.71\pm1.23$& $0.22\pm0.03 $ & $0.29\pm 0.1$ \\
IC 2574  &LSB  &4.0    &0.345    &4.2    &0.19    & 0.35 &$ 13.70\pm 1.30$ & $0.12\pm 0.028$ & $0.3 \pm 0.2$     \\
NGC 0925 &LSB  &9.2    &1.444    &3.9    &0.41    & 0.95 & $13.10\pm 1.7$ & $0.11 \pm 0.03$  & $0.28\pm0.11$  \\
\hline
\end{tabular}
\end{center}
\end{table*}

The HI Nearby Galaxy Survey (THINGS) catalogue contains nearby
galaxies with high resolution observations of rotation velocities
and distributions of matter. These observations provide high quality
HI rotation curves together with the column density of
hydrogen~\cite{Blok2008,walter2008,leroy2008}. Here, we also use
stellar distributions for this sample of galaxies and the $3.6\mu$m band
images from SINGS~\cite{ken2003}. From the colour of stars in each
part of a galaxy, we can measure the mass of the stellar components
of the disk obtained from the correlation between the colour and the
stellar mass-to-light ratio~\cite{oh}. We also use Near Infra Red
(NIR) information in addition to the shorter wavelengths such as B
band, for which the mass-to-light ratio is mainly dominated by
the young stellar population. We note that NIR mainly probes the old
stellar populations~\cite{bell2001}.

The THINGS catalogue contains $19$ galaxies, but we choose a
sub-sample of nine galaxies, which have full coverage of rotation
curves from the centre to the edges of the galaxies. Table (1) shows
the list of galaxies in the sub-sample of the THINGS catalogue. As we
noted before, we let the parameters of the effective potential and
the stellar mass-to-light ratio ($M/L = \Upsilon_\star$) change
during the fitting of the theoretical rotation curves to the galaxy
data. Figure (1) shows the best fit to the data with the
corresponding marginalized likelihood functions for the three free parameters of
the model. The best values for the modified gravity parameters
$\alpha$ and $\mu$ , with the corresponding mass-to-light ratios
$M/L$, are given in Table (\ref{tabl}).

In order to calculate an average value for $\alpha$ and $\mu$, we
use the combined likelihood functions of nine different galaxies in
the THINGS catalogue. Since the data for the galaxies are independent,
the overall likelihood function is the multiplication of each of the
distributions, $P = \prod_{i=1}^9 P_i$. Hence for our set of
galaxies the overall likelihood function can be written as
\begin{equation}
P(\chi^2) \propto \exp\left(-\frac12\sum_{i=1}^9 [\chi^2 -
\chi_i^2({\rm min})]\right).
\end{equation}
From the combined likelihood functions, we obtain the best fit
parameters of MOG: $\alpha = 8.89\pm0.34 $ and $\mu = 0.042 \pm
0.004~{\rm kpc}^{-1}$.

\setcounter{figure}{1}
\begin{figure*}
\begin{center}
\begin{tabular}{ccc}
\includegraphics[width=60mm]{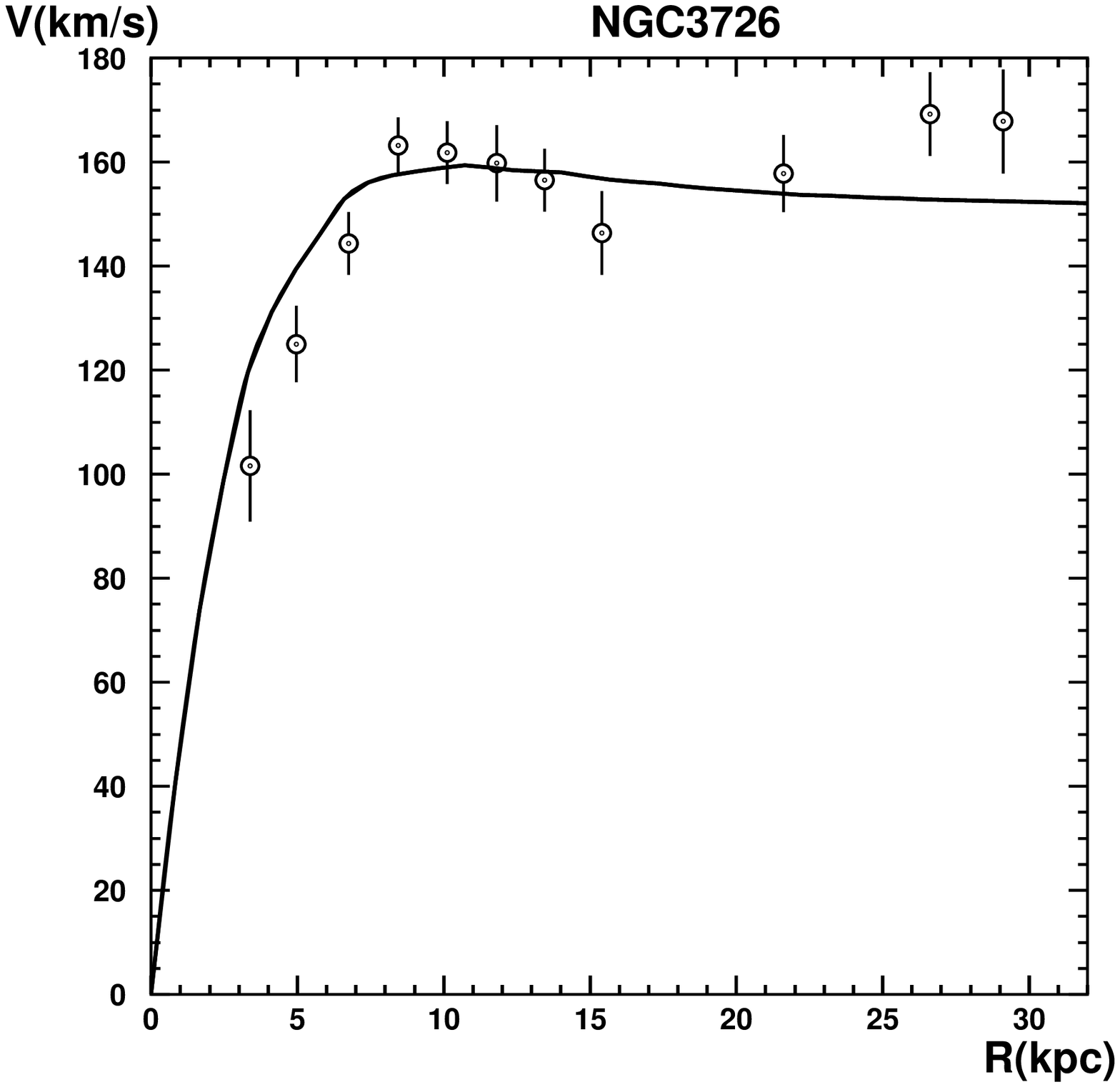}&
\includegraphics[width=60mm]{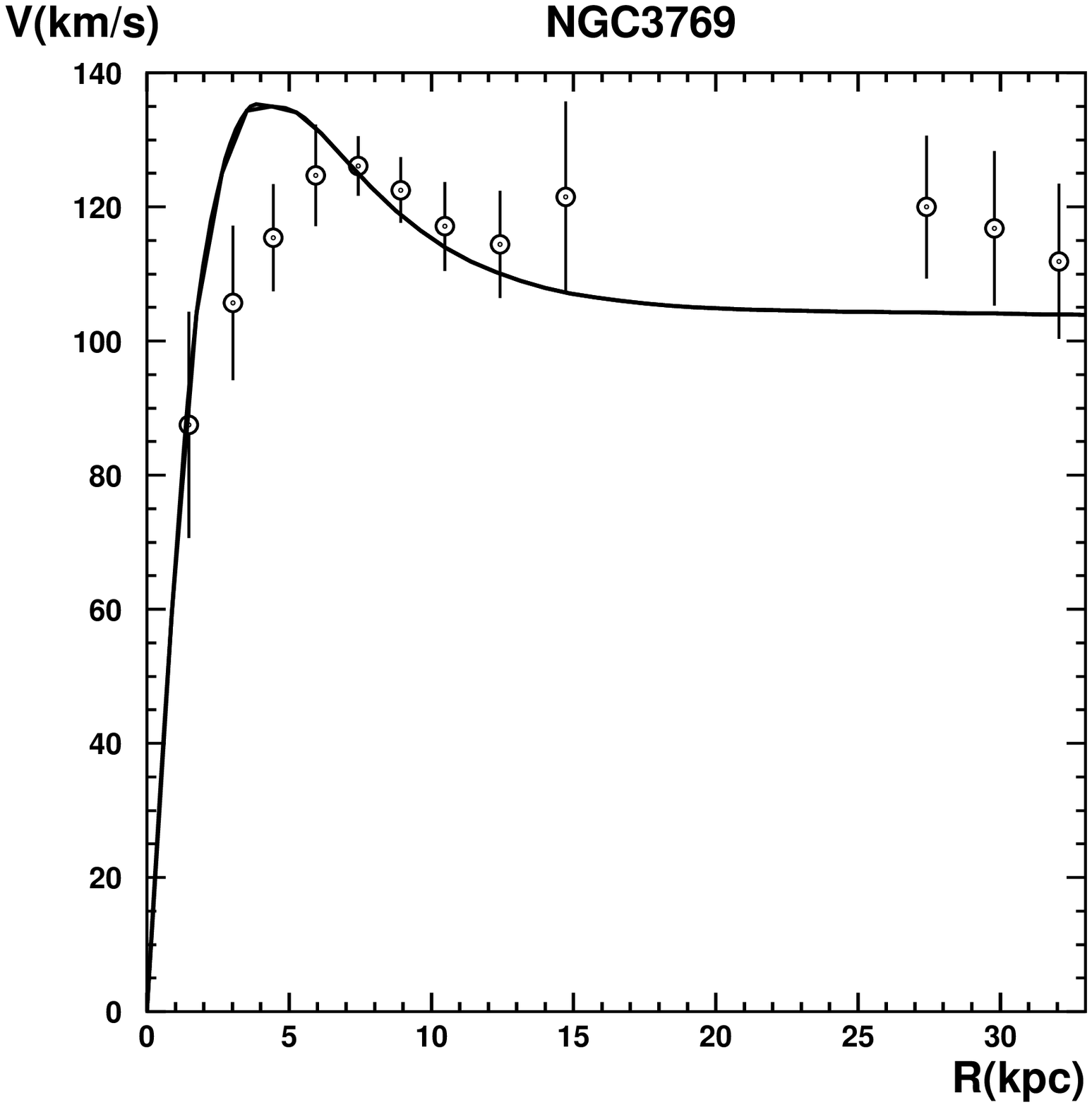}&
\includegraphics[width=60mm]{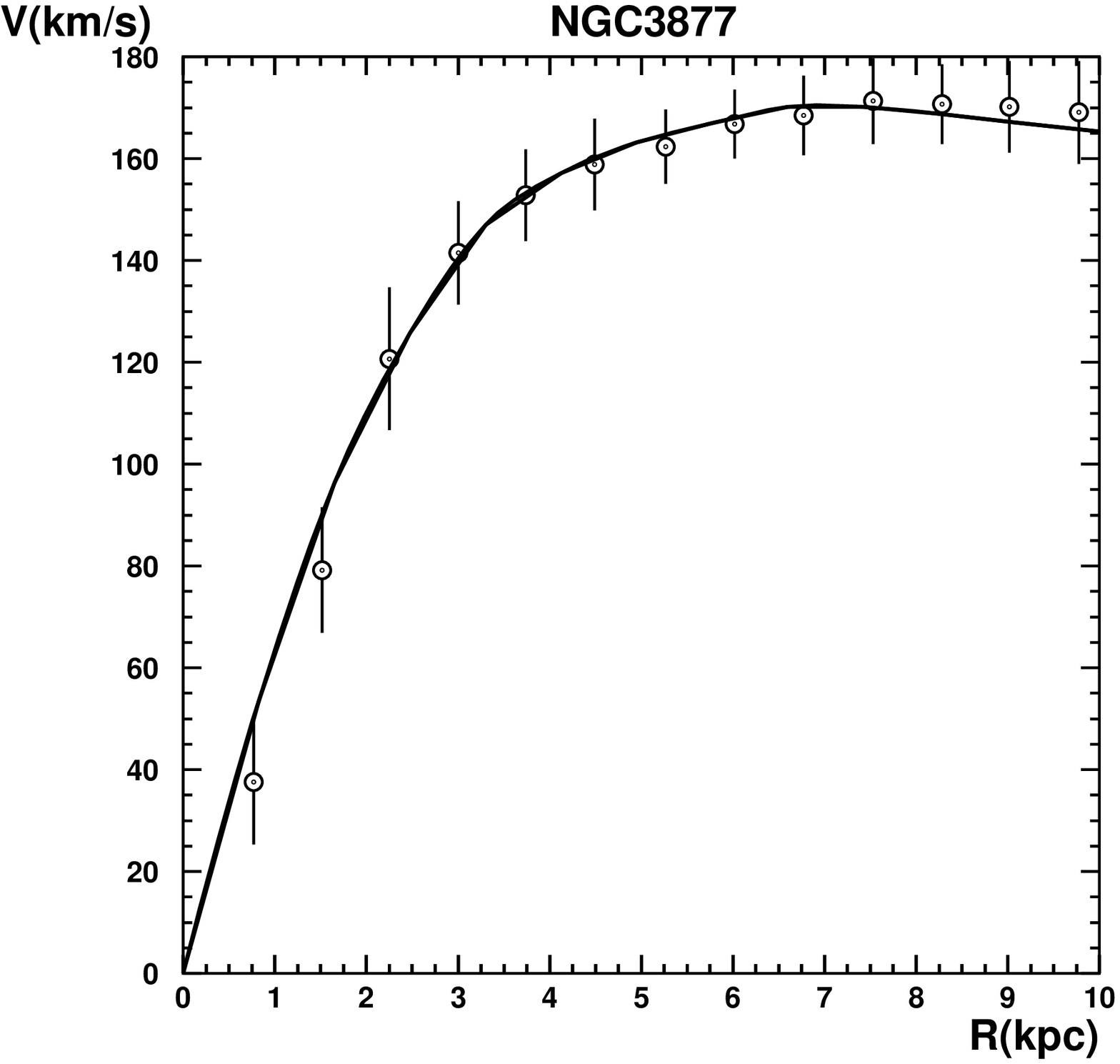} \\
\includegraphics[width=60mm]{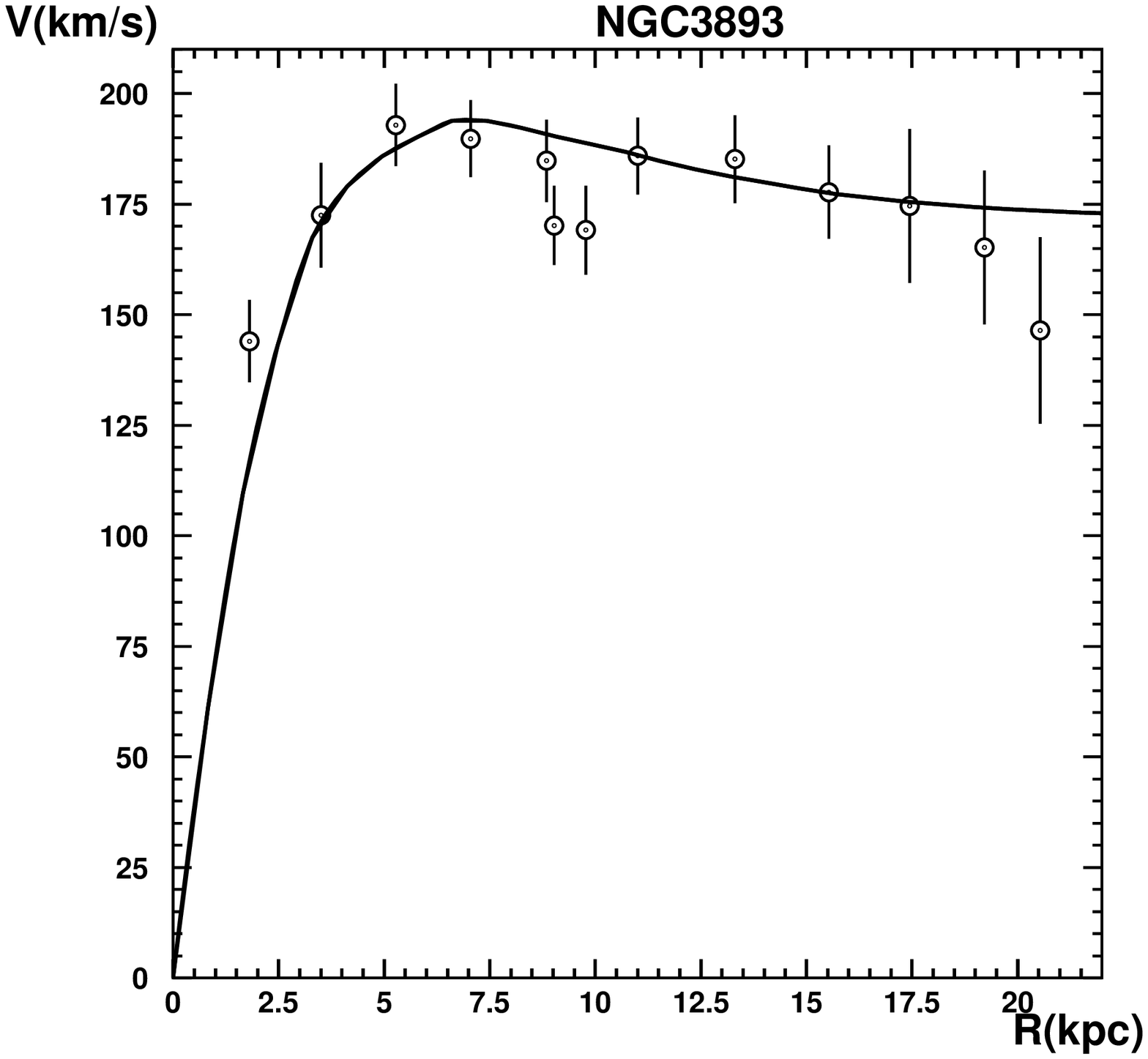} &
\includegraphics[width=60mm]{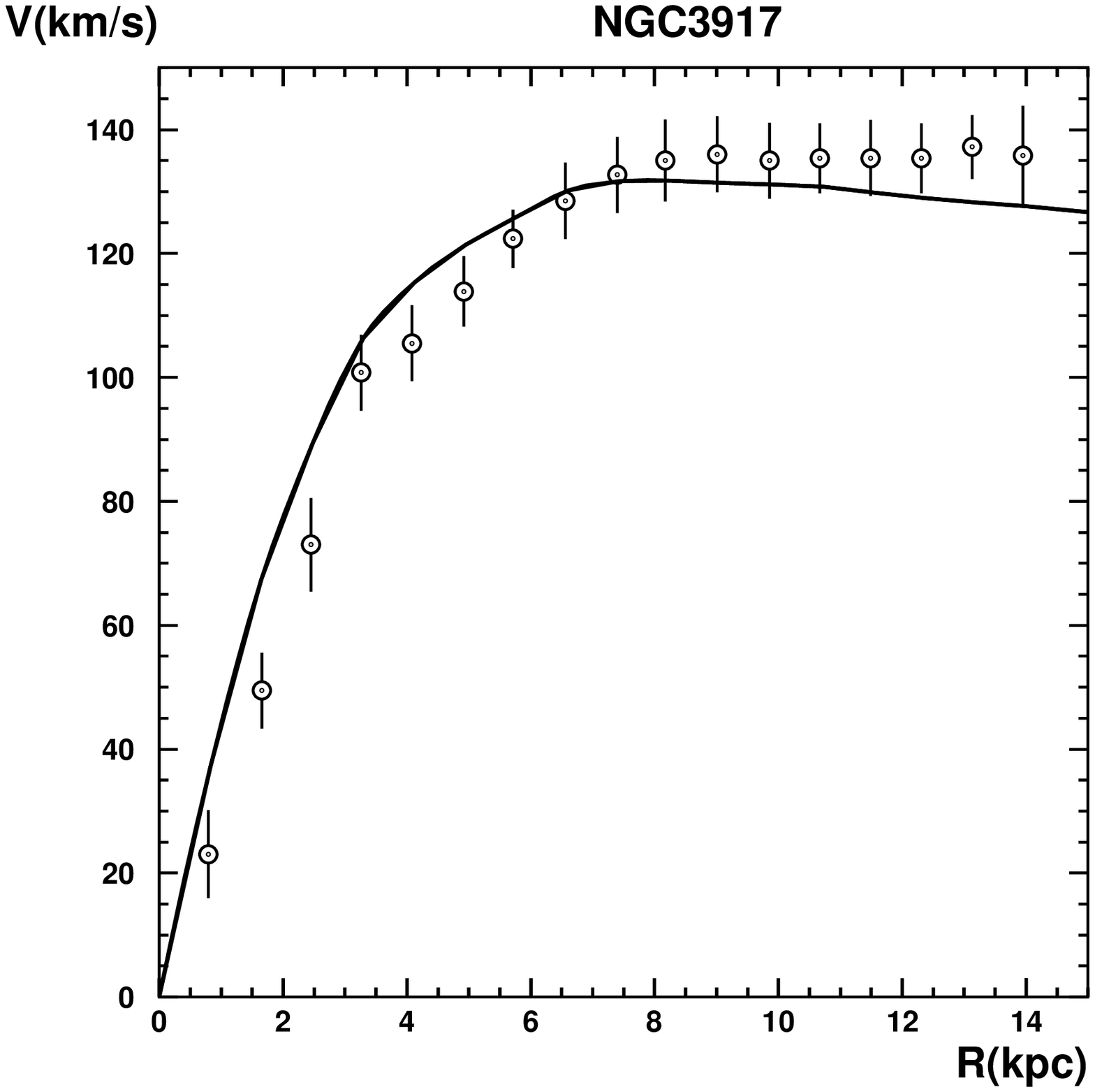}&
\includegraphics[width=60mm]{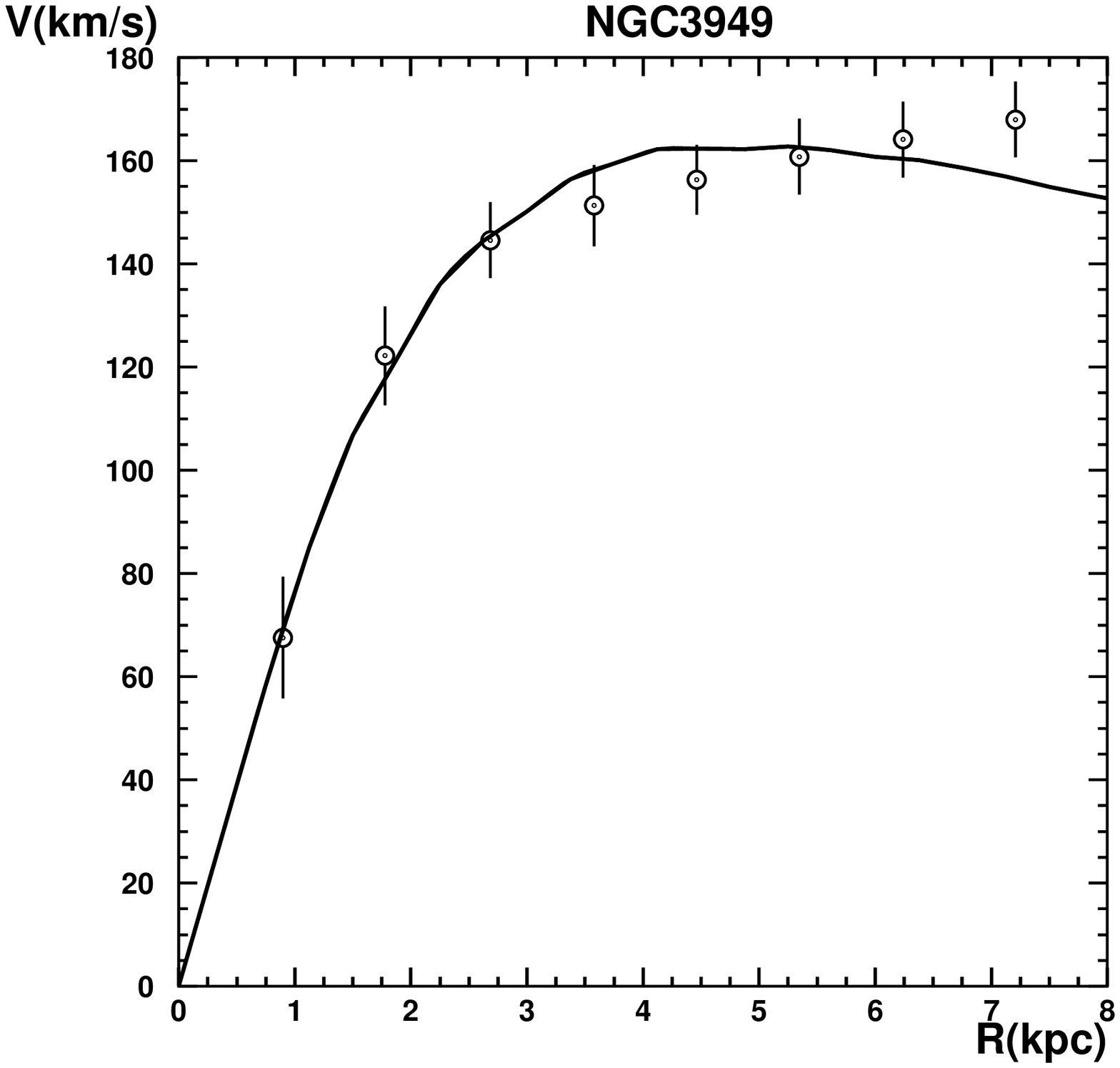}\\
\includegraphics[width=60mm]{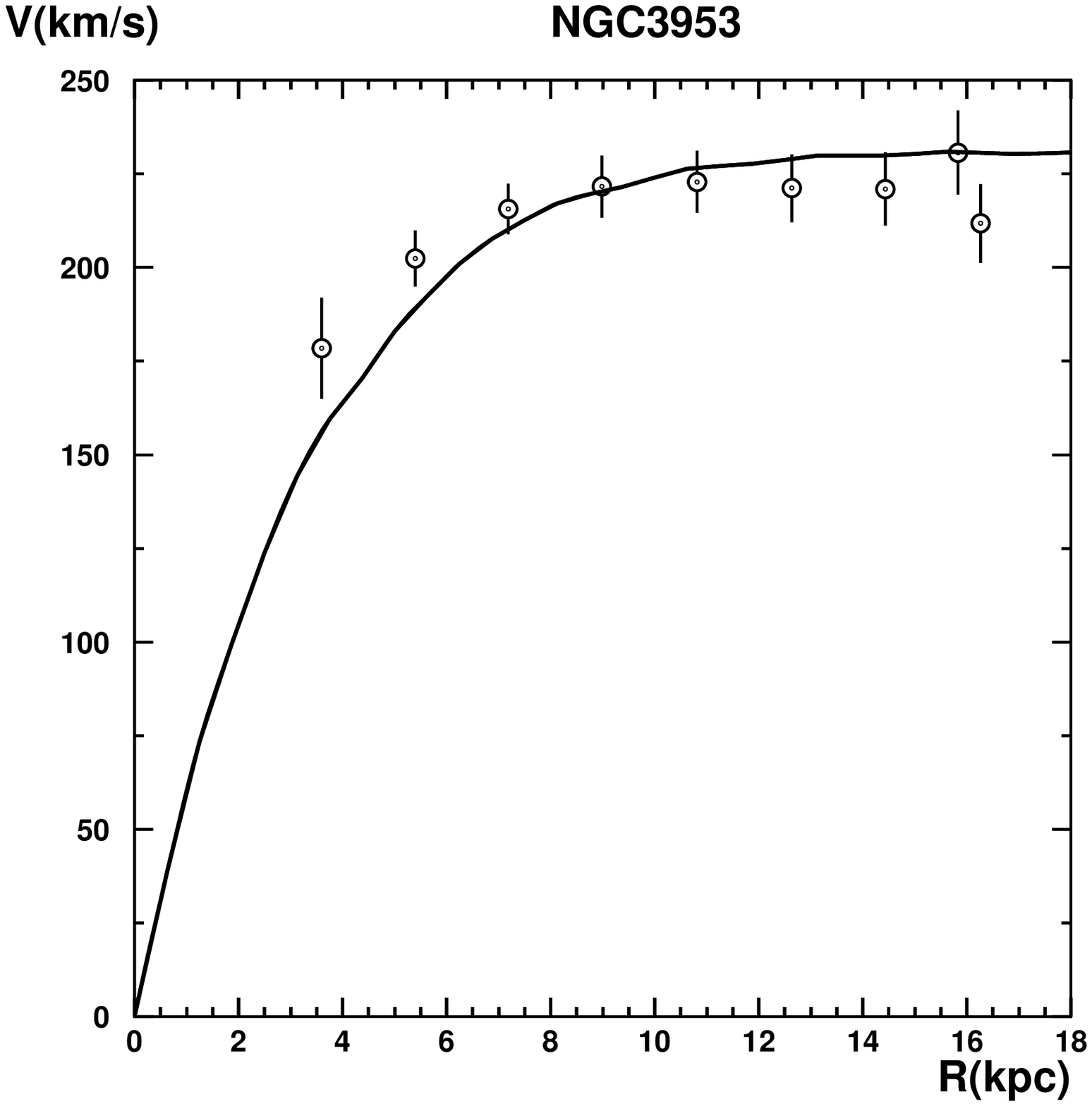}&
\includegraphics[width=60mm]{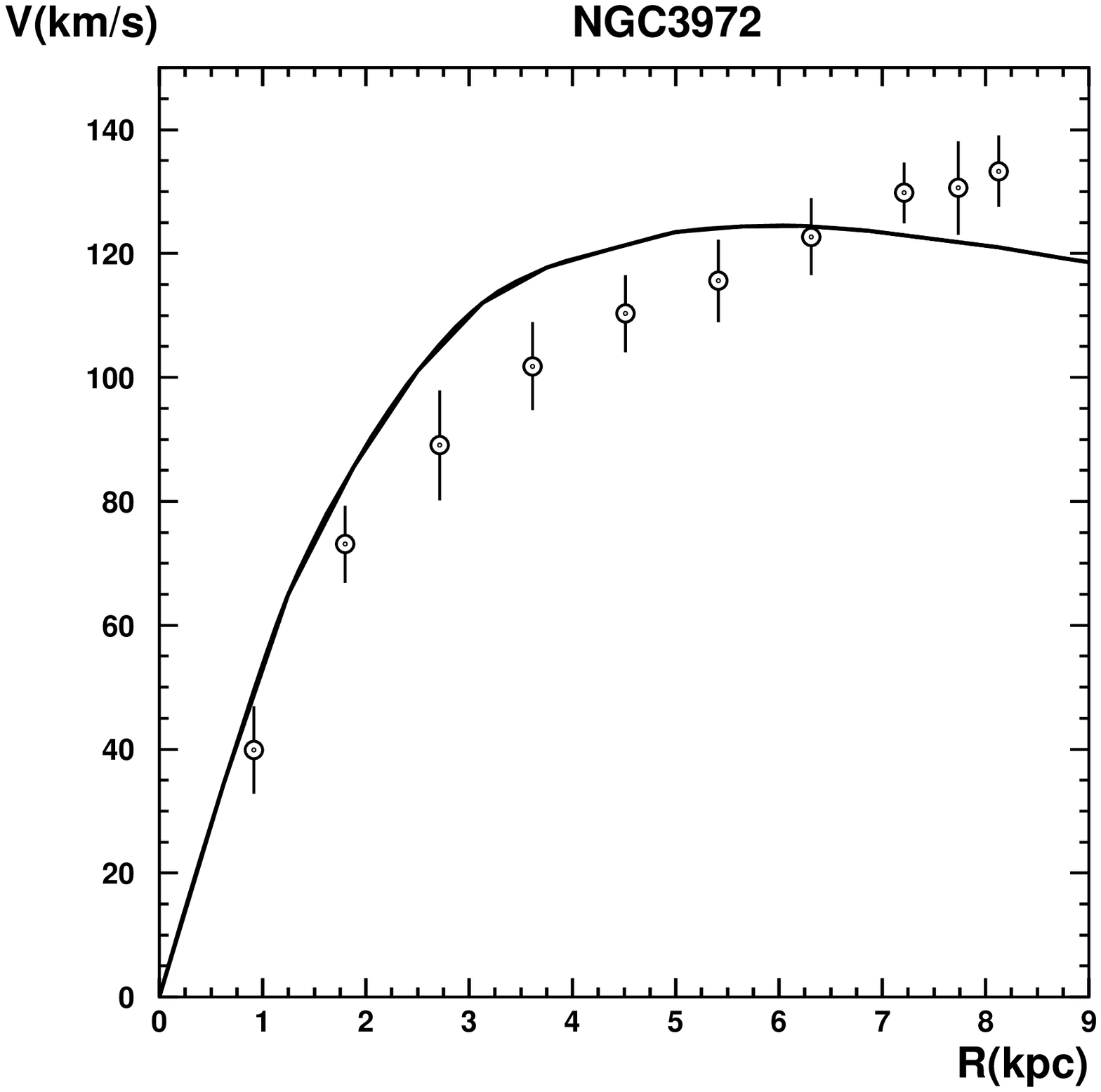}&
\includegraphics[width=60mm]{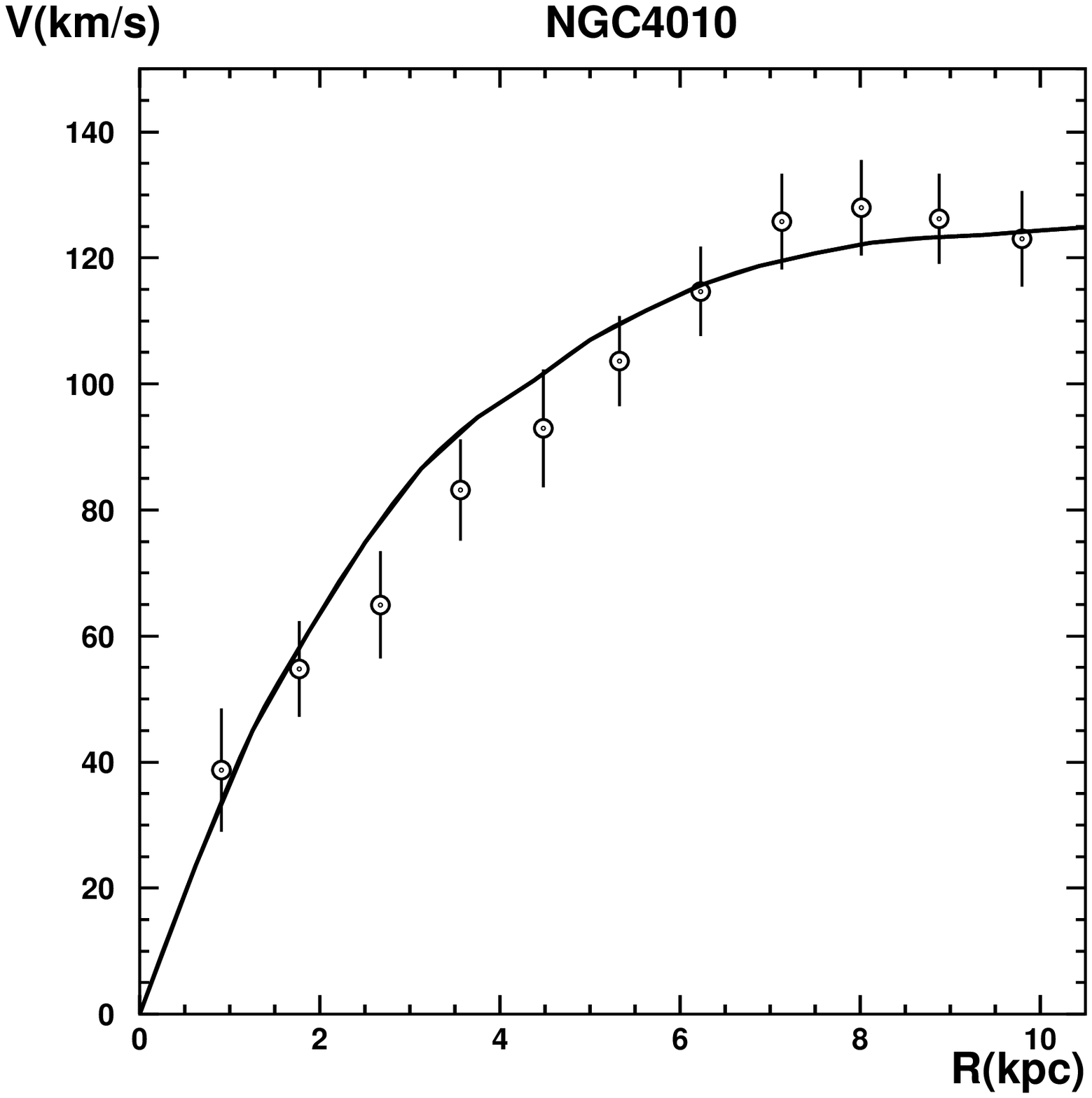}\\
\end{tabular}
\end{center}
\caption {The best fit to the rotation velocity curves of the
Ursa-Major sample. We fix $\alpha =8.89$ and $\mu = 0.042~{\rm kpc}^{-1}$
from the fits to the THINGS catalogue. We take the stellar
mass-to-light ratio $\Upsilon_\star$ as the  free degree of freedom.
\label{fig3} }
\newpage
\end{figure*}
%\clearpage

\setcounter{figure}{1}
\begin{figure*}
\begin{center}
\begin{tabular}{ccc}
\includegraphics[width=60mm]{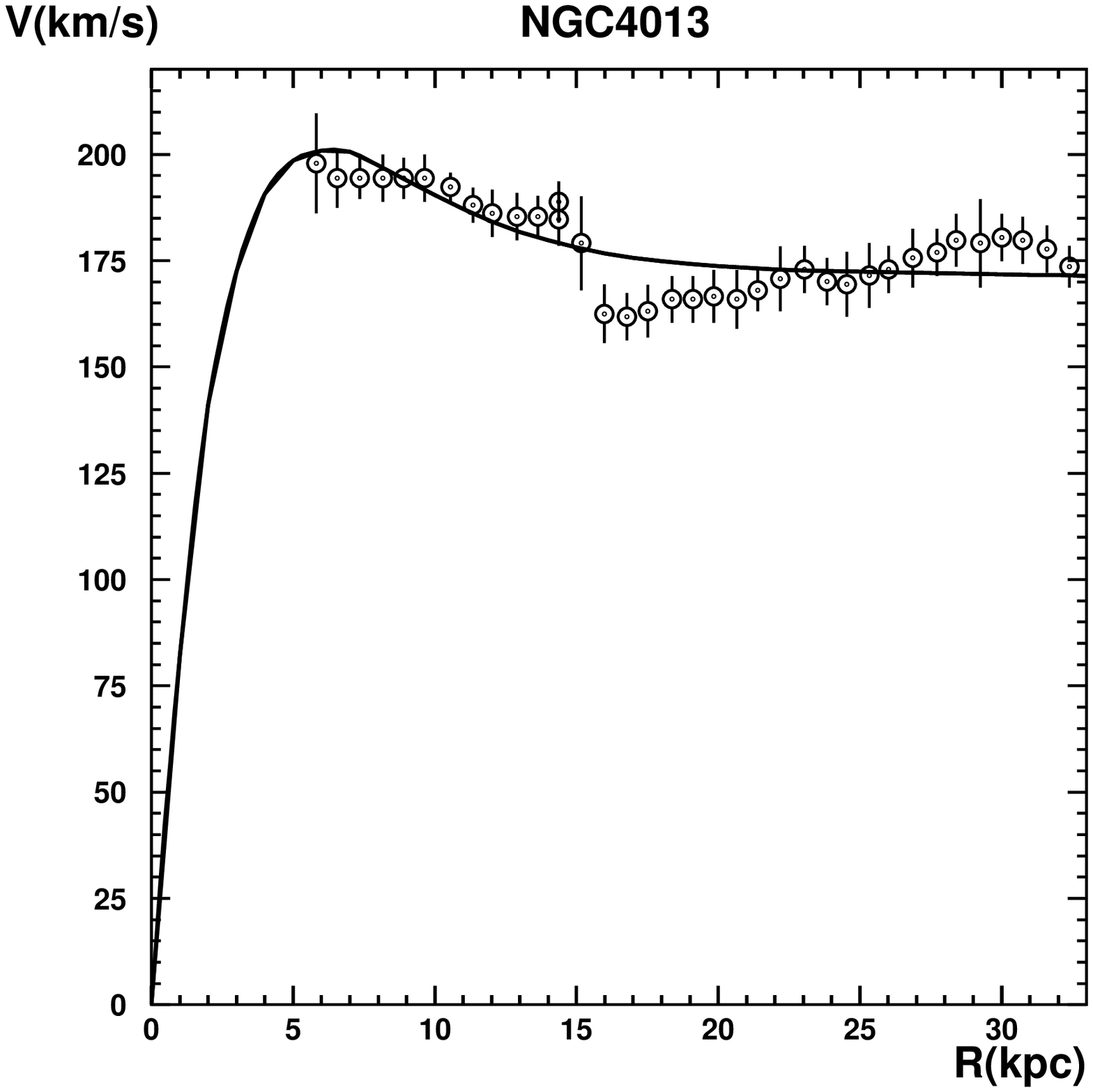} &
\includegraphics[width=60mm]{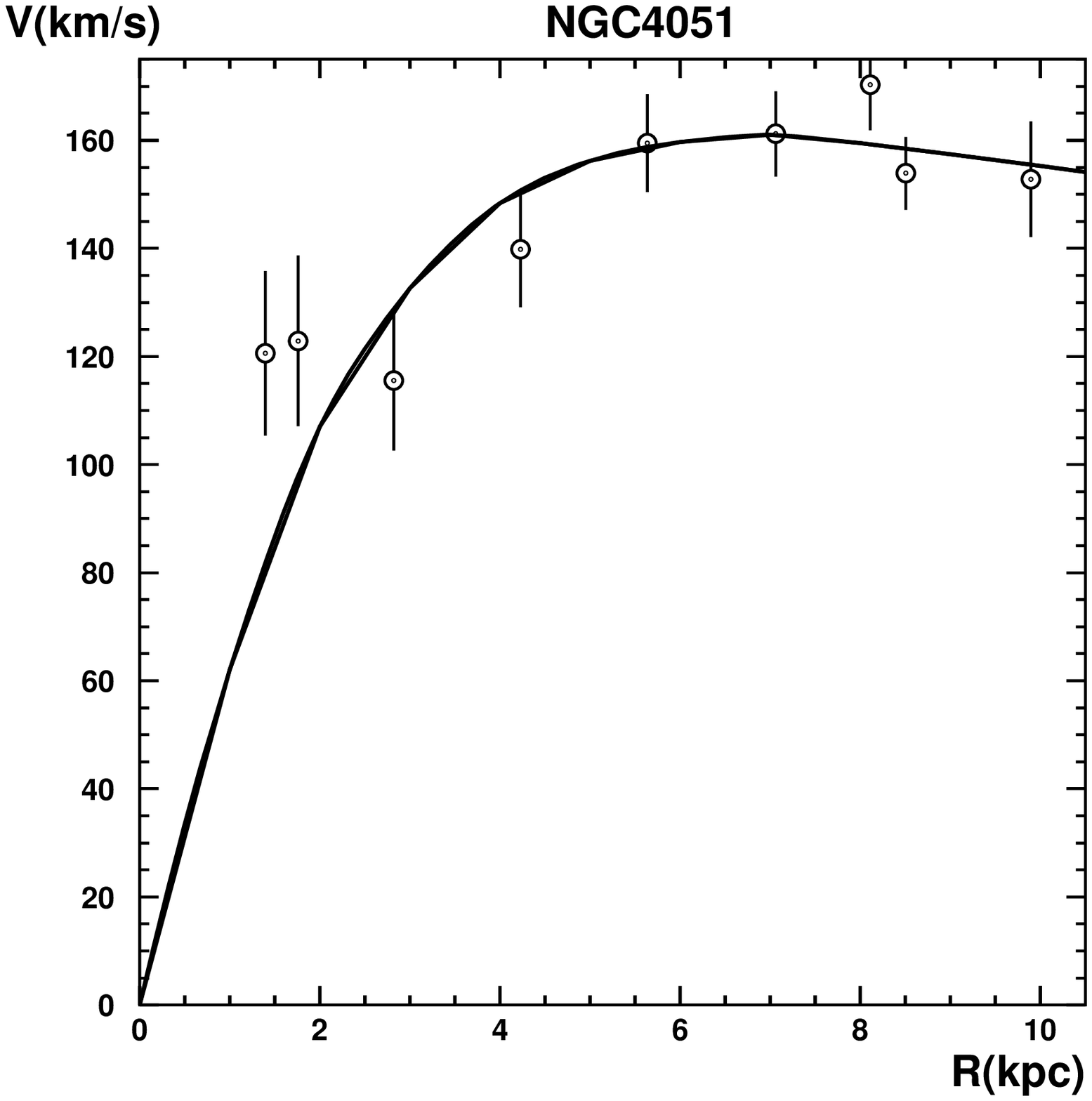}&
\includegraphics[width=60mm]{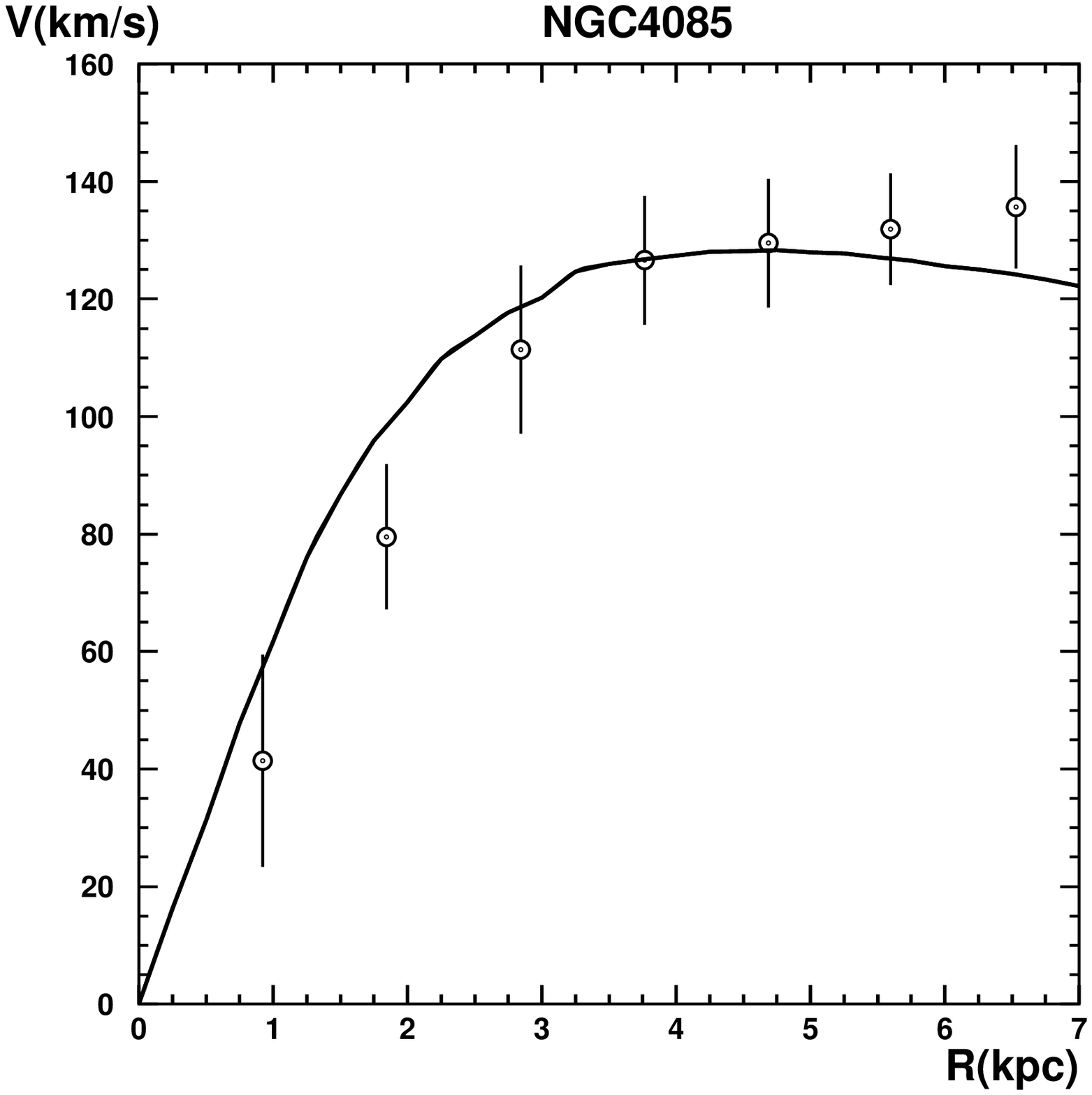} \\
\includegraphics[width=60mm]{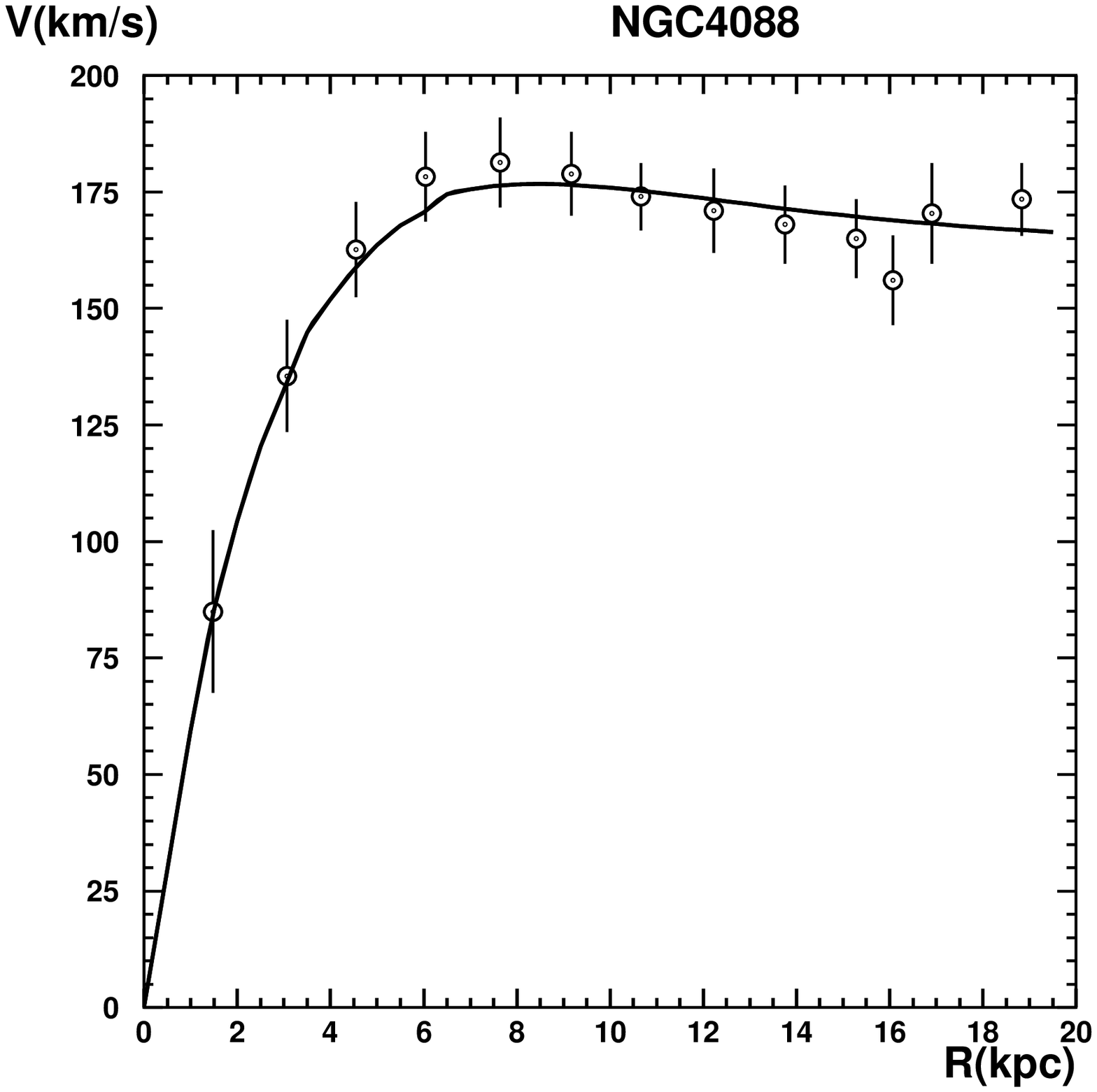} &
\includegraphics[width=60mm]{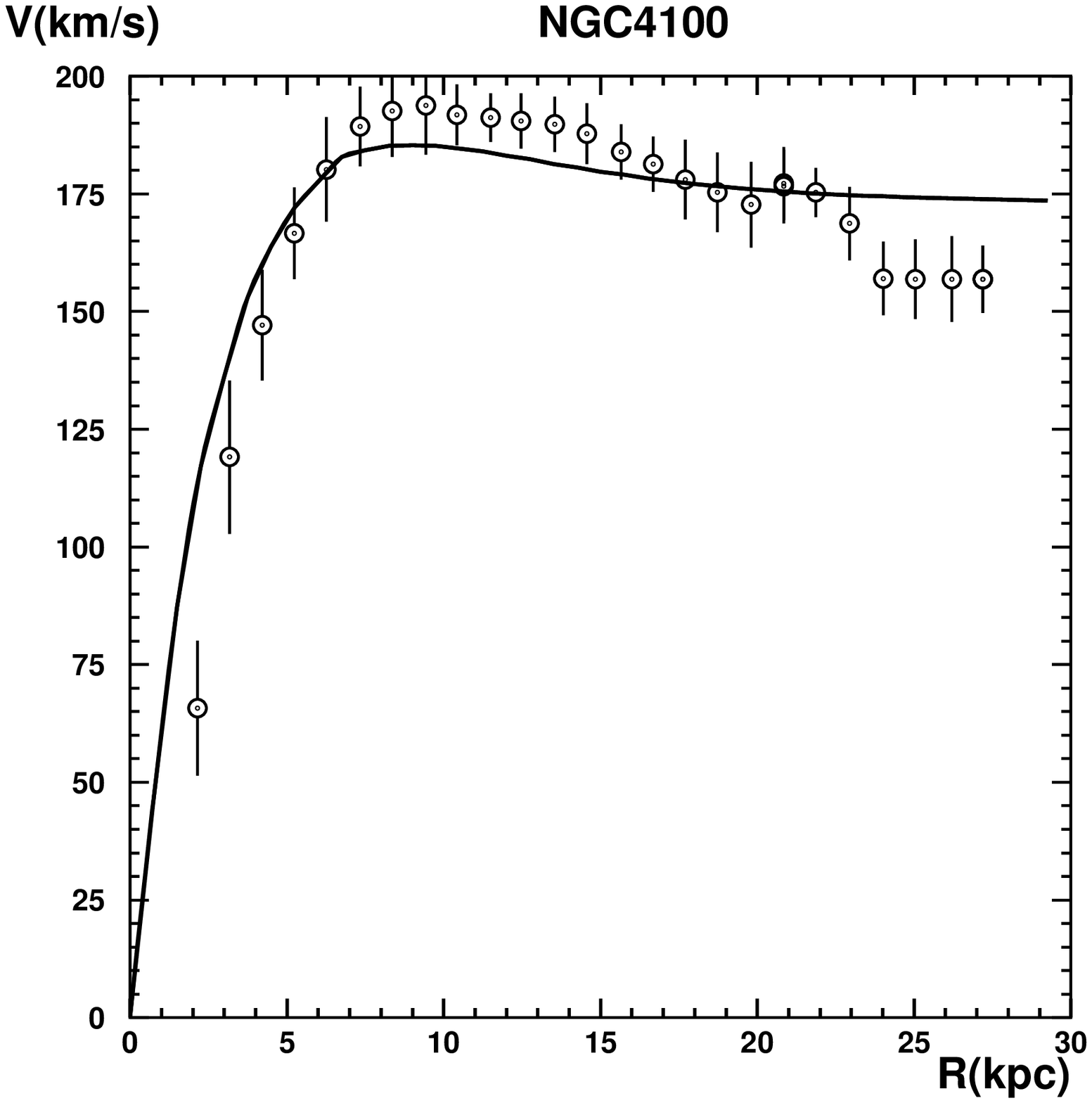}&
\includegraphics[width=60mm]{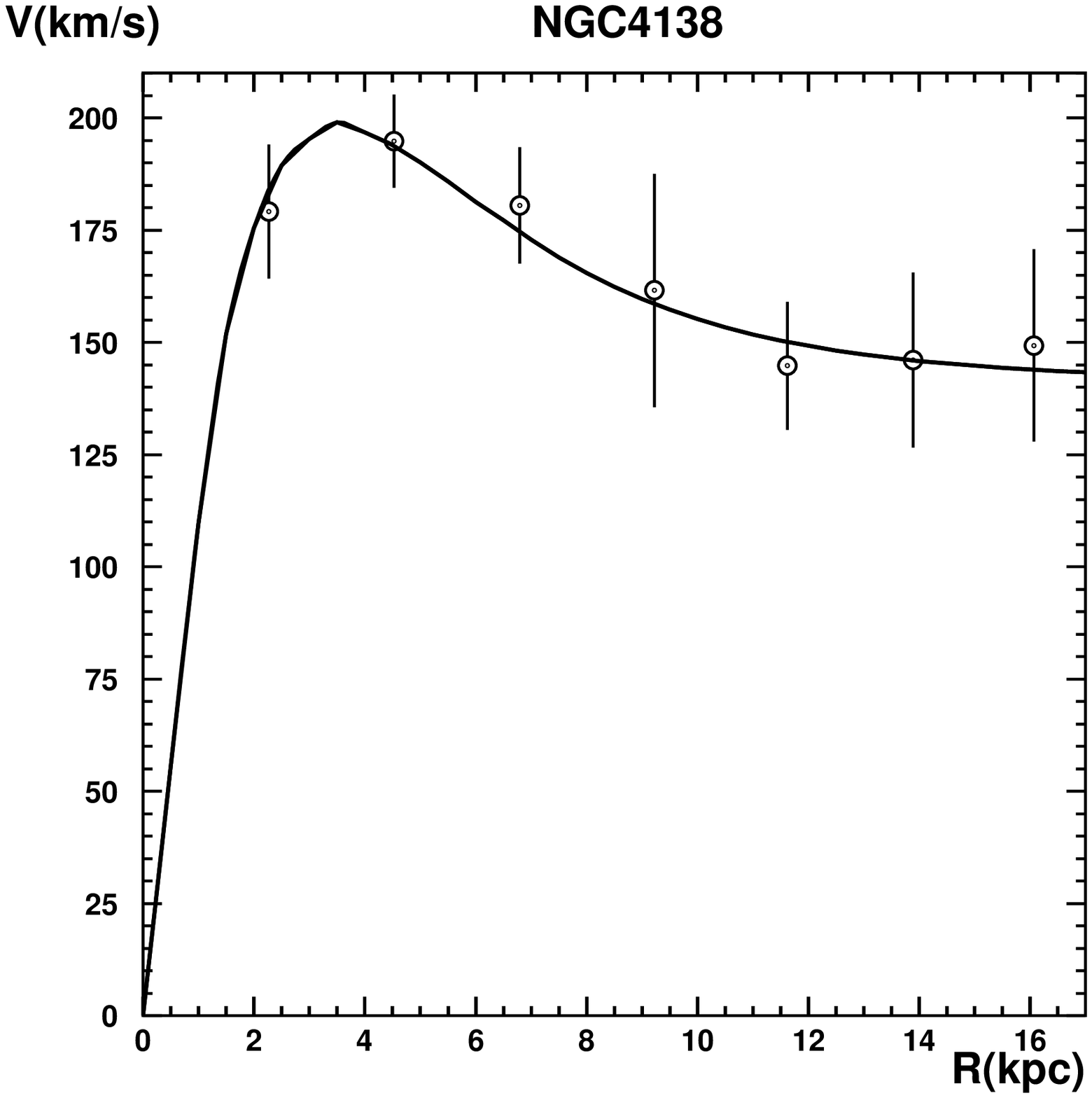}\\
\includegraphics[width=60mm]{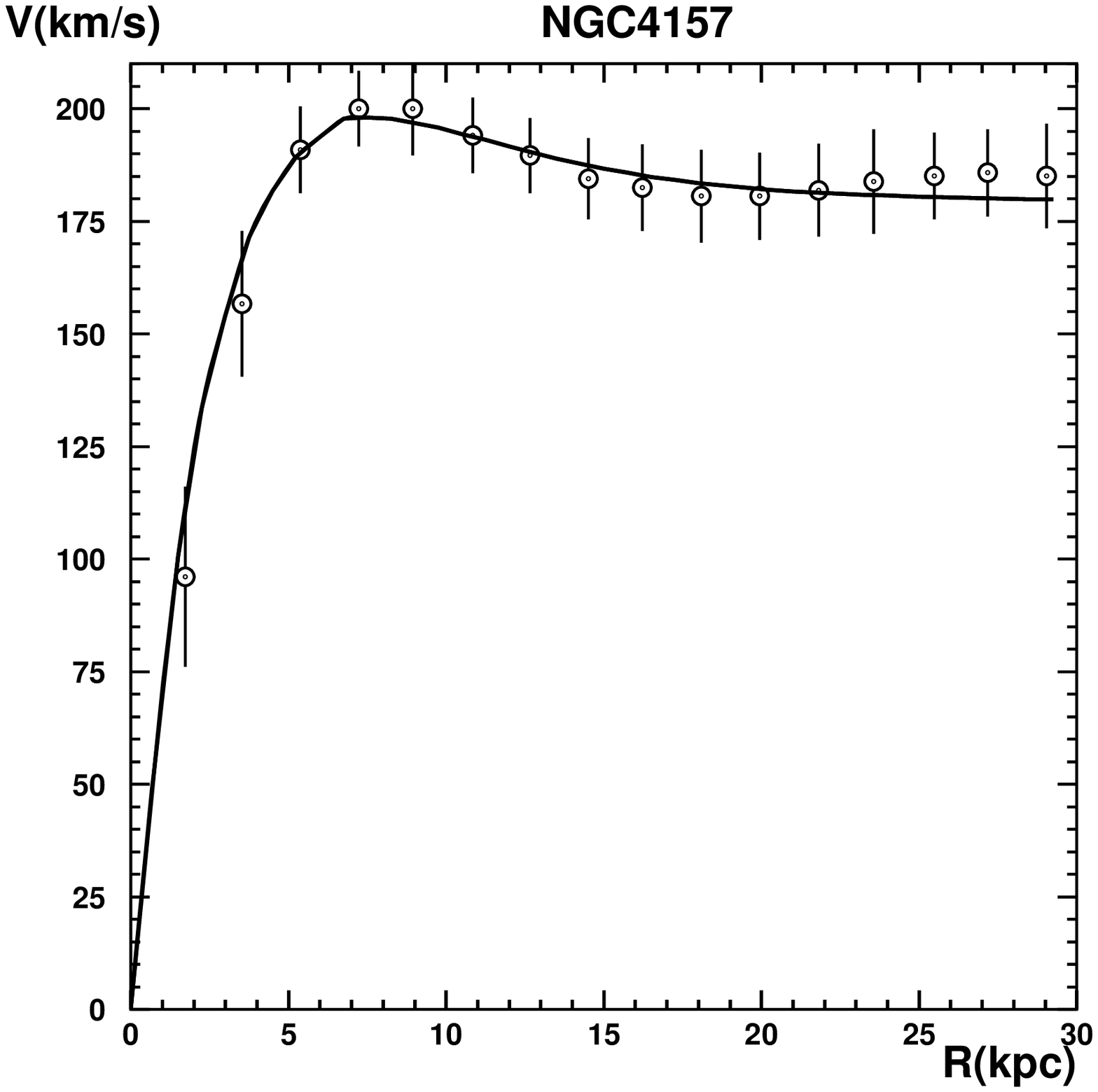}&
\includegraphics[width=60mm]{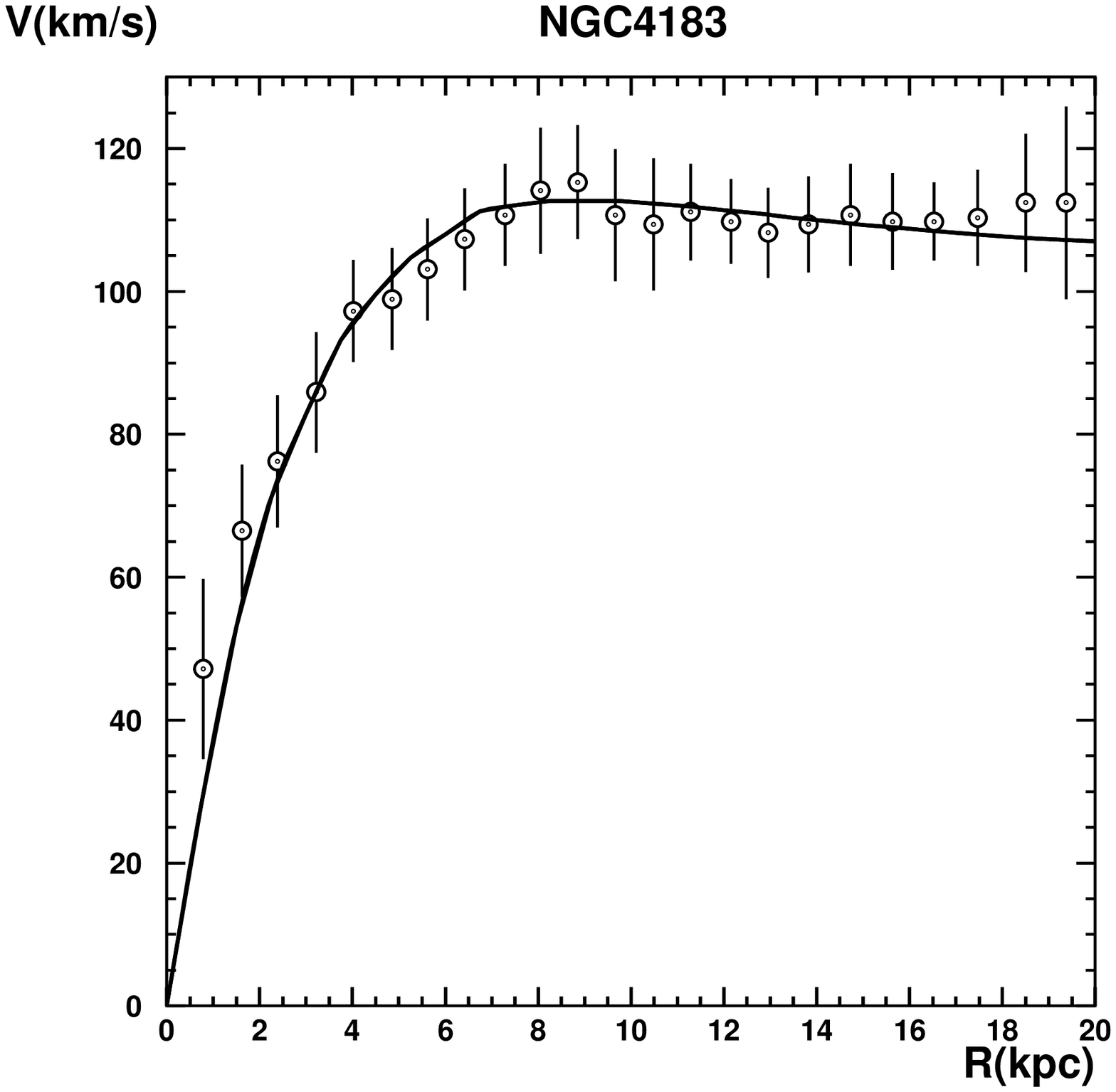}&
\includegraphics[width=60mm]{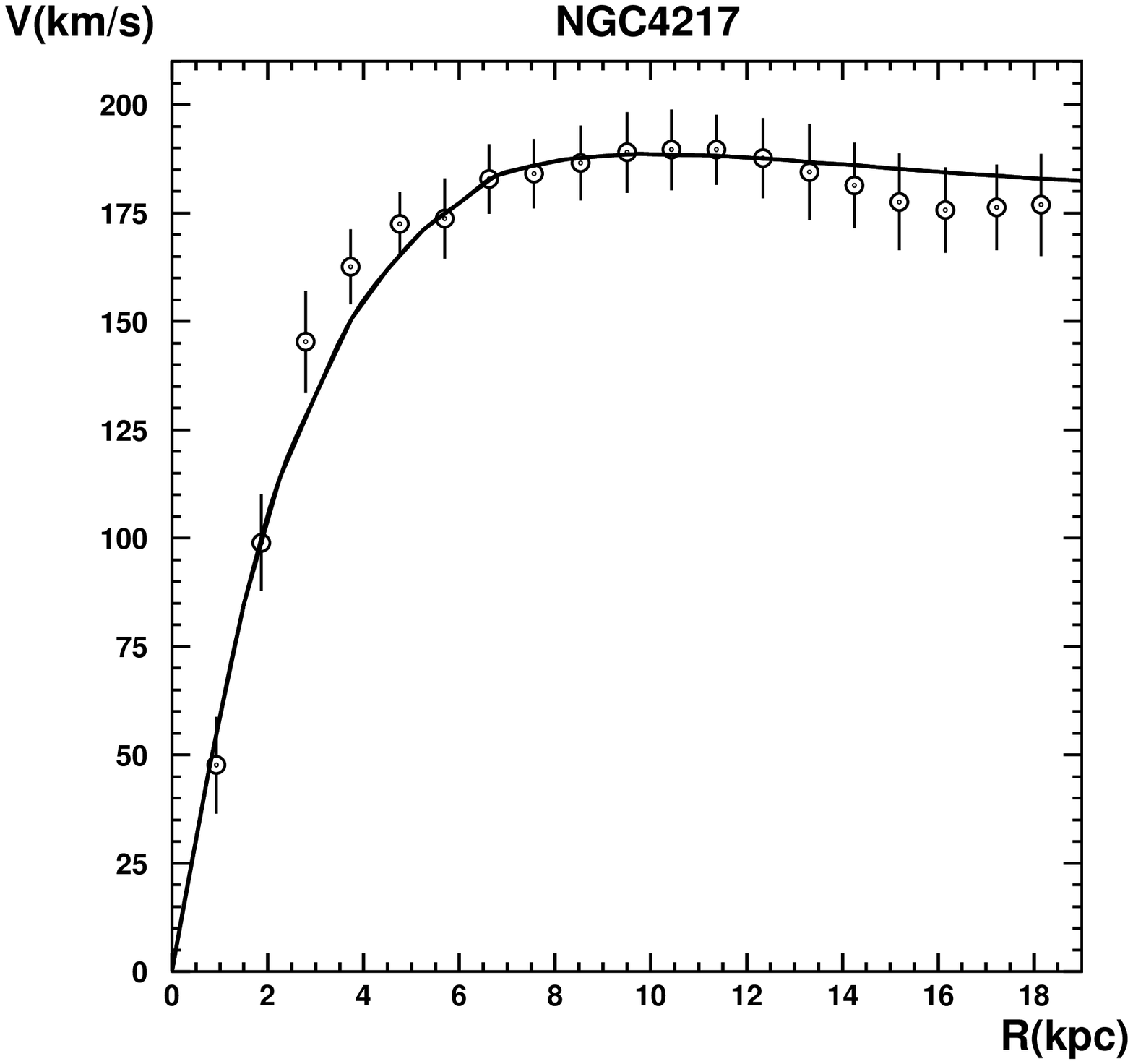}\\
\end{tabular}
\end{center}
\caption {--continued}
\newpage
\end{figure*}
%\clearpage

\setcounter{figure}{1}
\begin{figure*}
\begin{center}
\begin{tabular}{ccc}
\includegraphics[width=60mm]{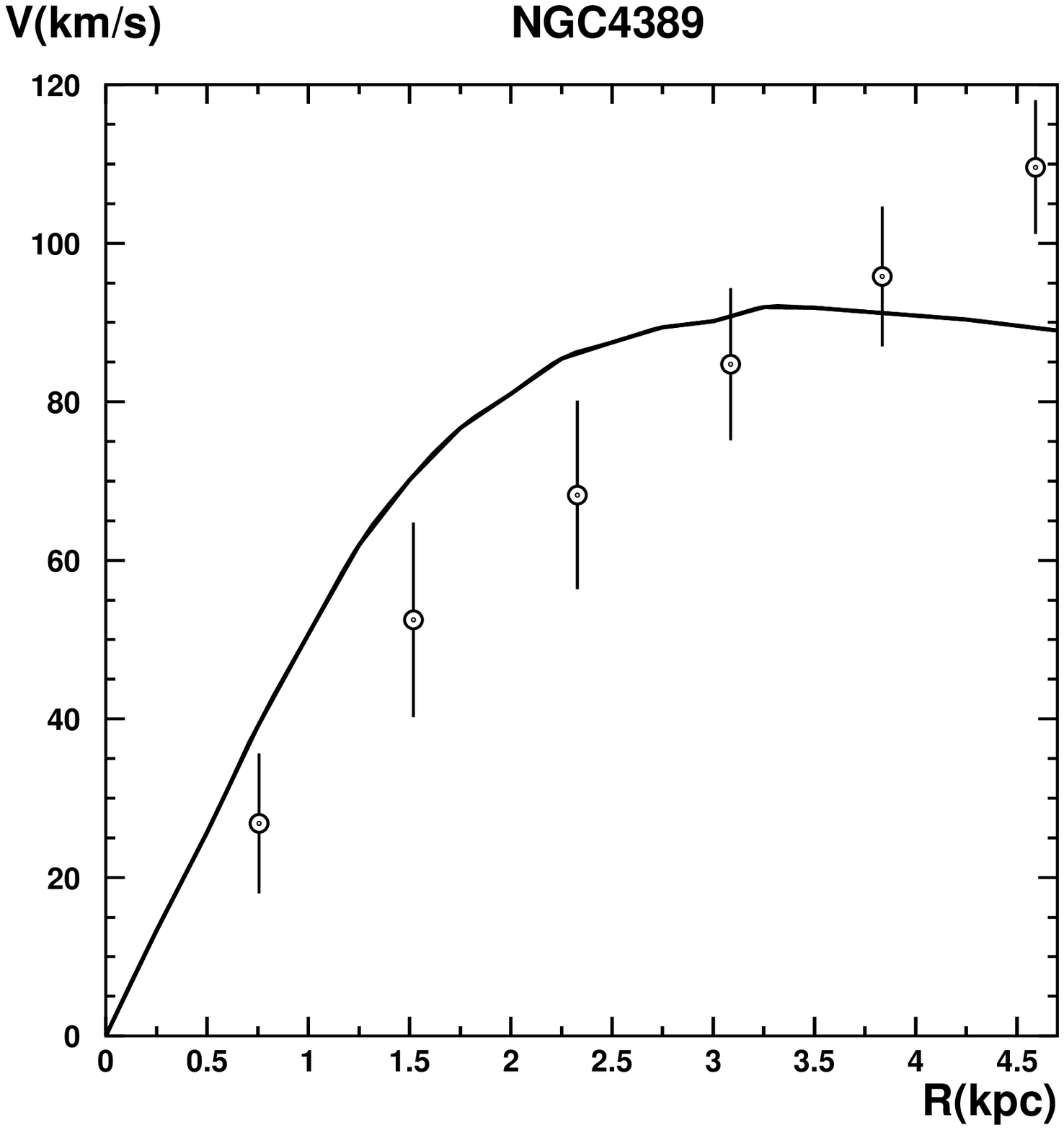} &
\includegraphics[width=60mm]{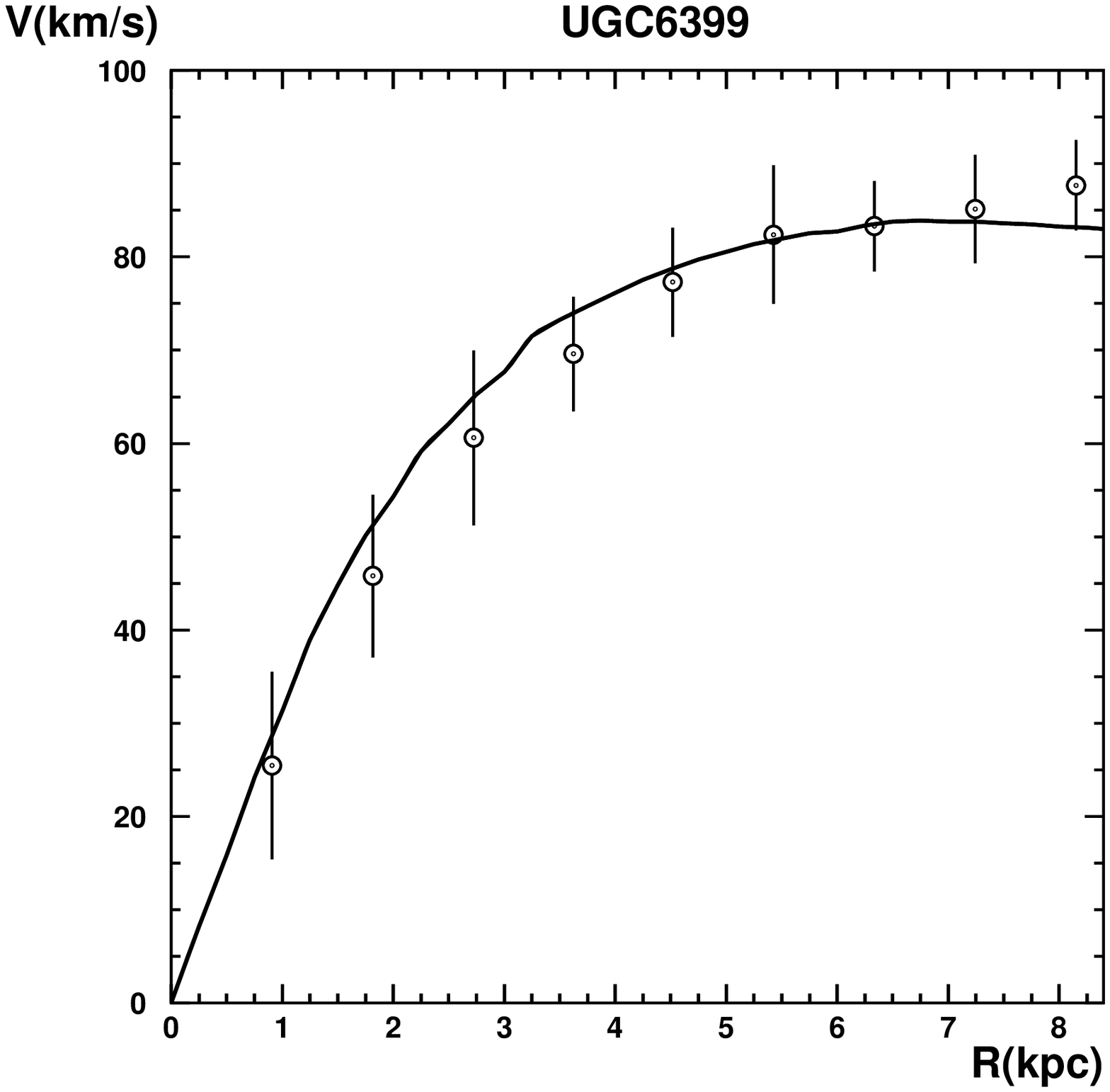}&
\includegraphics[width=60mm]{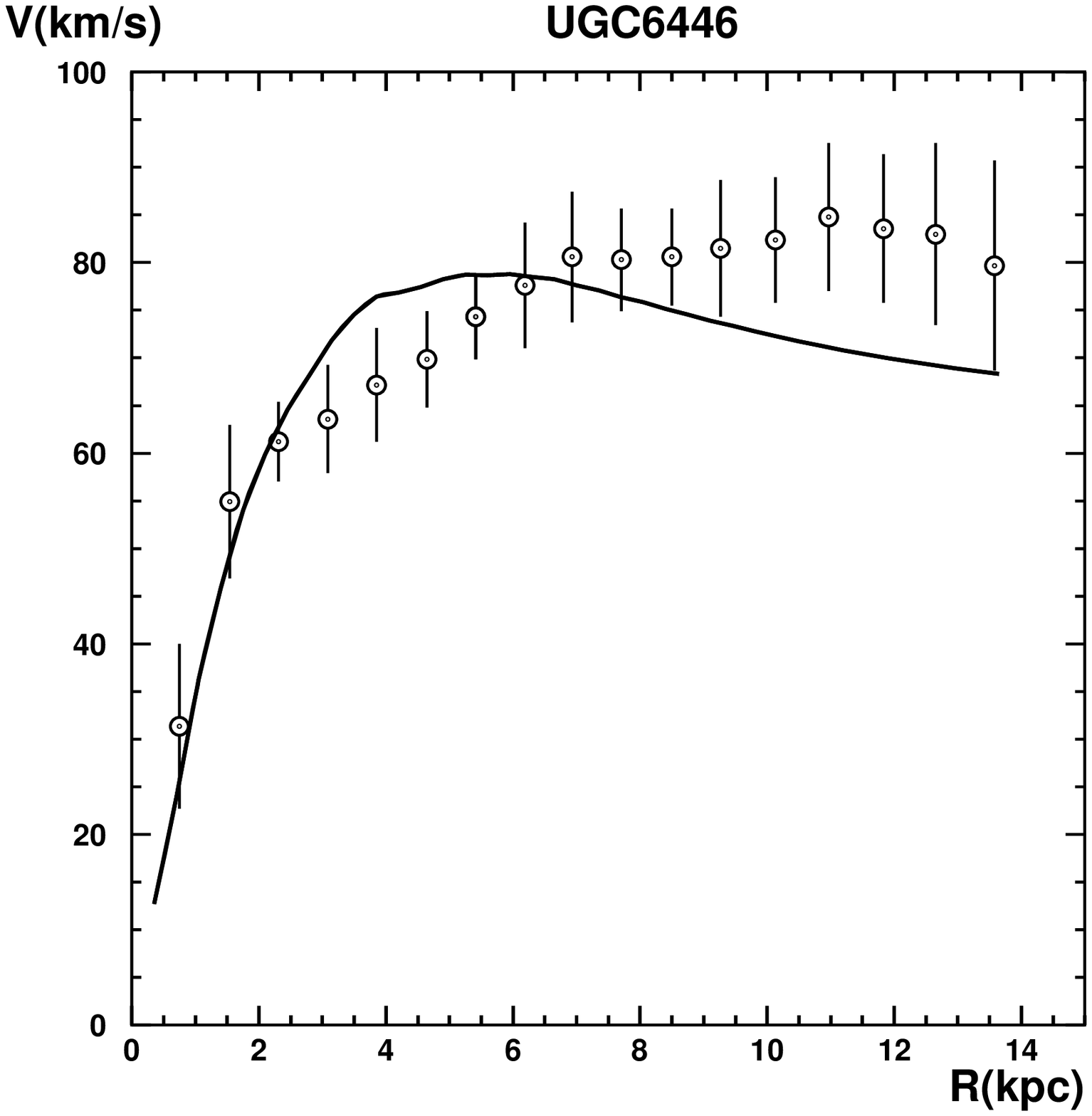} \\
\includegraphics[width=60mm]{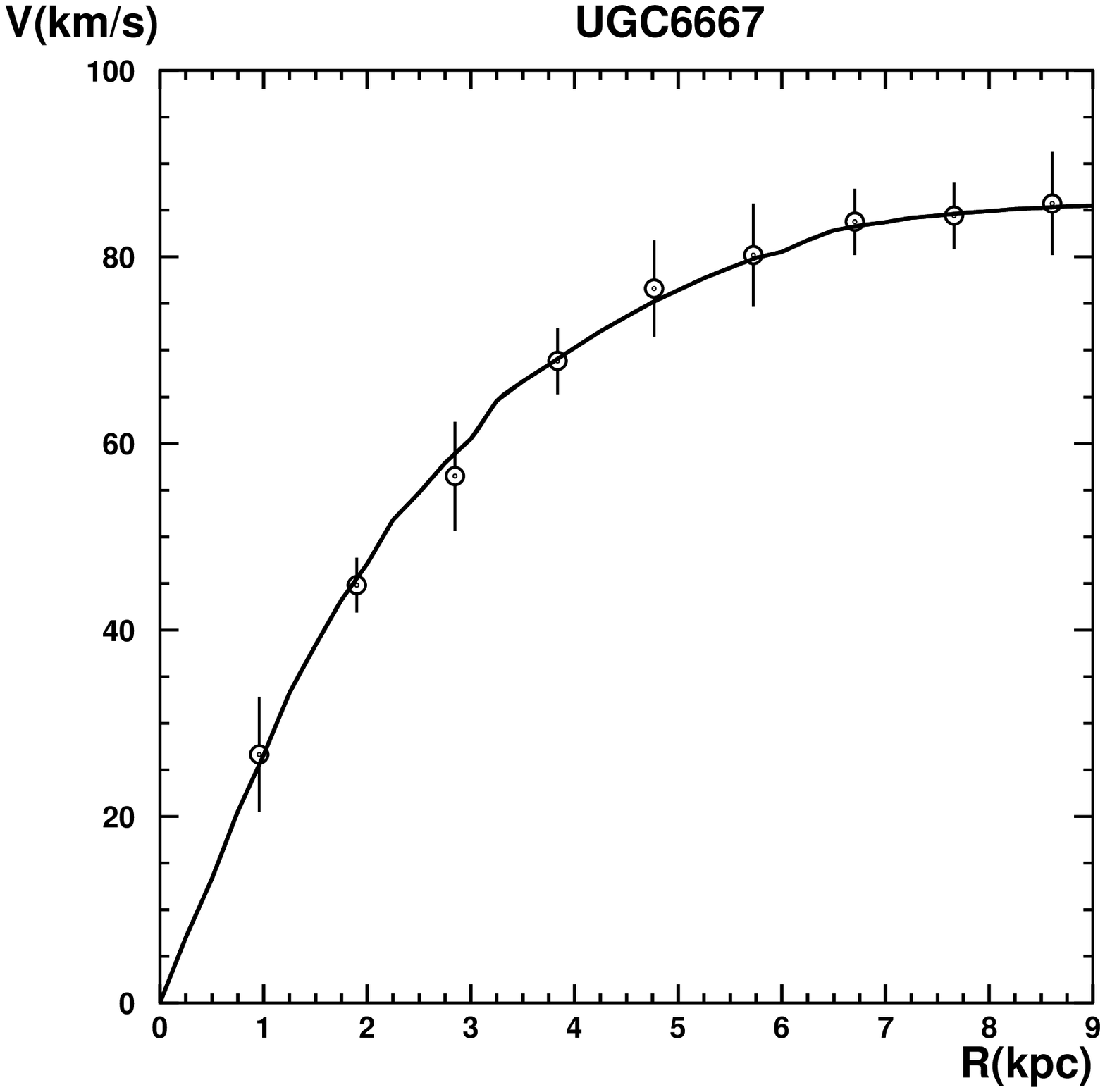} &
\includegraphics[width=60mm]{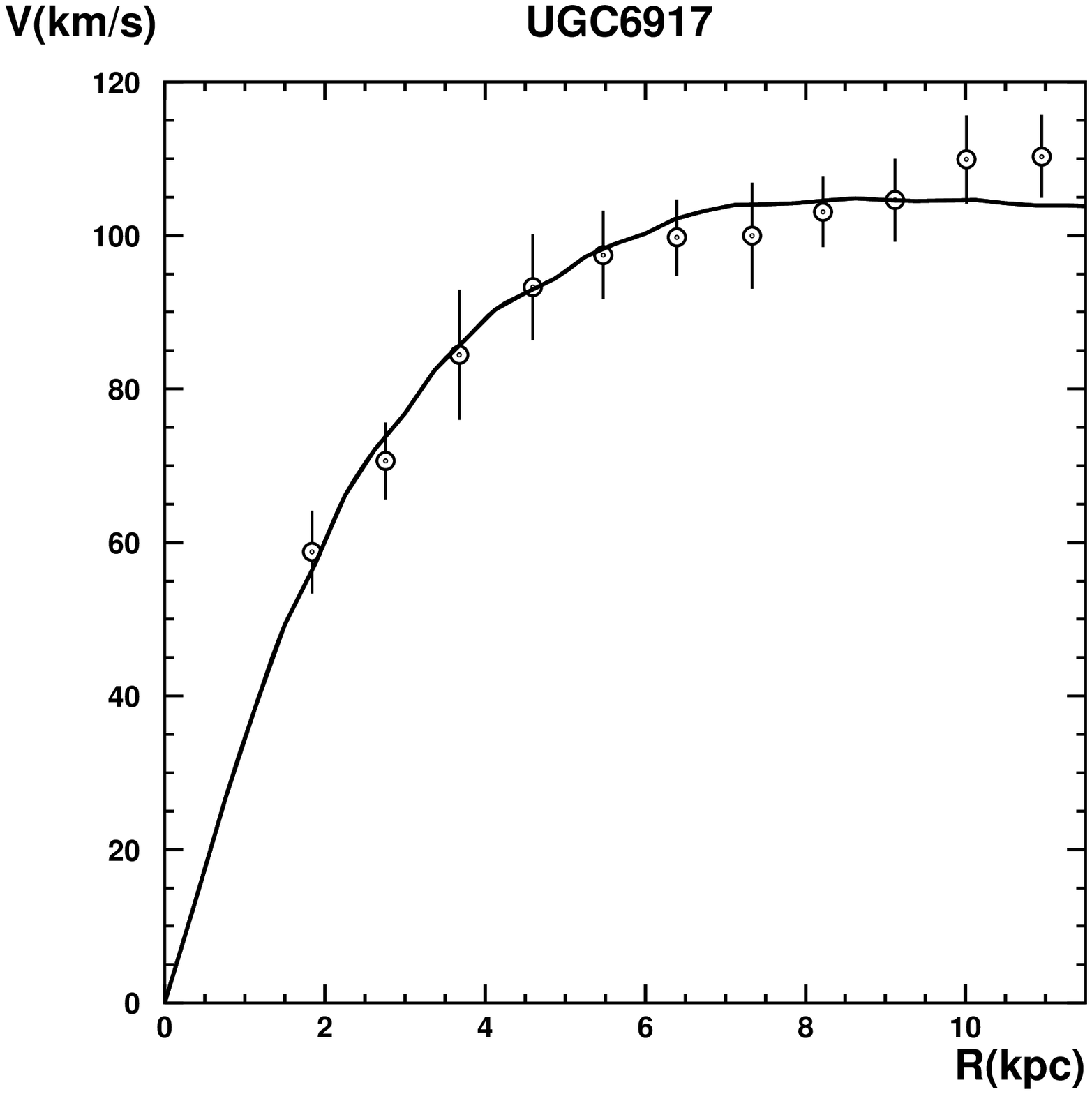}&
\includegraphics[width=60mm]{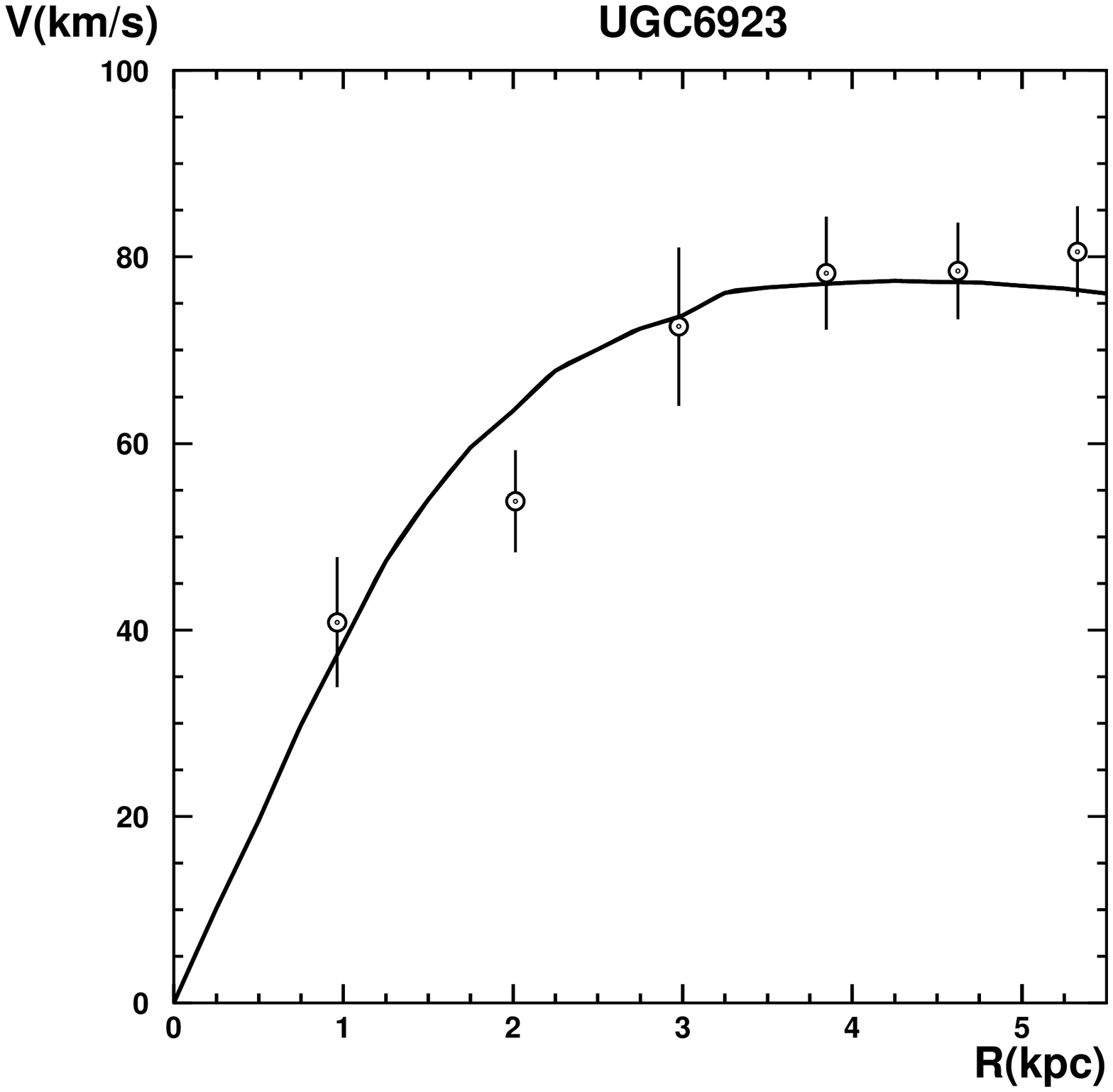}\\
\includegraphics[width=60mm]{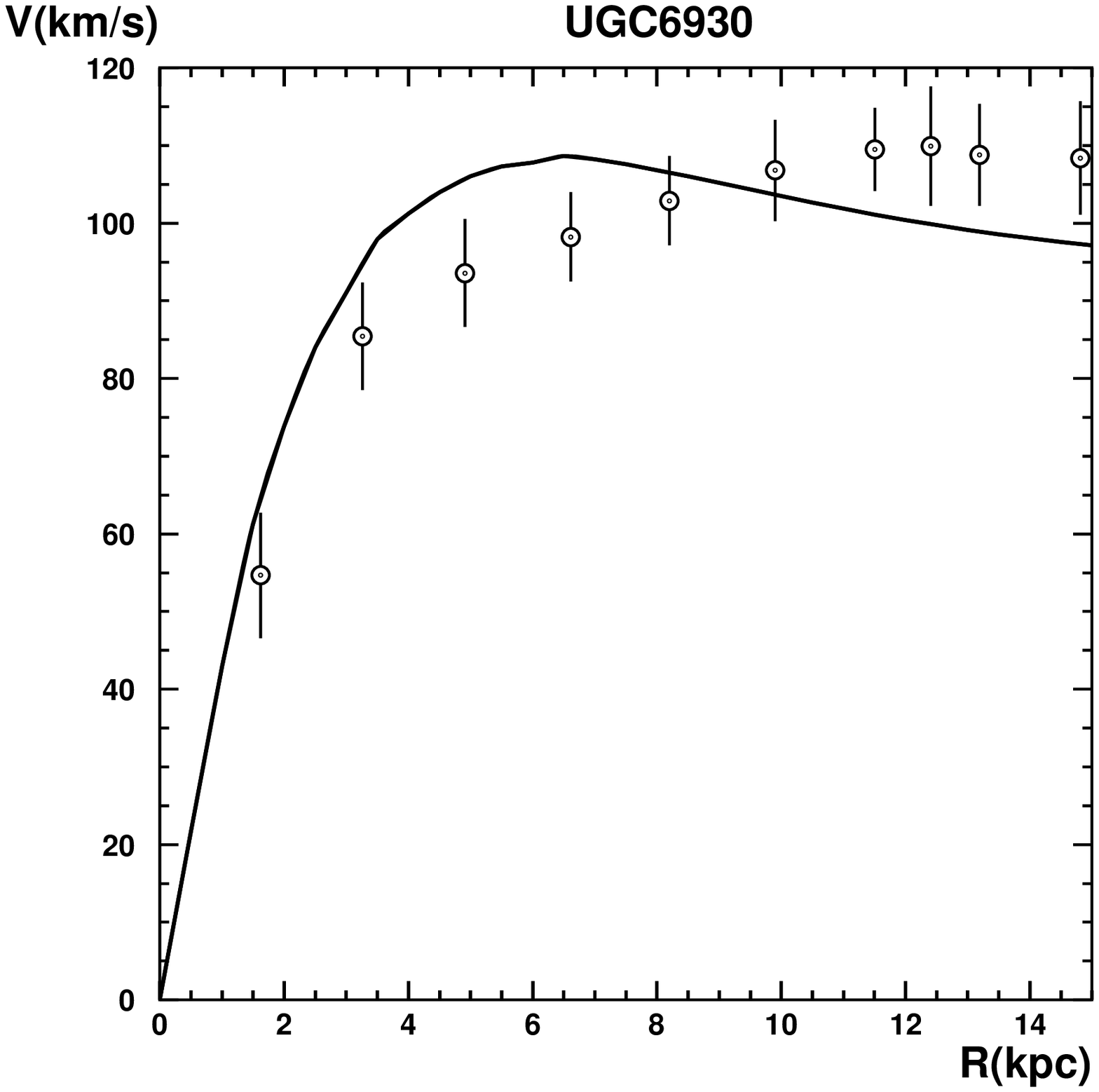}&
\includegraphics[width=60mm]{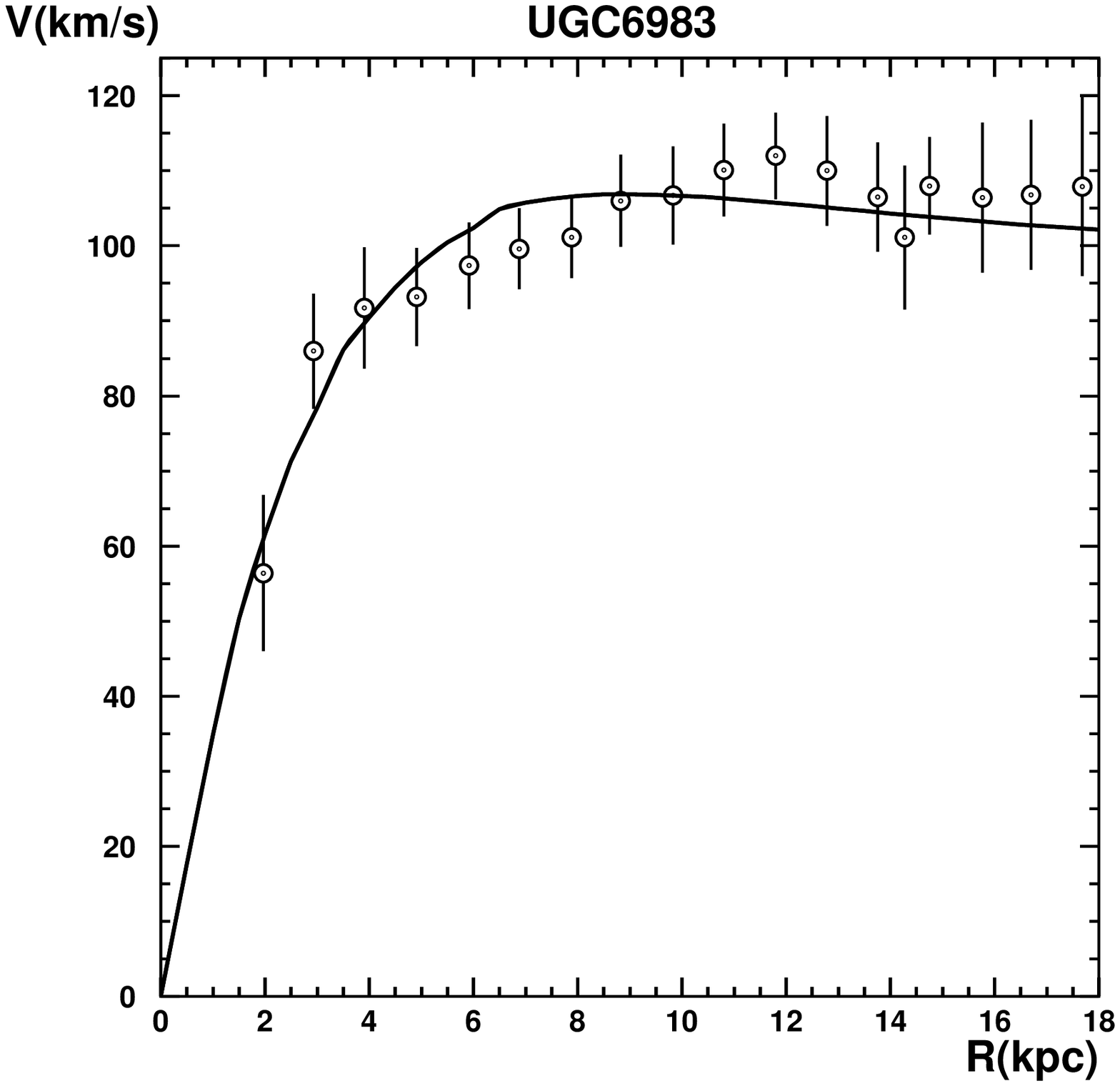}&
\includegraphics[width=60mm]{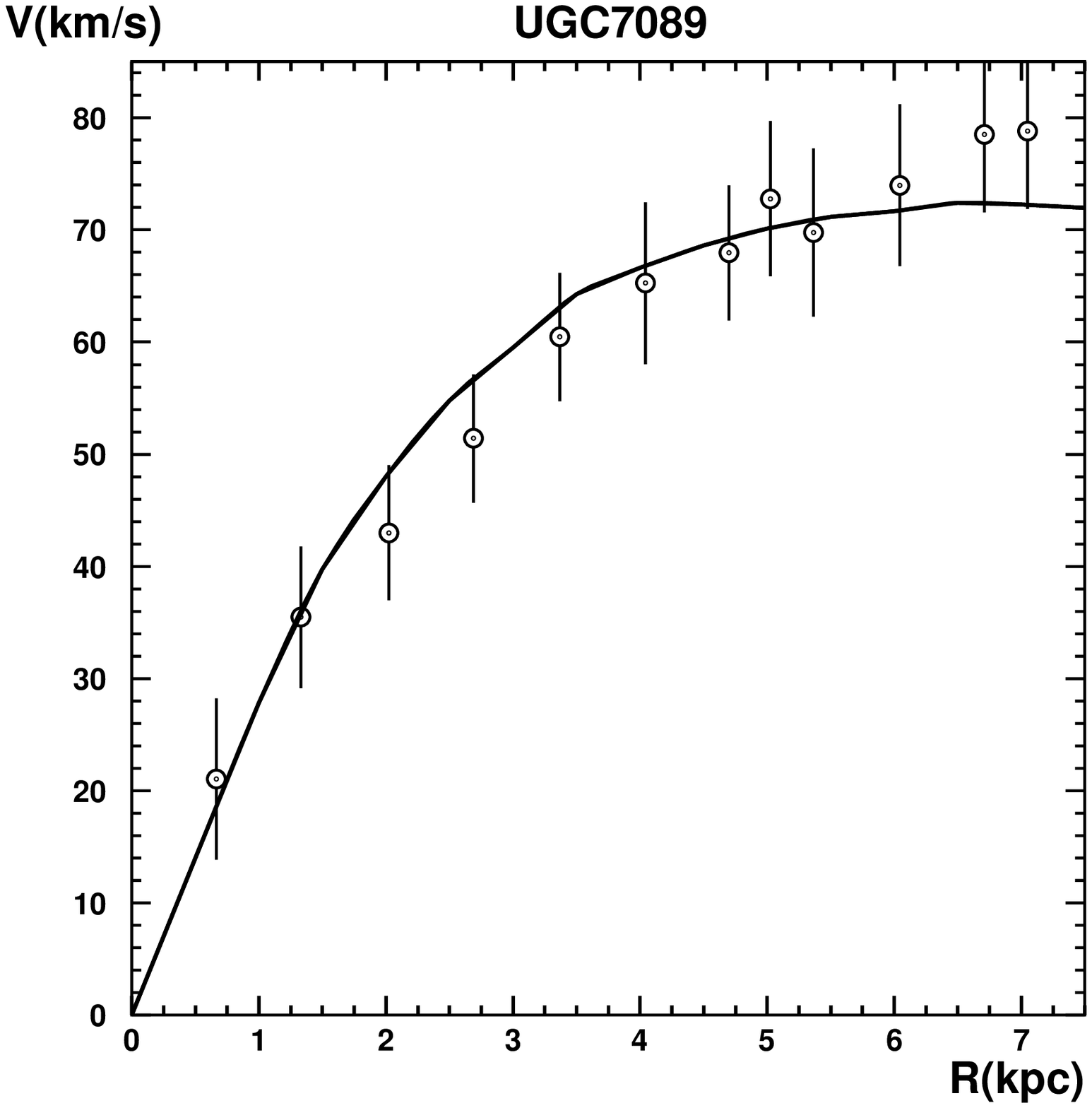}\\
\end{tabular}
\end{center}
\caption {--continued}
\newpage
\end{figure*}
%\clearpage

%\setcounter{figure}{3}
\begin{figure*}
\begin{center}
\begin{tabular}{ccc}
\includegraphics[width=55mm]{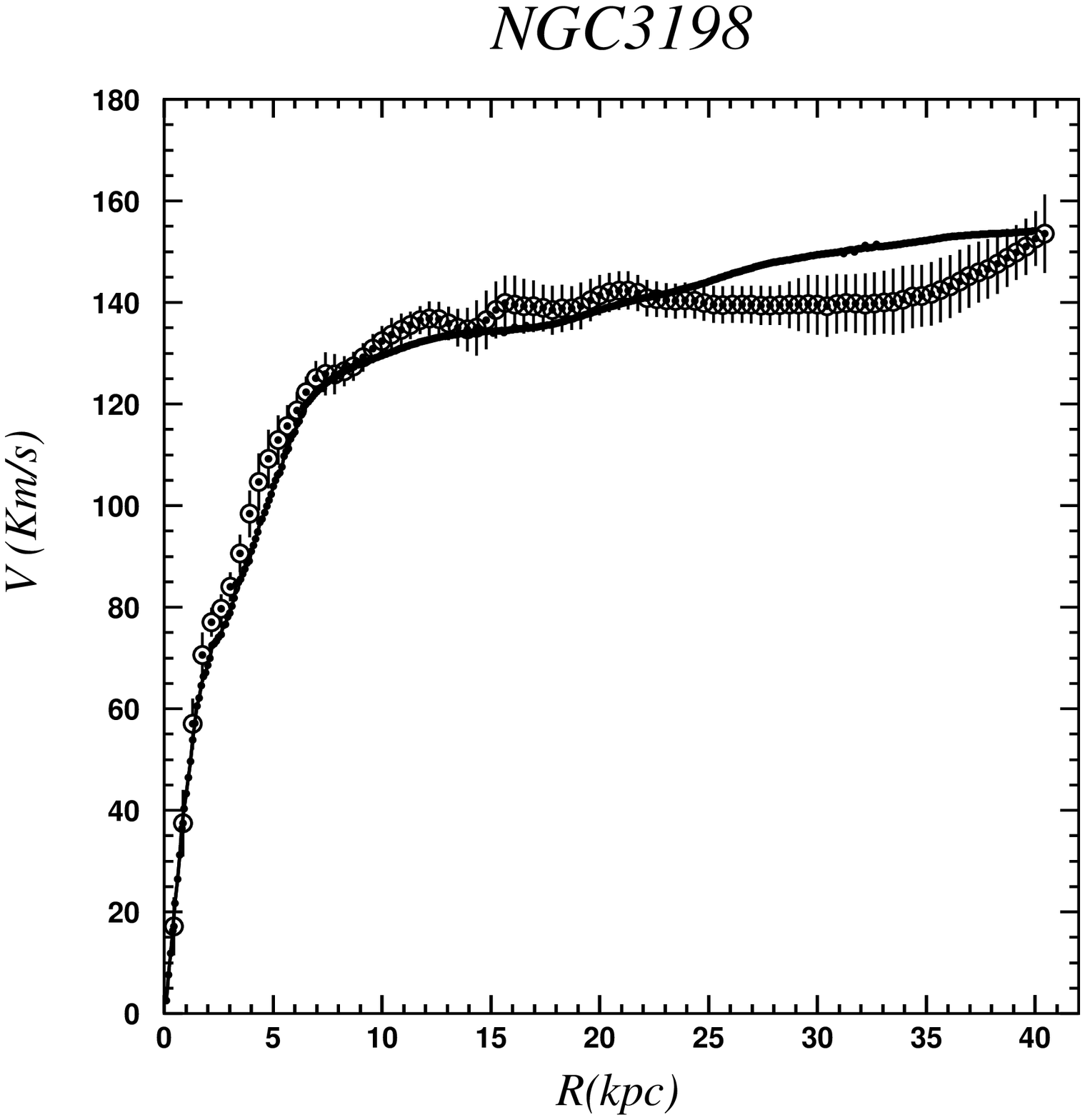} &
\includegraphics[width=55mm]{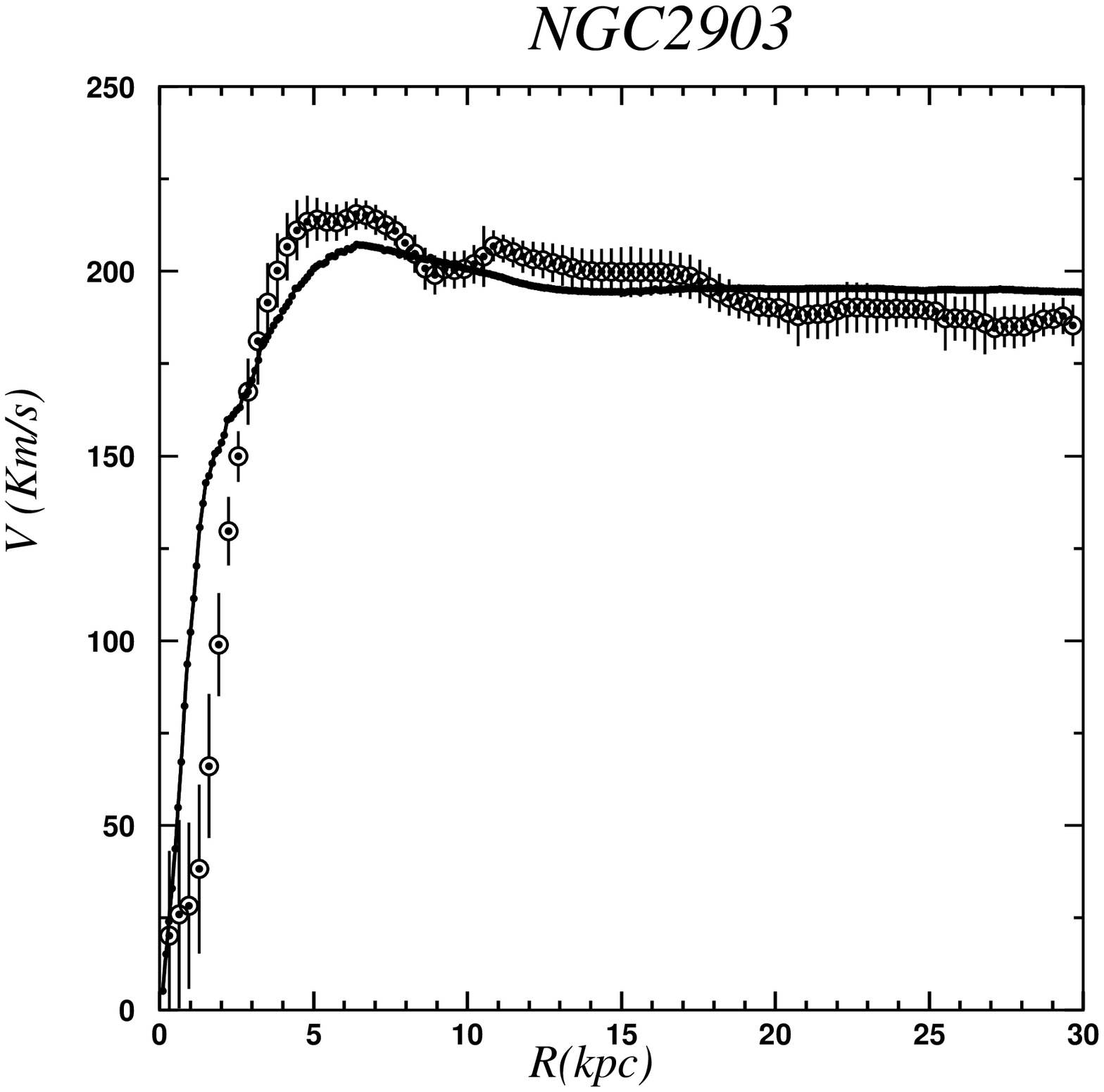}&
\includegraphics[width=55mm]{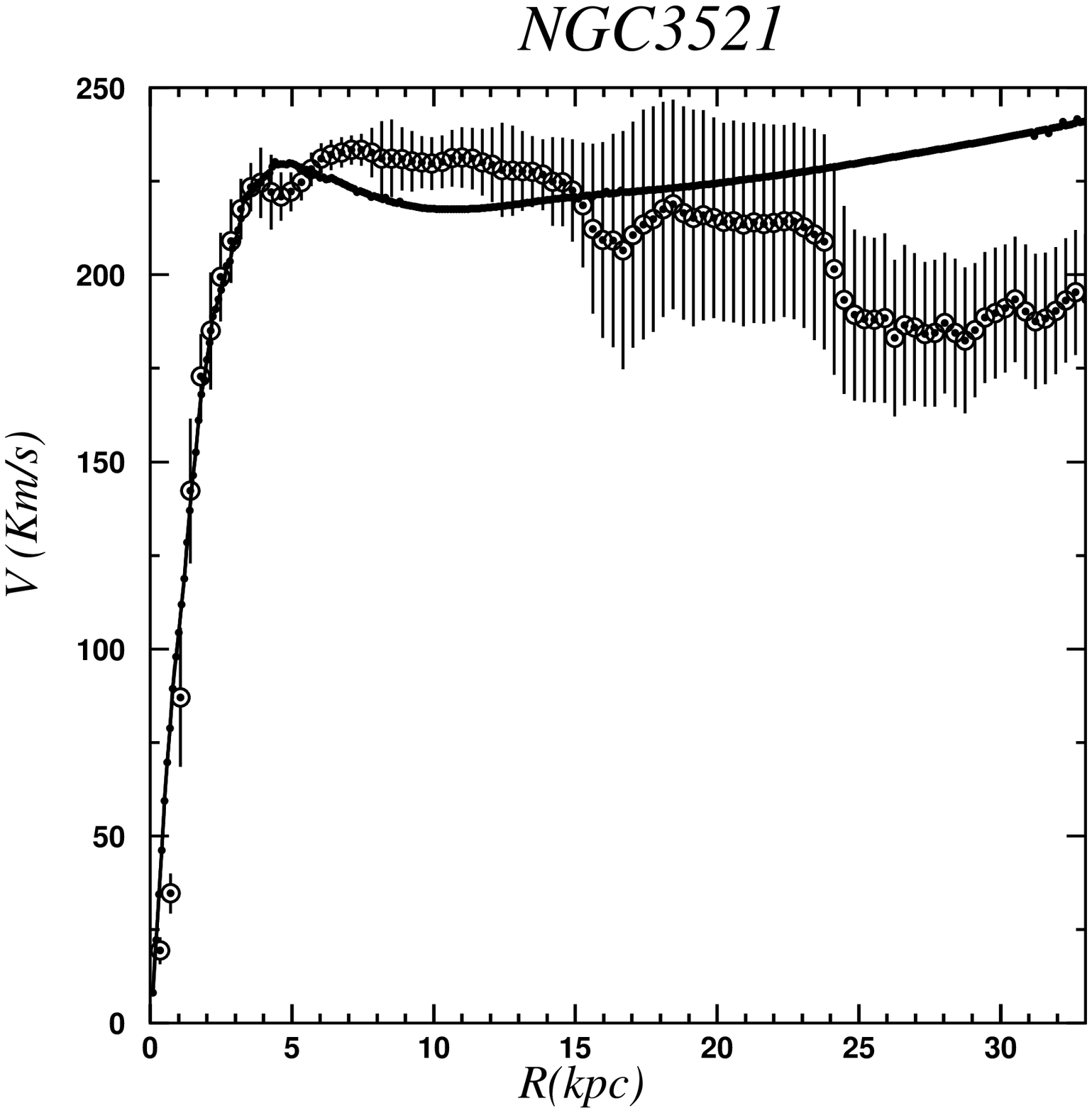} \\
\includegraphics[width=55mm]{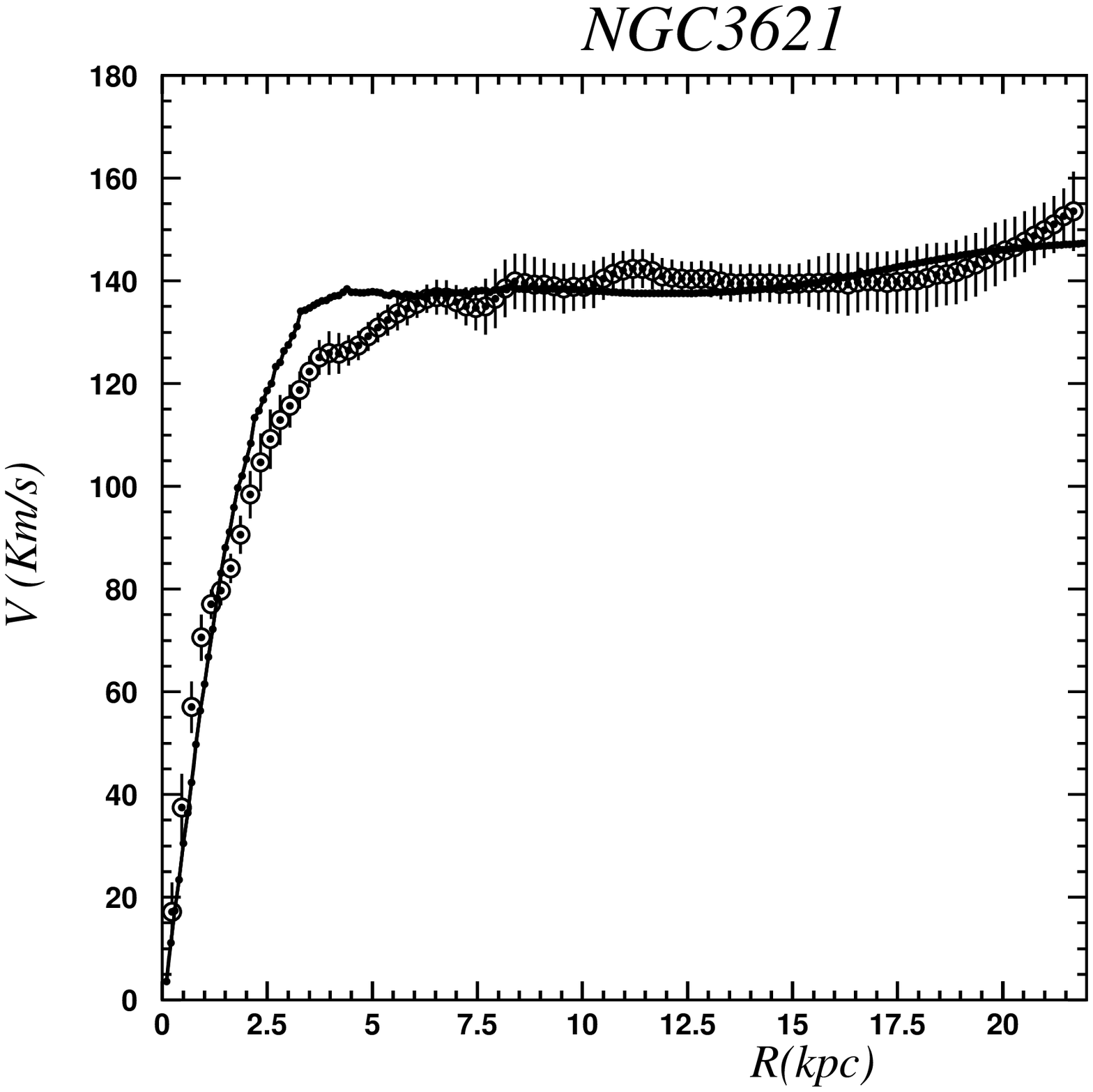} &
\includegraphics[width=55mm]{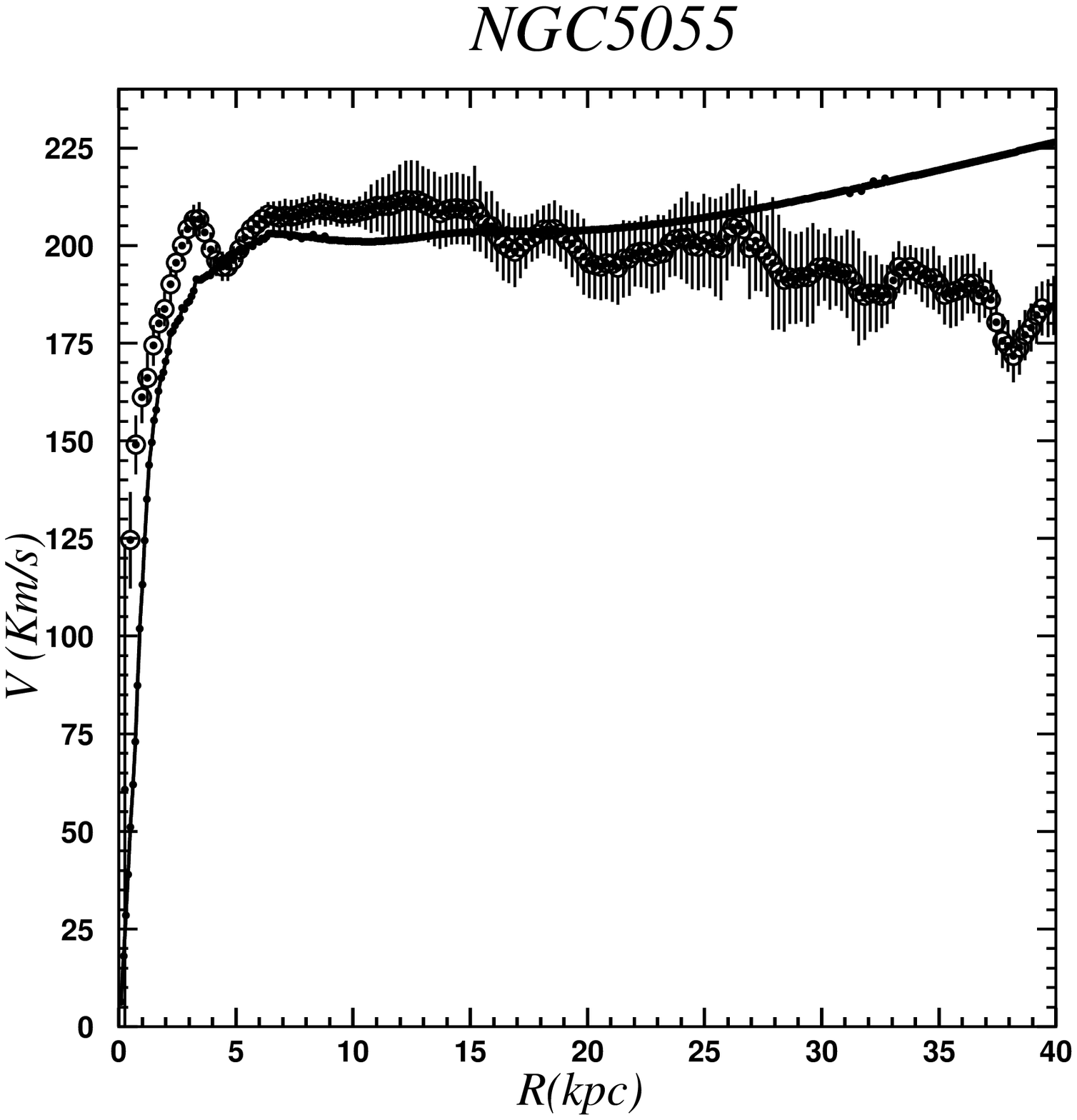}&
\includegraphics[width=55mm]{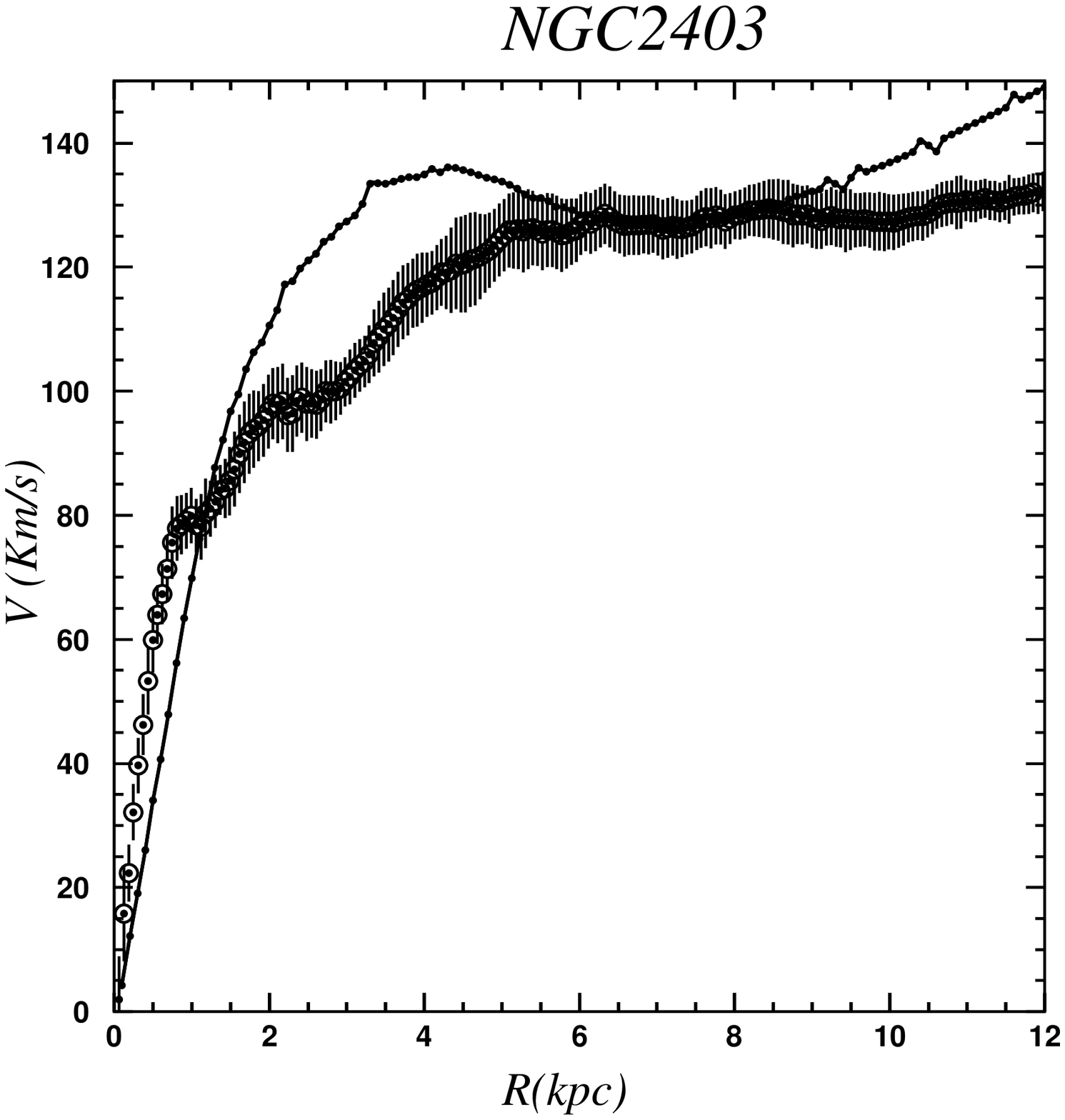}\\
\includegraphics[width=55mm]{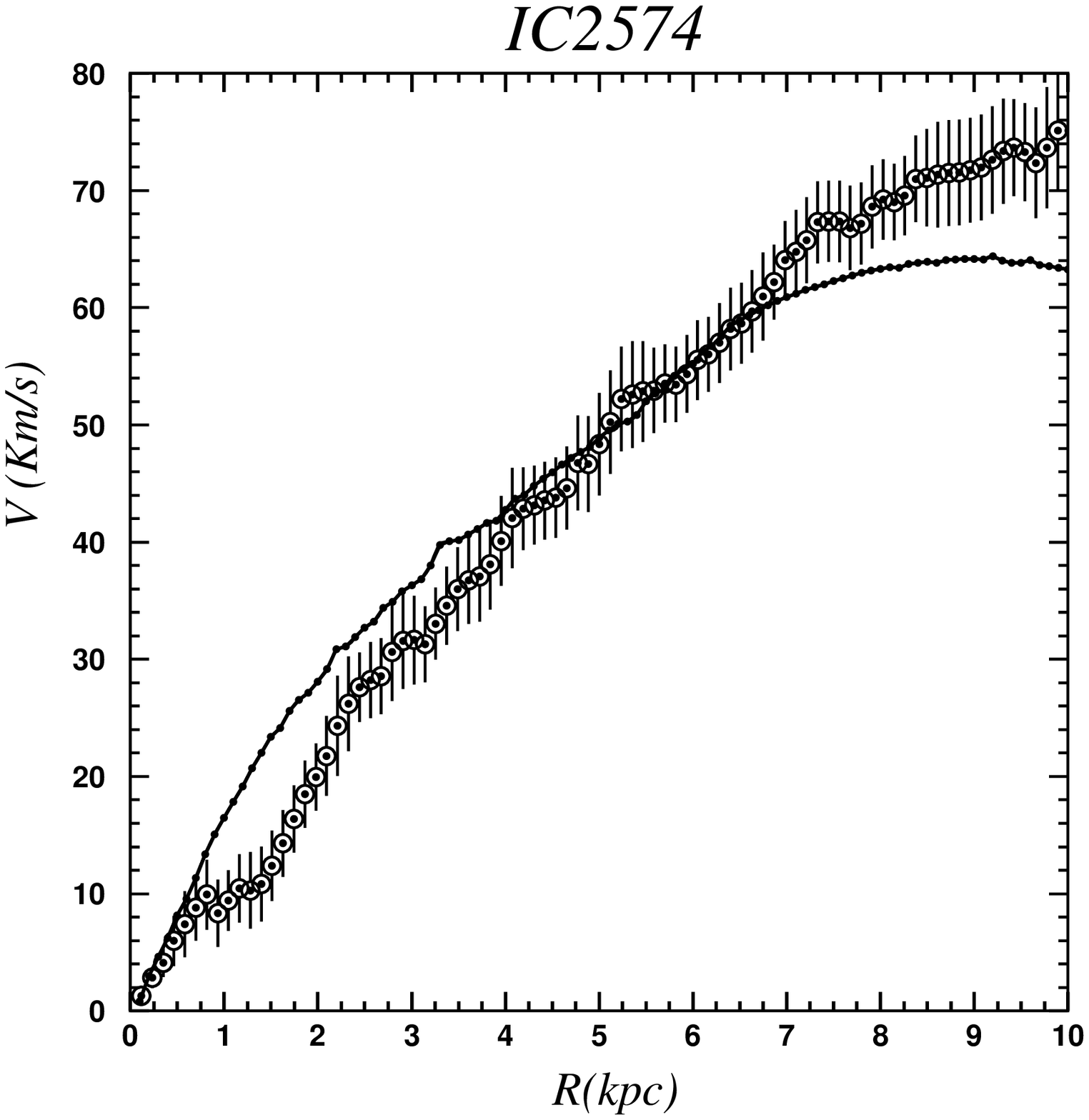}&
\includegraphics[width=55mm]{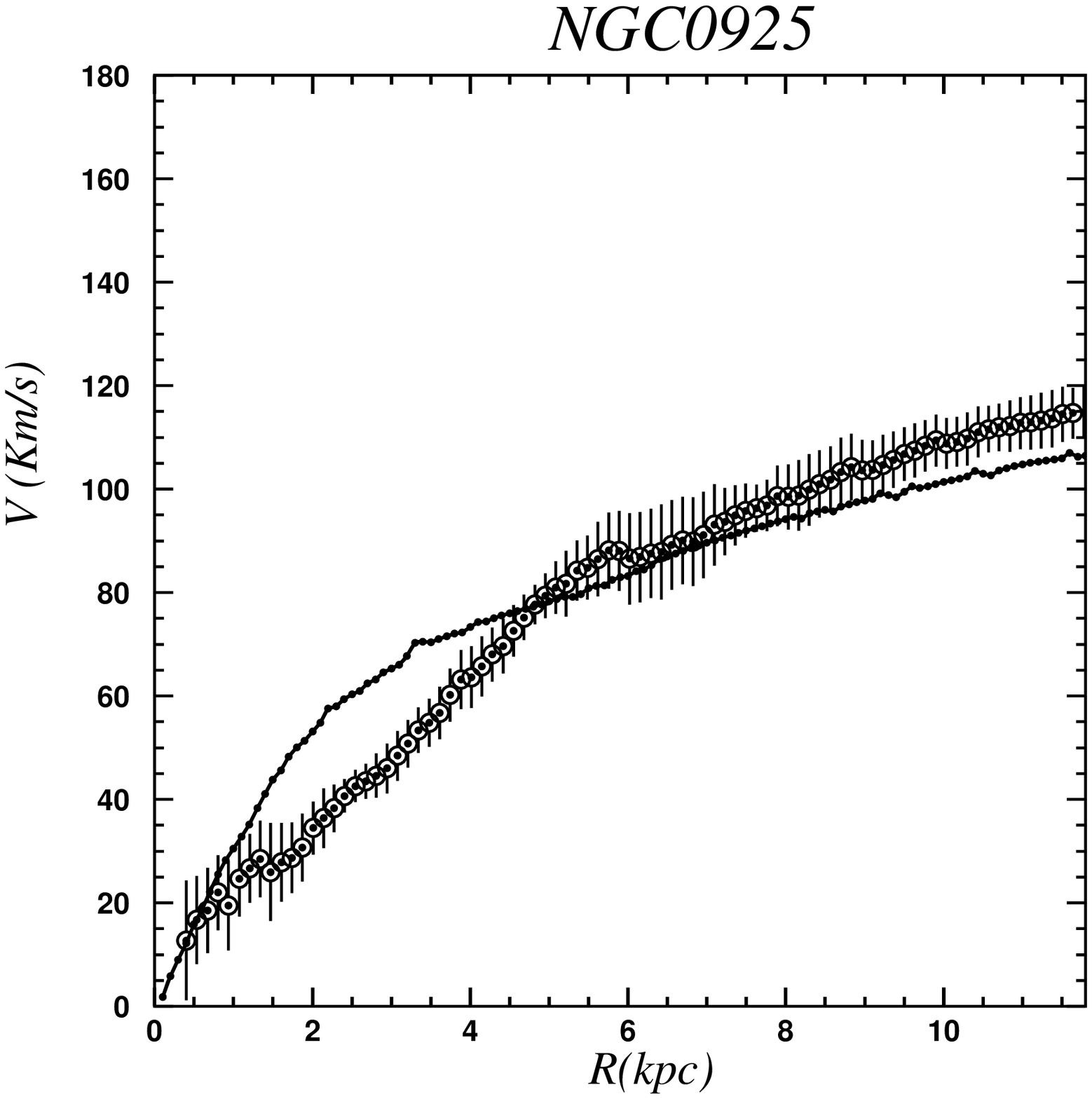}&
\includegraphics[width=55mm]{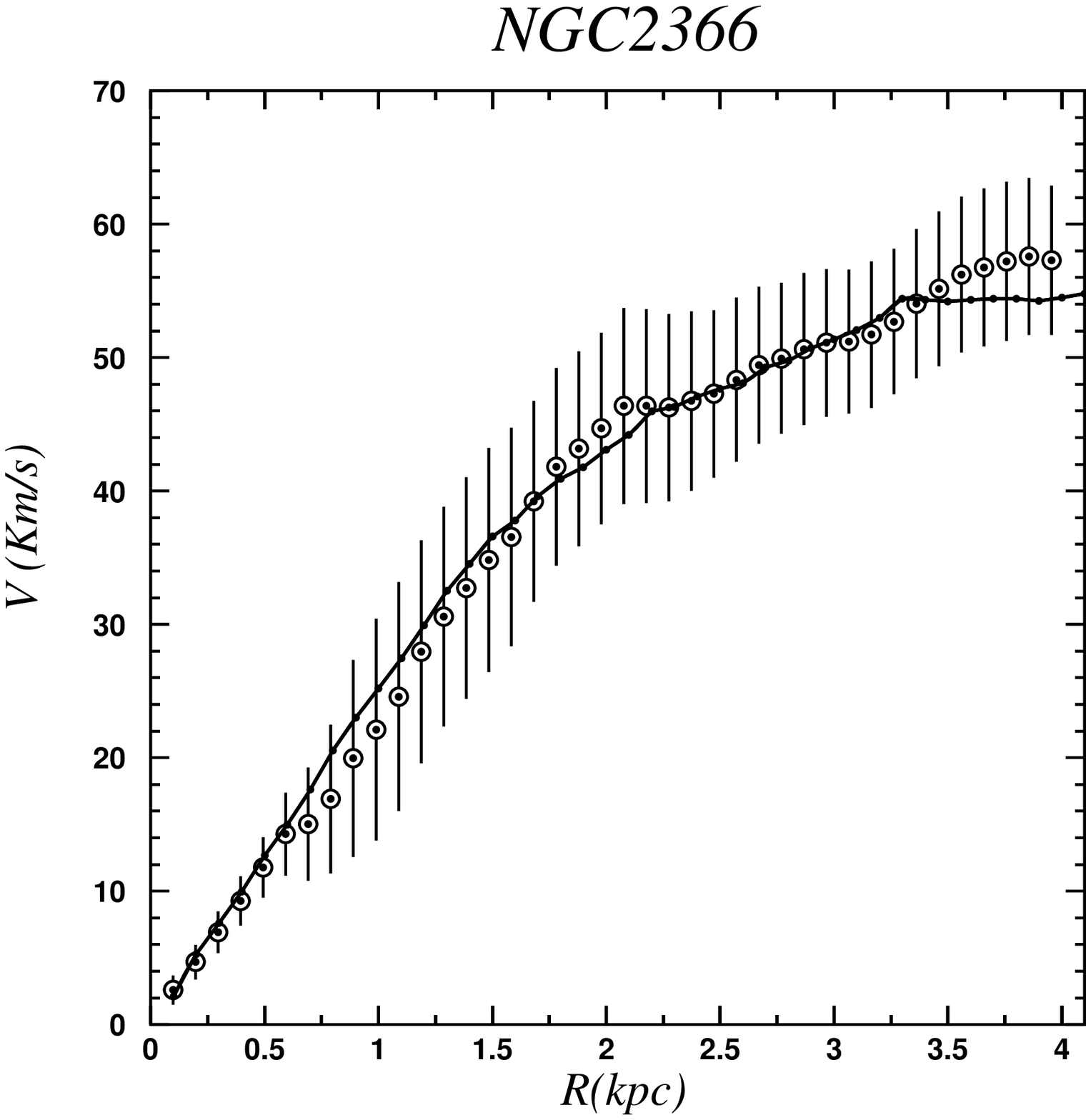}\\
\includegraphics[width=55mm]{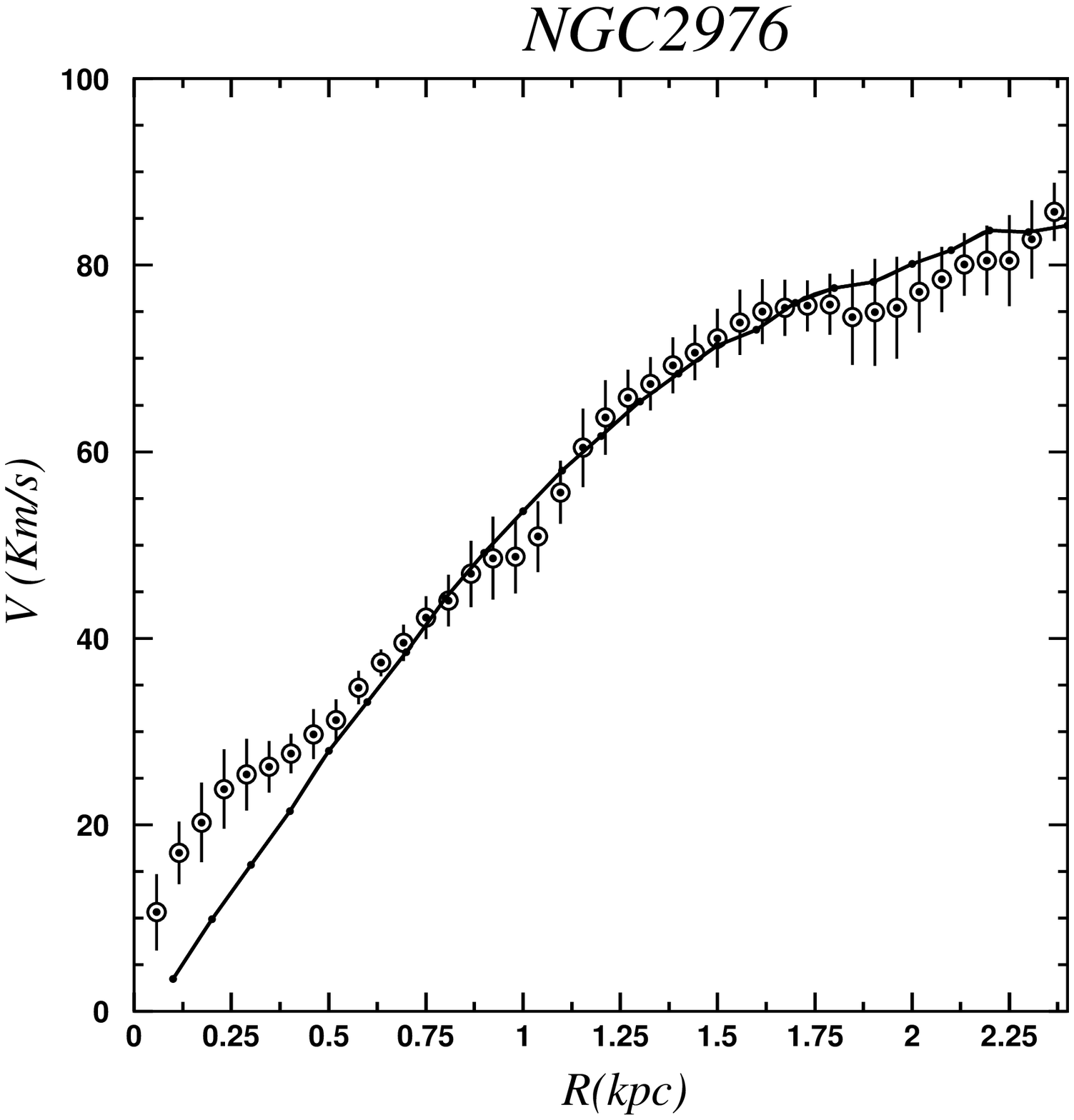}&
\includegraphics[width=55mm]{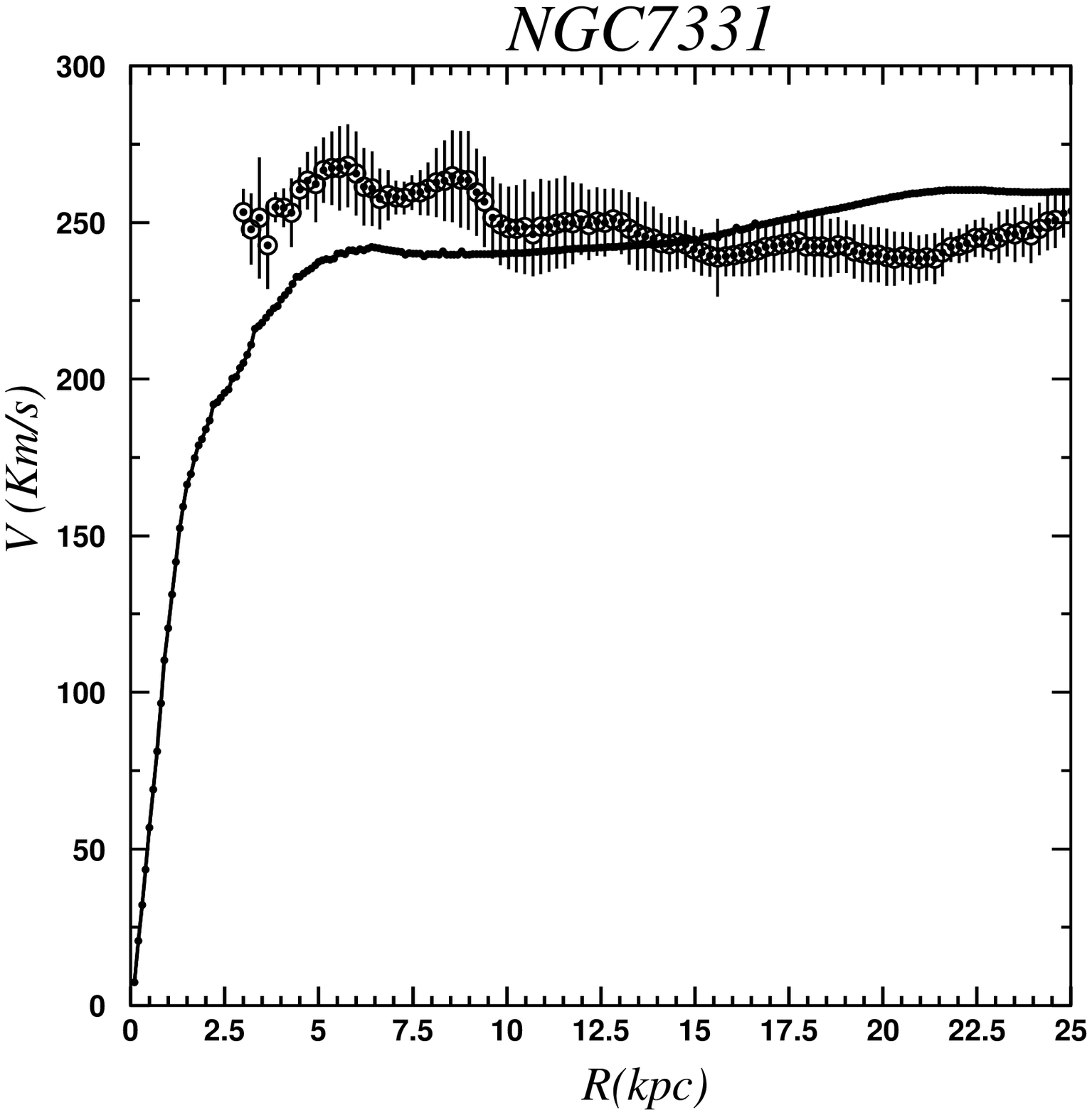} &
\includegraphics[width=55mm]{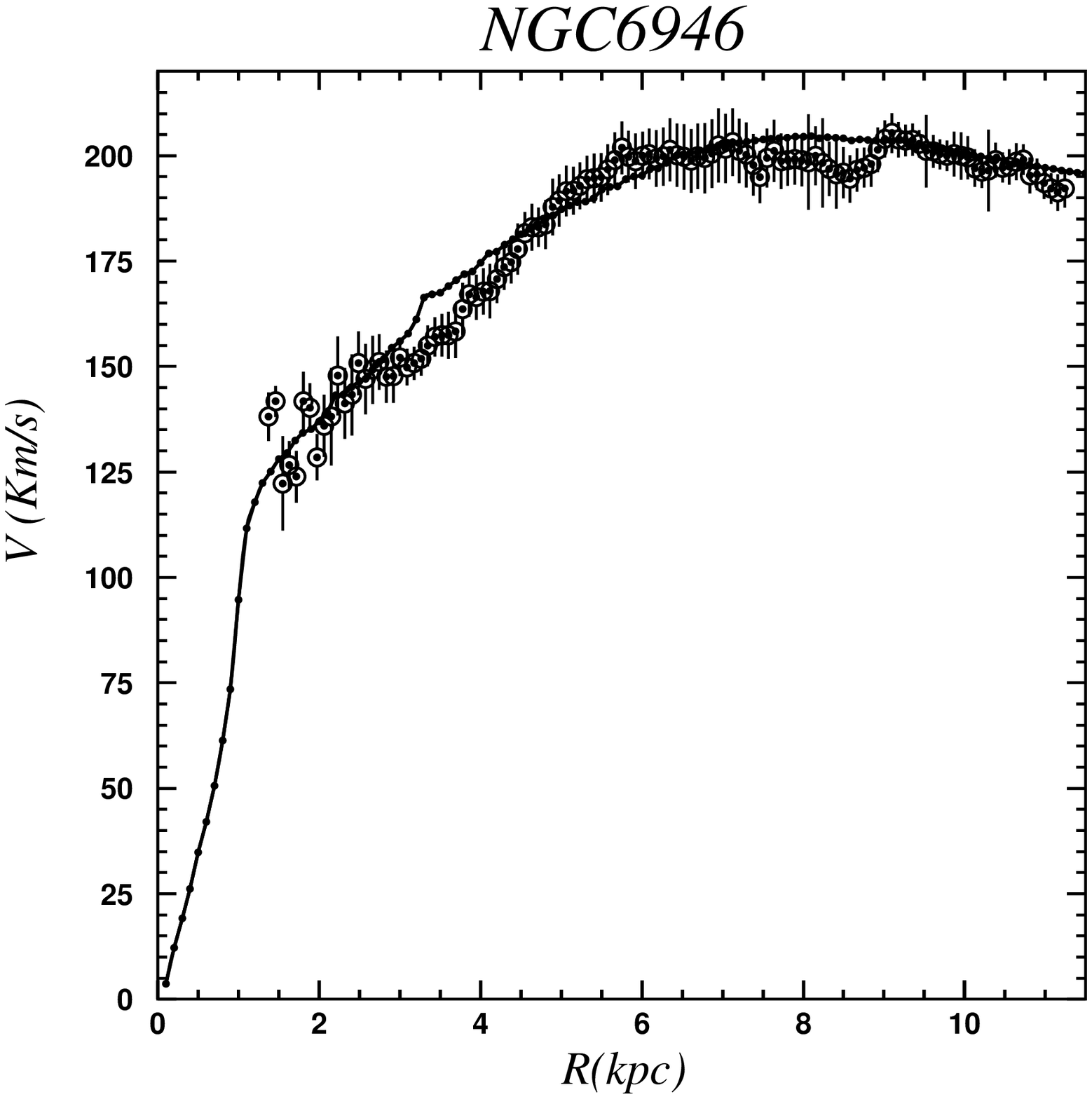}\\
\end{tabular}
\end{center}
\caption {The best fit to the rotation velocity curves of the
THINGs sample. We fix $\alpha =8.89$ and $\mu = 0.042~{\rm kpc}^{-1}$ and
let the stellar mass-to-light ratio $\Upsilon_\star^{3.6}$ be the
only free parameter. \label{fig5}}
\newpage
\end{figure*}

\begin{table*}
\begin{center}
\caption[]{ HSB and LSB galaxies from the set of Ursa-Major galaxies
\cite{Ursa1,Ursa2,Ursa3}. The columns are depicted as follows: (1)
name of the galaxy, (2) type of galaxy, (3) distance of the galaxy
from us, (4) the luminosity of the galaxy in the B-filter, (5) the
characteristic length of the galaxy, (6) mass of hydrogen, (7) the
overall mass of the galaxy calculated by $M_{disc} = \frac43 M_{HI} +
L_B\times\Upsilon_{\star}$, (8) the reddening-corrected
colour~\cite{sanders}, (9) internal extinction of galaxy in the B
band, (10) the best fit for the stellar mass-to-light ratio
$\Upsilon\star$, normalized to the solar value and, (11) the
normalized $\chi^2$ for the best fit to the data. \label{tab2} }
\begin{tabular}{|c|c|c|c|c|c|c|c|c|c|c|}
\hline\hline
       &      & Distance & $L_B$          & $R_0$ & $M_{HI}$          & $M_{disk}$         &  B-V        &  $A_B$          &$\Upsilon_{\star}$     &   $\chi^2$    \\
Galaxy & Type & (Mpc)    & ($10^{10}L_B$) &(kpc)  & $(10^{10}M_\odot)$& $(10^{10}M_\odot)$ & (mag)       &  (mag)        & $M_\odot/L_\odot$  &   $1/N.d.f$   \\
 (1)   & (2)  & (3)      & (4)            & (5)                       & (6)                &(7)          &(8)            & (9)                & (10)     & (11) \\
\hline
NGC 3726 &HSB & 17.4     & 3.340          &3.2    &0.60               &   4.00              &$0.45$      & 0.06          &$0.96^{+0.06}_{-0.06}$        & 1.66       \\
NGC 3769 &HSB & 15.5     & 0.684          & 1.5   &0.41               &   1.87              &  0.64      & 0.084           &$1.94^{+0.18}_{-0.18}$        & 1.60       \\
NGC 3877 &HSB &15.5      & 1.948          &2.4    &0.11               &   3.92              &$0.68$      & 0.084          &$1.94^{+0.12}_{-0.12}$        & 0.22       \\
NGC 3893 &HSB &18.1      &2.928           &2.4    & 0.59              &   5.09              & 0.56       & 0.077           & $1.47^{+0.12}_{-0.12}$       & 0.96       \\
NGC 3917 &LSB &16.9      &1.334           &2.8    &0.17               &   2.57             &$0.60$      & 0.077           &$1.76^{+0.09}_{-0.09}$        & 1.75       \\
NGC 3949 &HSB &18.4      &2.327           &1.7    & 0.35              &   2.93              &$0.39$      & 0.078          &$1.06^{+0.07}_{-0.07}$        & 0.63       \\
NGC 3953 &HSB &18.7     &4.236            &3.9    &0.31               &   8.97              &$0.71$      & 0.109          &$2.02^{+0.08}_{-0.08}$        & 1.63        \\
NGC 3972 &HSB &18.6     &0.978            &2.0   &0.13                &   1.89              &$0.55$      & 0.051          &$1.76^{+0.12}_{-0.12}$        & 3.43        \\
NGC 4010 &LSB &18.4     &0.883            &3.4  &0.29                 &   2.45             &  --        & 0.088          &$2.34^{+0.22}_{-0.22} $        & 0.8        \\
NGC 4013 &HSB &18.6     &2.088            &2.1  &0.32                 &   5.14              & $0.83$     & 0.060          &$2.26^{+0.06}_{-0.06}$       & 1.18        \\
NGC 4051 &HSB &14.6     &2.281            &2.3  &0.18                 &   3.45              &$0.62$      & 0.047          &$1.41^{+0.12}_{-0.12}$        & 1.59        \\
NGC 4085 &HSB &19.0     &1.212            &1.6  &0.15                 &  1.75              & $0.47$     & 0.066          &$1.28^{+0.18}_{-0.18}$       & 0.79        \\
NGC 4088 &HSB &15.8     &2.957            &2.8  &0.64                 &   4.66              & $0.51$     & 0.071          &$1.29^{+0.09}_{-0.09}$       & 0.59        \\
NGC 4100 &HSB &21.4     &3.388            &2.9  &0.44                 &   5.19              &$0.63$      & 0.084          &$1.36^{+0.05}_{-0.05}$        & 1.75        \\
NGC 4138 &LSB &15.6     &0.827            &1.2  &0.11                 &   3.45             &$0.81$      & 0.051          &$4.00^{+0.47}_{-0.47}$        & 0.10        \\
NGC 4157 &HSB &18.7     &2.901            &2.6  &0.88                 &   5.64              &$0.66$      & 0.077          &$1.54^{+0.10}_{-0.10}$        & 0.16        \\
NGC 4183 &HSB &16.7     &1.042            &2.9  &0.30                 &   1.92              &$0.39$      & 0.055           &$1.46^{+0.11}_{-0.11}$        & 0.25       \\
NGC 4217 &HSB &19.6     &3.031            &3.1  &0.30                 &   5.55              &$0.77$      & 0.063          &$1.70^{+0.07}_{-0.07}$        & 0.46        \\
NGC 4389 &HSB &15.5     &0.610            &1.2  &0.04                 &   0.73               & --        & 0.053          &$1.12^{+0.23}_{-0.23}$         & 2.59        \\
UGC 6399 &LSB &18.7     &0.291            &2.4  &0.07                 &   1.03              &--         & 0.061          &$3.24^{+0.32}_{-0.32}$         & 0.29        \\
UGC 6446 &LSB &15.9     &0.263            &1.9  &0.24                 &   0.72               & 0.39      & 0.059             &$1.54^{+0.19}_{-0.19}$        & 1.46        \\
UGC 6667 &LSB &19.8     &0.422            &3.1  &0.10                 &  1.14               & 0.65      & 0.058           &$2.40^{+0.20}_{-0.20}$         & 0.05        \\
UGC 6917 &LSB &18.9     &0.563            &2.9  &0.22                 &   1.70               & 0.53      & 0.098              & $2.52^{+0.18}_{-0.18}$        & 0.34        \\
UGC 6923 &LSB &18.0     &0.297            &1.5  &0.08                 &  0.61               & $0.42$    & 0.096          &$1.70^{+0.24}_{-0.24}$      & 0.85       \\
UGC 6930 &LSB &17.0     &0.601            &2.2  &0.29                 &  1.51                & 0.59      & 0.108            & $1.88^{+0.15}_{-0.15}$         & 2.20        \\
UGC 6983 &LSB &20.2     &0.577            &2.9  &0.37                 &   1.72               & 0.45      & 0.096              &$2.14^{+0.20}_{-0.20}$        & 0.44        \\
UGC 7089 &LSB &13.9     &0.352            &2.3  &0.07                 &   0.69                  & --     & 0.055              &$1.70^{+0.21}_{-0.21}$        & 0.35        \\
\hline
\end{tabular}
\end{center}
\end{table*}

\subsection{Ursa Major galaxies}
We adopt the best--fitting values of $\alpha$ and $\mu$, obtained from
the THINGS galaxies fitting process, as universal parameters and let
the stellar mass-to-light ratio $\Upsilon_\star$ of the galaxies be
the only free parameter. We find the best value of $\Upsilon_\star$
by fitting the observed rotation curves of galaxies with MOG. Figure
(\ref{fig3}) represents the observational data with the best fits to
the rotation curves of the galaxies. Table (\ref{tab2}) lists the
galaxies with the best stellar mass-to-the light ratio and the best
$\chi^2$ per degree of freedom for each galaxy. We have used in the fits to the observed
data $R$, $L_B$ and $M_H$ \cite{Ursa1,Ursa2,Ursa3} given in Table (\ref{tab2}). In this list
of results, we have three outliers: NGC3972 with $\chi^2 = 3.43$,
NGC4389 with $\chi^2 = 2.59$ and UGC6930 with $\chi^2 = 2.20$. For
the rest of the galaxies, we have very good results for the fitting
of the data with the average value of $\chi^2$ for all the galaxies
$\overline{\chi^2} = 1.07$. We also use again the THINGS catalogue and
let only the mass-to-light ratio of stars $M/L$ be the free
parameter. The best fits to the light curves are shown in Figure
(\ref{fig5}). The best value for $\Upsilon_\star^{3.6}$ with the
associated value of $\chi^2$ is shown in Table (\ref{tab3}).

\subsection{Stellar mass-to-light ratio and colour of galaxies}

In star formation scenarios, the stellar mass-to-light ratio is
related to the colour of galaxies~\cite{bell2001,bell2003}. This
relation depends on the details of the history of star formation and
the initial mass function (IMF). However, there are a number of
uncertainties due to the Stellar Population Synthesis (SPS) models
and the choice of IMF . Also due to dust in the interstellar medium
of galaxies, we may observe galaxies redder and fainter than their
actual colour and magnitude.

For the Salpeter mass function~\cite{salpeter}, the relation between
the mass-to-light ratio $\Upsilon_\star^B$ in the B band and for the colour of
galaxies is given by~\cite{bell2003}:
\begin{equation}
\log(\Upsilon_\star^B ) = 1.74(B-V )- 0.94 . \label{emlc}
\end{equation}
Using Kroupa's IMF, the slope of this function does not change.
However in equation (\ref{emlc}) the mass-to-light ratio shifts by
the amount $-0.35$ dex.

\begin{figure*}
\begin{center}
\begin{tabular}{cc}
\includegraphics[width=80mm]{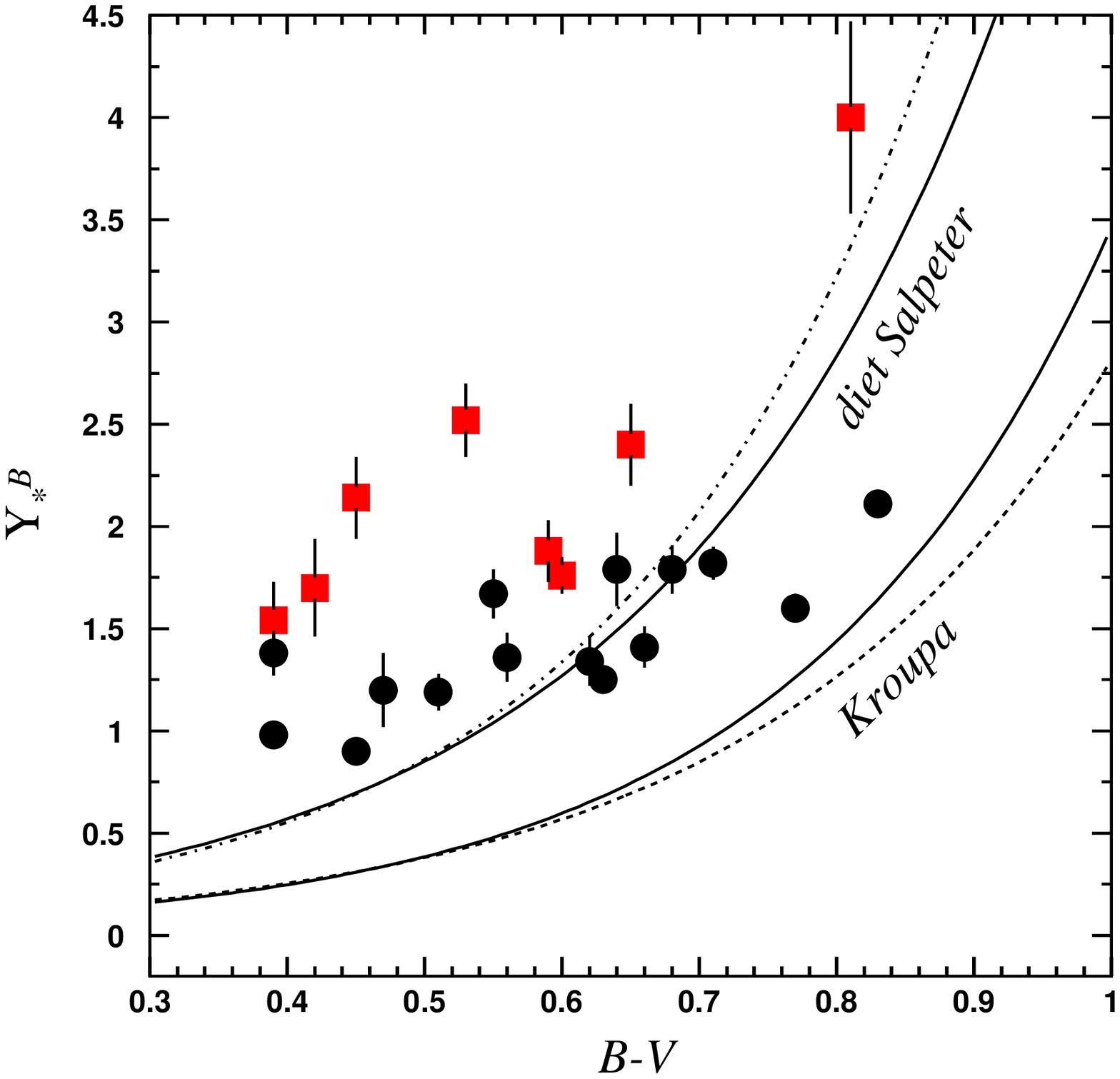} &
\includegraphics[width=80mm]{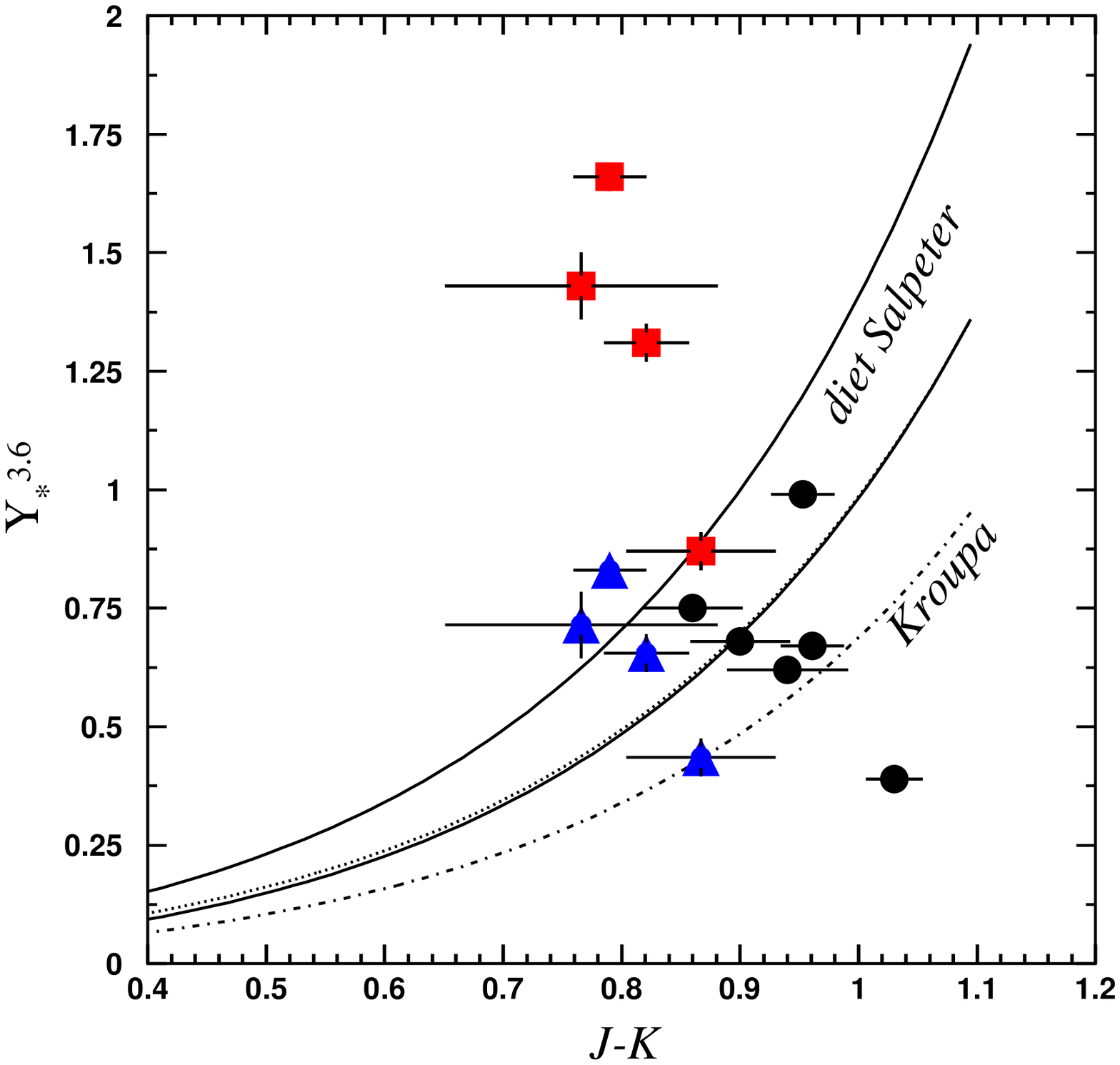}\\
\end{tabular}
\end{center}
\caption {Stellar mass-to-light ratio as a function of colour of the
galaxies in the Ursa-Major catalogue (left panel) and the galaxies in
the THINGS catalogue (right panel). For the Ursa-Major galaxies,
$\Upsilon_\star$ is given in the B band, while for the THINGS
galaxies it is in the $3.6\mu$m band. The theoretical models for
both the ``diet'' Salpeter and Kroupa IMF models are depicted as solid
lines with the margins to the models shown as dotted lines in both
panels. The black spots in both panels represent HSB galaxies and 
red squares represent LSB galaxies. In the right
panel (THINGS catalogue), we have normalized $\Upsilon_\star$ by factor of 2 for the
LSB galaxies represented by blue triangles. \label{colml} }
\end{figure*}

The relation between the mass-to-light ratios and the colour of
galaxies has been investigated in the longer wavelengths. The
advantage of longer wavelength is that the uncertainty in this
relation dramatically decreases near the infrared
(NIR). Here we adopt the results of the analysis of
the magnitudes of galaxies in the J, H and K bands~\cite{bell2001}
as well as the observations in the $3.6\mu$m band. From the SPS
models the relation between the mass-to-light ratio in the K band
and the colour in the $\rm {J-K}$ band is given by~\cite{bell2001}:
\begin{equation}
\label{mlc}
 \log(\Upsilon_\star^K) = 1.43 (J-K)-1.38.
\end{equation}
On the other hand, from the relation between $\Upsilon_\star^K$ and $\Upsilon_\star^{3.6}$~\cite{oh}:
\begin{equation}
\Upsilon_\star^{3.6} = 0.92\Upsilon_\star^K - 0.05, \label{kband}
\end{equation}
we can relate $\Upsilon_\star^{3.6}$ to the {\rm J-K} band. Again for the case of
Kroupa's IMF, we decrease the constant term in equation (\ref{mlc}) by
the amount of $0.15$.

Finally, in order to compare the mass-to-light ratio derived from
MOG with the stellar synthesis models, we plot in Figure
(\ref{colml}) the mass-to-light ratio both for the Ursa-Major
galaxies in the {\rm B} band and the THINGS catalogue in the $3.6\mu$m band and
compare the result with equations (\ref{emlc}) and (\ref{mlc}).  We
note that the colours and the magnitudes of galaxies in the
Ursa-Major galaxies are extinction corrected. Also in order to study
the behaviour of galaxies based on their types, we divide galaxies in
this plot into two classes of HSB and LSB galaxies. For the LSB
galaxies we have larger values of mass-to-light ratios compared to the
HSB galaxies. This effect also has been reported by fitting rotation
curves of galaxies with a dark matter model~\cite{verh99}. The
differences between the mass-to-light ratios for the HSB and LSB galaxies
has been studied in~\cite{zwaan}, where for the the LSB galaxies
$\Upsilon_\star$ is twice as big for the HSB galaxies.

While the physical correlation between $\Upsilon_\star$ and the
colour of galaxies has been proved, there are still uncertainties in
the analytical relation between these two parameters. One of the
uncertainties in equation (\ref{mlc}) is the initial mass function
(IMF) of the stars. Bell \& de Jong (2010) showed that to have
stellar disks consistent with the dynamics, the so-called ``diet''
Salpeter IMF has to have the stars' masses reduced below $0.35
M_\odot$. Relation (\ref{mlc}) corresponds to the diet Salpeter IMF.
In this case, the stellar mass has to be reduced by a factor of
$0.7$. Moreover, near infrared observations provide more reliable
values for the stellar mass-to-light ratio than the visual band
observations. In Figure (\ref{colml}), the stellar mass-to-light
ratio obtained from MOG in the $3.6\mu$m band is more compatible
with the theoretical model. By dividing $\Upsilon_\star$ for the LSB
galaxies by 2, we get results compatible with the theoretical model.

\begin{table*}
\begin{center}
\caption[]{Results obtained from fitting galaxies in the THINGS
catalogue with the MOG rotation curves for the case of the fixed
parameters $\alpha = 0.89$ and $\mu = 0.042~{\rm kpc}^{-1}$. The columns of this
table are as follows: (1) name of the galaxy, (2) the type of galaxy,
(3) distance of the galaxy, (4) colour of the galaxy in the $(J-k)$
band~\cite{Jar03}, (5) the stellar mass-to-light ratio $\Upsilon_\star^{3.6}$ in the $3.6\mu$m band,
derived from MOG, (6) the reduced
$\chi^2$. \label{tab3} }
\begin{tabular}{|c|c|c|c|c|c|c|}
\hline\hline
       &       & Distance & $J-K$   &  $\Upsilon_\star^{3.6}(MOG)$       & $\chi^2/N_{d.o.f}$  \\
Galaxy & Type  & $(Mpc)$  &         & $(M_\odot/L_\odot)$              &                     \\
 (1)  &  (2)   & (3)      & (4)     & (5) &(6) \\
\hline
NGC 3198 & HSB & $13.8 $  &  $0.940\pm0.051$   &     $0.63\pm0.01$          & 1.24                  \\
NGC 2903 & HSB & $8.9$    &  $0.915\pm0.024$   &     $2.37\pm0.03$           &  2.10                       \\
NGC 3521 & HSB & $10.7$   &  $0.953\pm0.027$   &     $0.99\pm0.02$              & 3.02                         \\
NGC 3621 & HSB & $6.6$    &  $0.860\pm0.042$   &     $0.76\pm0.01$           &   1.64                       \\
NGC 5055 & HSB & $10.1$   &  $0.961\pm0.027$   &     $0.67\pm0.01$          &    4.28                      \\
NGC 2403 & LSB & $3.2$    &  $0.790\pm0.031$   &     $1.68\pm0.01$          &    7.78                     \\
IC 2574  &LSB  &4.0       &  $0.766\pm0.115$    &     $1.43\pm0.07$            &  2.24                     \\
NGC 0925 &LSB  &9.2       &  $0.867\pm0.063$   &     $0.87\pm0.04$           &  3.67                        \\
NGC 2366 &LSB & 3.4       &  $0.667\pm0.146$    &     $2.76\pm0.23$               & 0.08    \\
NGC 2976 &LSB & 3.6       &  $0.821\pm0.036$    &     $1.31\pm0.04$               & 1.43     \\
NGC 7331 &HSB & 14.7      &   $1.03\pm0.024 $    &    $0.39\pm0.01 $              & 4.11    \\
NGC 6946 &HSB & 5.9       & $0.90\pm0.042$    &  $0.68\pm0.01$              &   1.20  \\
\hline
\end{tabular}
\end{center}
\end{table*}

\subsection{Tully--Fisher Relation}
Observations by Tully and Fisher (1977) showed that there is an
empirical relation between rotation curves of galaxies and their
luminosities, $v_c^4 \propto L$. We want to test whether this
relation is satisfied in MOG. The right-hand side of the
Tully-Fisher relation (absolute magnitude) can be obtained from
observations, using the apparent magnitude, distance and extinction
factor. On the other hand, the rotation curves of galaxies, as we
discussed in the pervious sections, are measured. Here we adopt the
rotation velocities of galaxies from the best fits of MOG to the
data, given in the last sections. For the the Tully-Fisher relation,
we can use both the maximum rotation curve of the galaxies,
$V_{max}$, and the flat rotation curve, $V_{flat}$.

\begin{figure}
\begin{center}
%\begin{tabular}
\includegraphics[width=80mm]{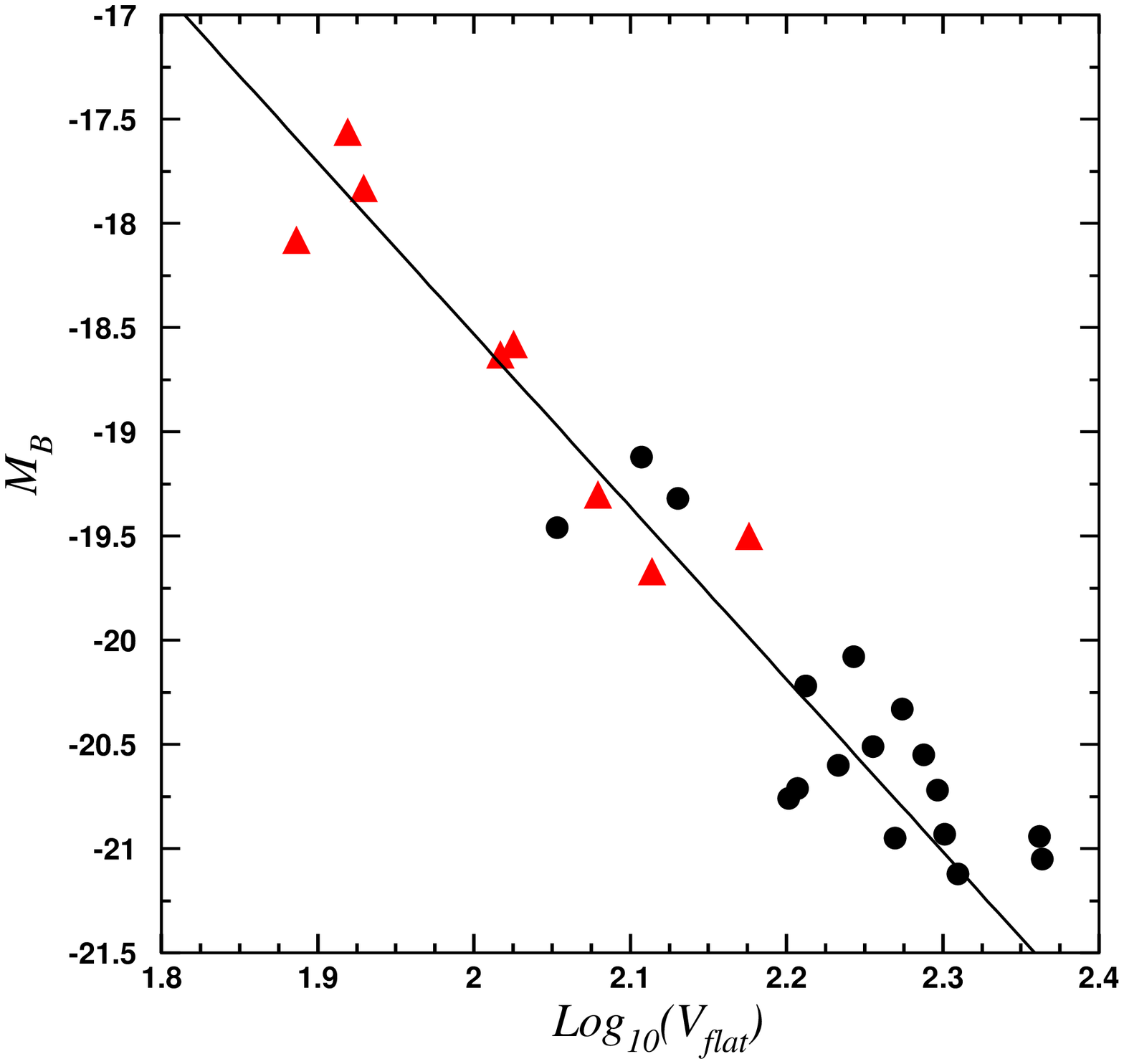}
%\end{tabular}
\end{center}
\caption{The absolute magnitude of galaxies in the Ursa-Major
catalogue as a function of the logarithm of the flat rotation curve.
This is the Tully-Fisher relation. The black spots
represent the HSB galaxies and the red triangles represent the LSB
galaxies. The flat rotation curves in this figure are taken from the
MOG fits to the data. The best fit to the data is given by $M =
-8.27 \times \log(V_{flat}) - 1.99$. \label{figtf}}
\end{figure}

In order to calculate the flat rotation curve, we adopt the
convention in~\cite{ver2001}: (a) for the galaxies with a rising
rotation curve, $V_{flat}$ cannot be measured, (b) for the galaxies
with a flat rotation curve, $V_{flat} = V_{max}$, (c) for galaxies with
a declining rotation curve $V_{flat}$ is calculated from averaging the
outer parts of the galaxy. Figure (\ref{figtf}) displays the
distribution of galaxies in terms of apparent magnitude versus
the logarithm of the flat rotation curve. The best fit to the data is given
by the apparent magnitude:
\begin{equation}
M_b = - 8.27 \times \log_{10}(V_{flat}) - 1.99. \label{tuequation}
\end{equation}
The slope of this function is compatible with the observational data
analysed by Verheijen (2001). For various samples of galaxies the
slope of the observed data in equation (\ref{tuequation}) changes
from $-8.7\pm 0.3$ to $-9.0\pm 0.4$.

\section{Conclusions}
We have developed the weak field approximation of MOG as an
alternative to dark matter models. MOG is a covariant modified
gravity theory which contains tensor, vector and scalar fields in
the action. The non-linear field equations in MOG and the equation
of motion for a massive test particle were applied to the study of
the dynamics of galaxies. The modified equation of motion contains
an extra contribution from the gradient of the vector field
$\phi_\mu$, proportional to the fifth force charge $q_5$, which is
related to the inertial mass, $q_5=\kappa m$.

We expanded the fields in the MOG action around Minkowski space--time
and combined the test particle equation of motion with the field
equations. An effective potential for an arbitrary distribution of
matter was obtained.  For any extended object this effective potential is
composed of an attractive and a repulsive Yukawa contribution. It contains two
free parameters $\alpha$ and $\mu$, where $1/\mu$
is the characteristic length scale associated with the vector field $\phi_\mu$.
We have shown that the effective potential at small and large
scales is given by the Newtonian potential, but with different
effective gravitational constants.

To test the observational consequences of the effective potential,
we used the well measured THINGS catalogue of galaxies to fit the
theoretical rotation curves predicted by MOG to the observed data.
This catalogue of galaxies contains both LSB and HSB galaxies. For
this set of galaxies, we let the three parameters of the model,
$\alpha$, $\mu$ and the stellar mass-to-light ratio, $\Upsilon_\star$
be the free parameters of the theory. The best values from the
combined likelihood functions yielded $\alpha = 8.89\pm 0.34 $ and
$\mu = 0.042 \pm 0.004~{\rm kpc}^{-1}$. As for the second step, we used a larger
set of Ursa-Major galaxies and with the fixed universal parameters $\alpha =
8.89\pm 0.34 $ and $\mu = 0.042 \pm 0.004~{\rm kpc}^{-1}$, we let the mass-to-light ratio $\Upsilon_\star$ of
the galaxies be the only free parameter in the fitting of the data.
We obtained excellent fits for the velocity rotation curves to the observational data with the average value $\chi^2/N_{\rm d.o.f}=1.07$.

As a prediction of MOG, we compared the deduced stellar
mass-to-light ratios of galaxies with their colours both for the THINGS and
the Ursa-Major galaxies. Depending on the Initial Mass Function
(IMF) the theoretical relation between these two parameters can
change. In addition there is more uncertainty in the shorter
wavelengths data compared to the Near Infrared wavelengths data. Our
MOG prediction in the infrared wavelengths ($3.6\mu$m) was
consistent with the predictions of theoretical astrophysical models.
The advantage of this result is that, knowing the colour of galaxies
in the infrared ($J-K$), we obtain valuable information about the
stellar mass to light ratios $M/L$. On the other hand, from the
luminosity of galaxies in the visual band and the HI radio emission
data, we can obtain the baryonic masses of the galaxies. From this
information, we are able to calculate the dynamics of galaxies and
other large-scale systems, such as clusters of galaxies and the
merging of galaxies {\it without any free parameters}. This means
that there will be no free degrees of freedom in MOG when
calculating the dynamics of astrophysical systems.

In dark matter models in which galaxies are fitted with a
dark matter spherical halo, it has not been possible so far to
obtain parameter-free fits to rotation velocity curve data. Dark
matter profiles require at least two free parameters for each galaxy, in addition to the
stellar mass-to-light ratio, to enable fits to
rotation curve data. This is in contrast to our results for MOG
which yields excellent fits to rotation curve data
with only one free parameter, $M/L$.

Moreover, we can successfully predict the Tully-Fisher relation for
galaxies because of the direct relationship between rotation curves
and luminous matter.  This is not possible in standard dark matter
models because there is no relation between the dominant dark matter and the stellar luminosity of galaxies.
We used the values of the flat rotation curves of galaxies
predicted by MOG and plotted the absolute magnitudes of the galaxies
in terms of $\log(V_{flat})$. The best fit to the data has a slope
of $-8.27$, which is in good agreement with the observed data. Our
analysis from the solar system scale to the galactic scales showed
that MOG is a consistent covariant modified gravity theory without
exotic dark matter, and for these scales we can replace the
non-linear MOG field equations with an effective weak field
gravitational potential, which can be easily adapted to any
astrophysical system.

\label{conc}

\section*{Acknowledgements}
The John Templeton Foundation is thanked for its generous support of
this research. The research was also supported by the Perimeter
Institute for Theoretical Physics. The Perimeter Institute was
supported by the Government of Canada through Industry Canada and by
the Province of Ontario through the Ministry of Economic Development
and Innovation. We also would like to thank Niayesh Afshordi, Viktor
Toth and Martin Green for helpful discussions and comments.  This work made 
use of THINGS, "The HI Nearby Galaxy Survey" \cite{walter2008}.

\end{document}